\documentclass{article}
\newcommand\eq[1] {(\ref{#1})}

\newcommand{\beqa}{\begin{eqnarray}}
\newcommand{\eeqa}[1]{\label{#1}\end{eqnarray}}
\newcommand{\bequ}{\begin{equation}}
\newcommand{\eequ}[1]{\label{#1}\end{equation}}

\newcommand{\beq}{\begin{equation}}
\newcommand{\eeq}{\end{equation}}
\newcommand{\overliner}{\begin{eqnarray}}
\newcommand{\earr}{\end{eqnarray}}
\newcommand{\beqn}{\begin{equation*}}
\newcommand{\eeqn}{\end{equation*}}
\newcommand{\overlinern}{\begin{eqnarray*}}
\newcommand{\earrn}{\end{eqnarray*}}

\usepackage{graphics,graphicx}
\usepackage{amsmath}
\usepackage{mathtools}
\usepackage{epsf}
\usepackage{array}
\usepackage{xcolor}
\usepackage{amsfonts}
\usepackage{amssymb}
\usepackage[english]{babel}
\usepackage[applemac]{inputenc}            
\usepackage{color}
\usepackage{multirow}
\usepackage{bm}
\usepackage{enumitem}
\usepackage[cal=boondoxo]{mathalfa}
\usepackage{caption}
\usepackage{subcaption}
\usepackage{parskip}
\usepackage{physics}
\usepackage{float}

\setlist[itemize]{leftmargin=*}
\usepackage[toc,page]{appendix}

\usepackage[hidelinks]{hyperref}

\usepackage[toc,page]{appendix}

\PassOptionsToPackage{hyphens}{url}\usepackage{hyperref}

\usepackage[export]{adjustbox} 
    \usepackage{etoolbox}

\usepackage{tabu}

\numberwithin{equation}{section}

\usepackage{mathtools}

\usepackage[numbers,sort&compress]{natbib}

\DeclareMathAlphabet{\mathpzc}{OT1}{pzc}{m}{it}

\usepackage{calrsfs}
\DeclareMathAlphabet{\pazocal}{OMS}{zplm}{m}{n}

\usepackage{caption}
\usepackage{color}

\usepackage[symbol]{footmisc}

\begin{document}

\title{{\bf Coupled dynamics of chiral waves and gyroscopic systems with applications to atmospheric phenomena
 }}

\author{{\bf Alessio Kandiah\thanks{Corresponding author: sgakandi@liverpool.ac.uk}, Ian S. Jones, Natasha V. Movchan \& Alexander B. Movchan}
\\
\\
 {University of Liverpool, Department of Mathematical Sciences,  Liverpool, L69 7ZL, UK}}

\maketitle

\begin{abstract}
\noindent The present paper introduces the notion of chiral gravitational elastic waves and explores their connections to equatorial and planetary waves. The analysis of the gravity-induced waveforms in gyroscopic systems composed of gyropendulums provides important insights into the dynamics of waves in the vicinity of the equatorial belt. We show that the direction of motion of the chiral waveforms can be controlled by choosing the orientation of the spinners. The presence of gravity is shown to affect the stop band frequencies for such structures, providing an additional control parameter for the chiral waveguides. The effect of the Coriolis force is demonstrated for chiral continuum models describing waves in the equatorial region and the polar regions on a rotating sphere. Additional asymptotic features of equatorial waves are presented in this paper. We show that the shape of a ridge of a polar vortex can be approximated by the governing equations of a gyropendulum. The theoretical work is accompanied by illustrative examples. 
\end{abstract}

\section{Introduction}\label{asiodhad1zcaoish}

 In this paper, we introduce a discrete model of a chiral elastic lattice strip consisting of gyroscopic pendulums. We demonstrate that gravity-induced waves can propagate through the discrete strip linking them to equatorial waves. In atmospheric dynamics, the Coriolis force arising from the rotation of the planet is crucial in the analysis of wave motions; an analogy between the Coriolis force in continuous media and the gyroscopic force in discrete structures is provided in this paper. A novel analysis of equatorial waves is presented in the context of the shallow water equations, which is applicable to studying waveforms on a thin strip. The interplay of chirality and gravity emerges as significant factors in studying atmospheric wave phenomena. We also demonstrate that the motion of a single gyropendulum, which was studied in \cite{kandiah2023effect, kandiah2024controlling, CMDS14}, can follow similar patterns to that of atmospheric flows on the poles of rotating planets; this analogy arises from the fundamental combined characteristics of gyroscopic and gravitational forces.

The paper is organised as follows. In Section \ref{asdjpasdj08212} we provide a preliminary introduction to discrete lattice systems subjected to gravity. In Section \ref{3dring1}, we introduce a three-dimensional finite chiral structure subjected to gravity, and examine its eigenfrequencies and eigenmodes using an independent finite element model built in COMSOL Multiphysics $6.1.$ Section \ref{discretelatstrip} presents a new mathematical model of a gyro-elastic strip with gyroscopic pendulums, accompanied by the illustrative examples. Here, we explore how the combined presence of gyroscopic spinners and gravity determine the strip vibrations for two different boundary conditions applied to the outer boundaries. The detailed analysis of the nodal point trajectories for the Dirichlet lattice strip is provided in Section \ref{akjshdfg}. Moreover, the discussion of the dispersion properties of the Dirichlet strip is supplied in Section \ref{newdispersionproper1}. In Section \ref{sdoublepend} we give some analytical details of the dispersion characteristics for an active Dirichlet lattice strip. Following the analysis of the strip with the Dirichlet boundary conditions considered in Sections \ref{discretelatstrip} and \ref{newdispersionproper1}, we study the case of Neumann boundary conditions in Section \ref{class}. In Section \ref{nochiralitysec}, the standing and propagating waveforms in the non-chiral Neumann lattice strip are discussed. We examine the dispersion diagrams for various configurations of the strip, and the important applications to atmospheric wave phenomena in equatorial regions are discussed. In Section \ref{rotatingshall1} we present asymptotic analytical solutions for the shallow water equations in a narrow band, and relate the resulting wave motions to equatorial waves. The assumptions of the boundary conditions align with the formulation of the governing equations. We also examine equatorial wave dispersion diagrams and provide examples of eigenfunctions in the vicinity of the equator. The asymptotic analysis of the meridional velocity modes in the equatorial region is presented in Section \ref{rotatusabcd1}, which includes the examination of the corresponding dispersion relations. The analogies between equatorial waves in a continuum and elastic waves in the discrete lattice strip are detailed in a new framework. Lastly, Section \ref{temporinterfacegyro} provides a formal description of the gyropendulum dynamics linked to the shallow water equations, with an emphasis on the approximate polygonal trajectories related to atmospheric flows in polar regions of rotating planets.

\section{Discrete lattice systems in equatorial regions}\label{asdjpasdj08212}
 The analysis of waveguides in multi-scale structured systems yields the emergence of chiral vortex waves \cite{carta2019wave, carta2020one, carta2020chiral, garau2018interfacial, carta2017deflecting}, highlighting the role of gyroscopic elements in discrete chiral structures. In this case, the spinners create a gyroscopic coupling of the velocity components, providing an approximation of the rotational behaviours in atmospheric wave phenomena. The work presented in this section provides a new mathematical framework for modelling wave propagations within narrow equatorial bands. We show that large-scale chiral wave motions in the equatorial region can be approximately analysed using a discrete lattice strip consisting of gyropendulums. The Bloch-Floquet method is used to examine wave propagation in the discrete lattice models.

 \subsection{Finite element model of the chiral belt}\label{3dring1} 
We discuss vibration modes of a finite three-dimensional belt composed of nodal points, connected by elastic links, and subjected to gyroscopic and gravitational forces;  the simulation being produced in COMSOL Multiphysics $6.1$. 
The main feature of this structure is the variable action of the gyroscopic forces within the belt. The structure consists of three horizontal circular rings of nodal points. There are no gyroscopic actions on the nodal points in the central ring, while the gyroscopic effects acting on the nodal points in the upper and lower rings are the same in magnitude and opposite in sign.  In nature, there is an example of a similar gyroscopic system: it is the pair of  Hadley cells in the equatorial region of the planet Earth, where the Coriolis force changes sign when crossing the equator.

In the finite element model, the gyroscopic action in the structure is taken into account by introducing a force proportional to the nodal point displacement at each junction. The gyroscopic force acting on the $i$-th nodal point is defined by ${\bf F}_{i}= m_{i}({\bf \Omega}_{i} \times {\bf v}_{i}),$ where $m_{i}$ is the mass, ${\bf v}_{i}$ is the velocity vector and ${\bf \Omega}_{i}= \Omega (\cos(\theta_{i}), \sin(\theta_{i}), 0)$ is the gyricity vector with $\Omega$ representing the gyricity parameter. Here we describe the gyroscopic effect through the generic gyricity parameter $\Omega.$ However, in Sections \ref{rotatingshall1}-\ref{temporinterfacegyro}, the notation $\Omega$ will be used for the angular speed of the Earth and will be connected to the Coriolis force by ${\bf F} = - 2 m ({\bf \Omega} \times {\bf v}'),$ with $m$ and ${\bf v}'$ denoting the mass and velocity vector of the moving particle relative to the non-inertial system of coordinates, respectively, and ${\bf \Omega}$ representing the vector of rotation of the body. The angle $\theta_{i}$ denotes the azimuth angle in the $xy$-plane of the $i$-th nodal point relative to the centre of the corresponding ring from the $x$-axis, with $0 \leq \theta_{i} < 2 \pi.$ In particular, since there are no gyroscopic forces acting on the nodal points in the central ring, then $\Omega = 0$ rad/s for such points. The gyricity parameters for the nodal points in the lower and upper rings are $\Omega$ and $-\Omega,$ respectively, as discussed above.  Moreover, gravity is modelled as an additional force on each nodal point acting radially inwards towards the centre of each horizontal ring. In the FE model, each nodal point is connected to its nearest neighbouring nodal point by massless and elastic links. We compute the eigenfrequencies and eigenmodes of the three-dimensional structure and show the existence of linearly tangential oscillations along the central ring, with either eastward or westward propagations of the individual wave components. These modes are linked to the standing modes of an infinite discrete two-dimensional chiral strip, with zero group velocity and non-vanishing phase velocity. This characteristic resembles the zonally propagating equatorial waves detailed in Section \ref{rotatingshall1}, which includes the analysis of waves in narrow equatorial bands described by the shallow water equations.  

\begin{figure}[H]
  \centering
  \begin{minipage}[b]{0.46\textwidth}
    \hspace{-0.15cm}\includegraphics[width=1\linewidth]{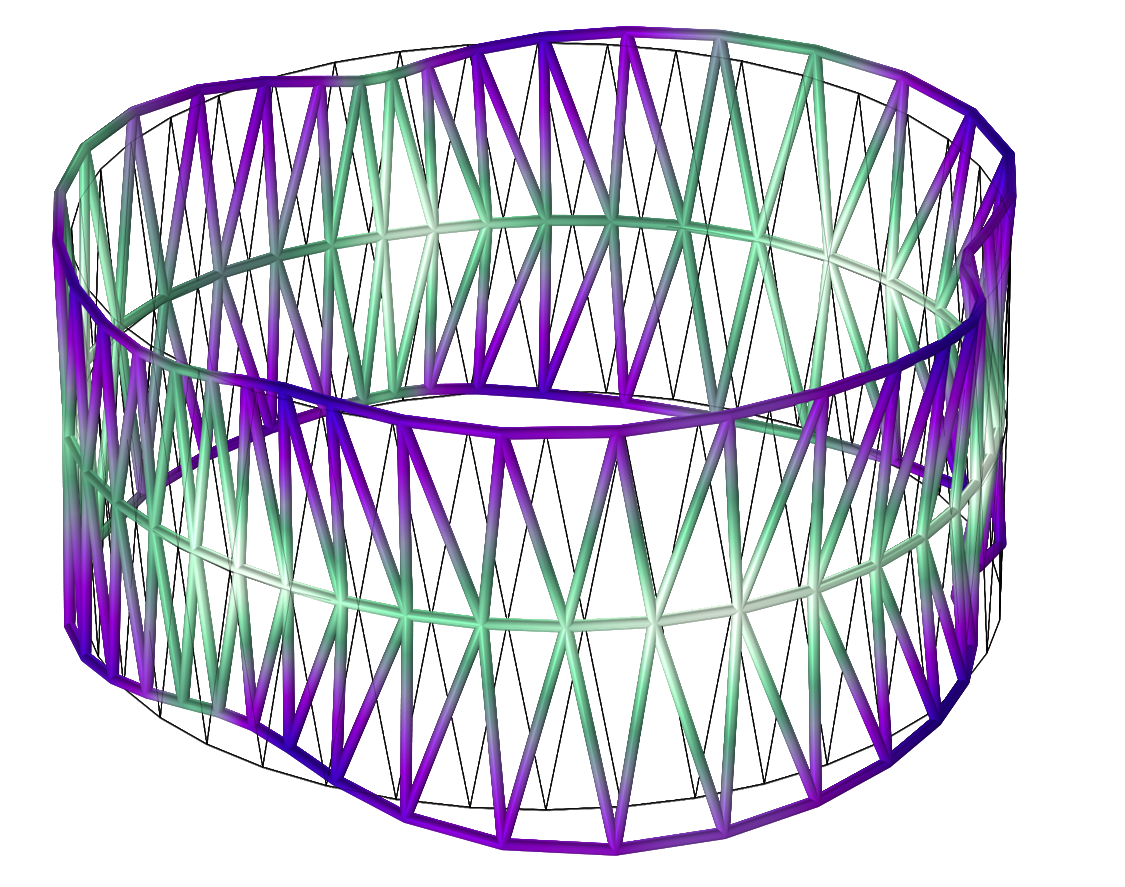}
    \centering\caption*{\footnotesize (a) $\omega = 1.4845$ rad/s.}
  \end{minipage}
  \begin{minipage}[b]{0.46\textwidth}
    \hspace{-0.1cm}\includegraphics[width=1\linewidth]{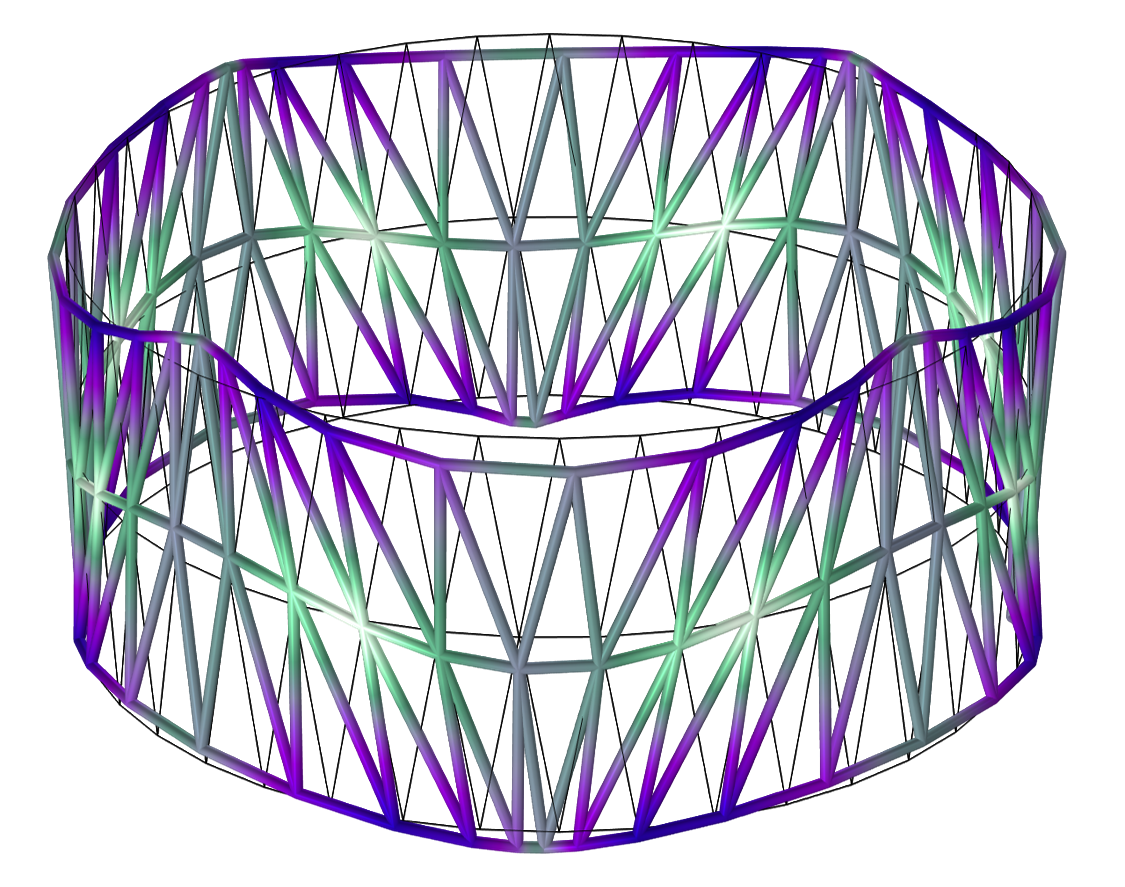}
    \centering\caption*{\footnotesize (b) $\omega =  2.2597$ rad/s.}
    \end{minipage}
  \caption{\footnotesize Eigenmodes of the chiral belt with (a) horizontal and (b) vertical oscillations of the nodal points in the central ring. Gyroscopic forces are present for the nodal points in the upper and lower rings, while there are no gyroscopic actions on the nodal points in the central ring. Gravitational forces are present at each nodal point and act uniformly. The magnitudes of the gyroscopic and gravitational parameters are both chosen to be unitary. The eigenfrequencies of the modes are: (a) $\omega = 1.4845$ rad/s and (b) $\omega=  2.2597$ rad/s.}
  \label{eigenmode12}
\end{figure}

 In Fig. \ref{eigenmode12}, we present examples of the chiral belt eigenmodes for the case where each nodal point has a mass of $1$ kg, connected by links of length $1$ m, and the gyricity parameter is chosen as $\Omega=1$ rad/s. In the illustrative examples, each ring of the chiral belt consists of $40$ nodal points, and the gravitational force acts uniformly and radially inwards for each ring with the unit magnitude. The eigenmodes of the belt are shown for two different eigenfrequencies. The horizontal vibrations of the nodal points positioned along the central ring, inducing eastward-propagating waveforms are shown in Fig. \ref{eigenmode12}(a) with $\omega=1.4845$ rad/s. Conversely, the mode presented in Fig. \ref{eigenmode12}(b) corresponds to vertical tangential motions with localised horizontal vibrations of the nodal points in the central ring for $\omega= 2.2597$ rad/s. The latter frequency also displays eastward-moving oscillations along the structure. Additionally, the nodal points in the upper and lower rings shown in Fig. \ref{eigenmode12} trace elliptical trajectories, which move in a clockwise and anticlockwise direction, respectively. This occurs due to the gyroscopic forces imposed on the masses in the upper and lower rings, together with the linear harmonic motion of the nodal points in the central ring.

  In the context of equatorial waves, inertia-gravity waves can exhibit  horizontal (zonal) and vertical (meridional) motions (see Section \ref{rotatingshall1}), which is a similar feature of the nodal points presented in the eigenmodes of the FE model. The three-dimensional chiral belt provides a simple yet effective approximate representation of atmospheric flows in a thin layer at the Earth's equator. In particular, the model takes into account the Coriolis effect within the equatorial region of the Earth, through the gyroscopic forces for each nodal point.

 \subsection{Elastic chiral lattice strip subjected to gravity}\label{discretelatstrip}
 
In this section, we present a structured study of dispersive waves in an infinite chiral elastic strip subjected to gravity, with two different types of boundary conditions (Neumann or Dirichlet conditions) prescribed on the outer layers of the strip. The effects of gravity and chirality are introduced through the gyroscopic pendulums embedded at the junctions of the strip. The dispersion properties for the elastic lattice strip are studied, and it is shown that one can control the preferential direction of the unidirectional waveforms and the locations of stop bands by tuning the gyricity and gravity parameters. The analysis of the waveforms corresponding to frequencies near the touching and crossing points of the dispersion curves is also provided, illustrating diverse standing wave patterns and interference phenomena. The gyroscopic lattice strip models presented here have no external energy flux, and are referred to as passive gyroscopic systems.

  \begin{figure}[H]
  \centering
  \begin{minipage}[b]{0.75\textwidth}
    {\hspace{-1.95cm}{\includegraphics[width=1.25\textwidth]{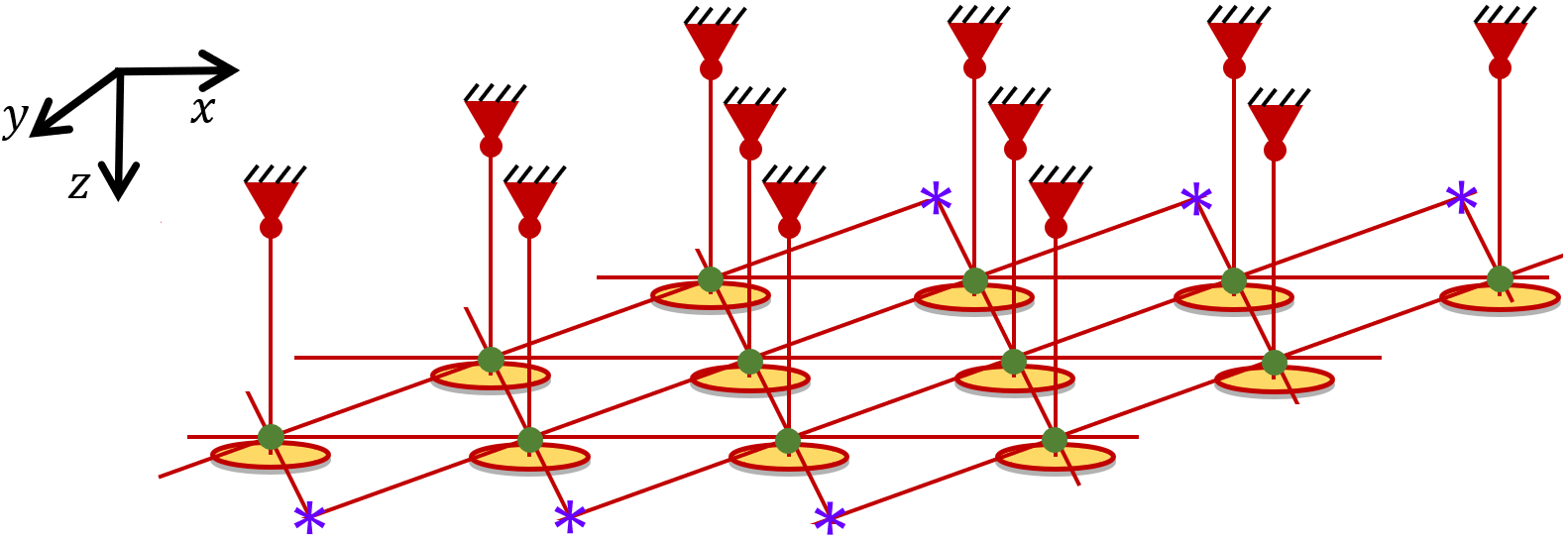}}}
    {\hspace{2cm}{\caption*{\footnotesize (a) Side view}}}
  \vspace{0.5cm}
  \end{minipage}
    \begin{minipage}[b]{0.5\textwidth}
    {\hspace{-1.3cm}{\includegraphics[width=1.25\textwidth]{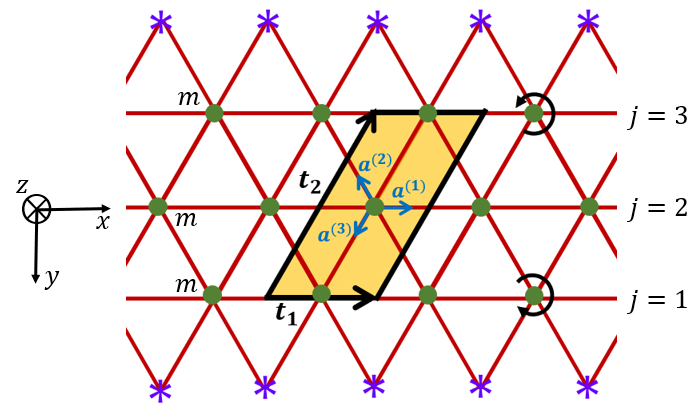}}}
    \hspace{1.8cm}{\caption*{\footnotesize (b) Top view}}
  \end{minipage}
  \caption{\footnotesize Chiral gravitational lattice strip consisting of nodal points connected by elastic springs, which are attached to gyropendulums (with gravity acting in the positive $z$-direction). Dirichlet boundary conditions are prescribed in the $y$-direction and the lattice strip is infinite in the $x$-direction; (a) side view and (b) top view.}
       \label{3dgyrops}
\end{figure}

\subsubsection{Chiral gravitational lattice strip with Dirichlet boundary conditions}\label{Dirichlet12a}
We consider an infinite two-dimensional chiral lattice strip of horizontal layers with Dirichlet boundary conditions, containing embedded gyroscopic pendulums. The Dirichlet boundary conditions are set on the outer horizontal boundaries of the five-layered strip shown in Fig. \ref{3dgyrops}. In contrast, the Neumann problem for the three-layer strip is investigated in Section \ref{class}. The pendulum feature of the structure allows for a description of the problem in the presence of gravitational forces, while the gyroscopic effect is taken into account by the properties of the spinners. For both the Dirichlet and Neumann problems, the lattice strip consists of three infinite horizontal rows of gyropendulums suspended from their pivot points so that they can swing freely under gravity, and each nodal point is connected to its nearest neighbours by elastic links. We formulate the governing equations for the nodal points, and analyse the types of propagating waves in such systems. The periodicity of the model allows for the application of Bloch-Floquet theory.

In the undisturbed configuration, the axis of each gyropendulum arm is perpendicular to the $xy$-plane. When the lattice nodes move, the spinners exert a force perpendicular to the displacement of the nodal point. In our analysis, the gyropendulums are assumed to undergo small-amplitude oscillations with minimal deviations from their equilibrium state, confining the lattice particles to move in the $xy$-plane.

\subsubsection{Governing equations}\label{governos1z}
We formulate the governing equations of the problem and describe the waves that can propagate in a discrete elastic Dirichlet strip. The nodes at the junctions of the structure are assumed to be the same and each nodal point is connected to its nearest neighbour by massless springs of length $l$ and stiffness $c$. The displacements at each point are assumed to be small so that the motion of the structure can be described by linear equations. 

The lattice strip, with the elementary cell shown in Fig. \ref{3dgyrops}(b), consists of three nodal points arranged in a parallelogram configuration, and the lattice vectors characterising the periodicity of the system are defined as follows
\begin{equation}\label{primos1}
{\bf t}_{1} = l(1,0)^T, ~~~ {\bf t}_{2} = l(1,\sqrt{3})^T.
\end{equation}

It is assumed that the nodal points in the inner layers of the strip, that is for $j=1, j=2$ and $j=3,$ have masses $m.$ 
We also introduce the three unit vectors ${\bf a}^{(j)},$ $j=1,2,3,$ representing the directions of links within the elementary cell:
\begin{equation}
{\bf a}^{(1)} = (1,0)^T, ~~~ {\bf a}^{(2)} = \Big(-\frac{1}{2}, \frac{\sqrt{3}}{2}\Big)^T, ~~~ {\bf a}^{(3)} = \Big(-\frac{1}{2}, -\frac{\sqrt{3}}{2}\Big)^T. \label{unitoa2}
\end{equation}
The position vector of the mass on the $j$-th layer ($j=1,2,3$) is given by 
\begin{equation}
{\bf x}^{(n,j)} = {\bf x}^{(0,j)} + n {\bf t}_{1}, \label{posFloquet}
\end{equation} 
where $n$ is the integer index, ${\bf x}^{(0,j)}$ is the position of a reference particle in the Dirichlet strip and the vector ${\bf t}_{1}$ is defined in \eq{primos1}.  

The lattice strip shown in Fig. \ref{3dgyrops} consists of gyroscopic spinners attached at every junction except those in the outer layers (shown by purple asterisks), which correspond to fixed points in accordance with the Dirichlet boundary conditions. Each gyropendulum consists of a massless rod, that is perpendicular to the plane of the two-dimensional lattice strip. The gyroscopic spinners generate a moment creating a ``vortex-type'' motion of the structure. The elementary cell of the periodic structure, shown in Fig. \ref{3dgyrops}(b), includes three particles, whose in-plane displacements are denoted by ${\bf u}^{(n, j)}$, for $j=1,2,3.$ The displacements are assumed to be time-harmonic with the radian frequency $\omega.$ Then the equations of motion for the three lattice masses in the time-harmonic regime are as follows:
\begin{equation}\label{gove1abc}
\begin{split}
- \frac{m \omega^2}{c} {\bf u}^{({ n}, 1)} =  {\bf a}^{(1)} \cdot \Big({\bf u}^{({n}+ 1, 1)} - {\bf u}^{(n, 1)}\Big){\bf a}^{(1)} + (-{\bf a}^{(1)}) \cdot \Big({\bf u}^{(n- {1}, 1)} - {\bf u}^{(n , 1)}\Big)(-{\bf a}^{(1)}) \\ + {\bf a}^{(2)} \cdot \Big({\bf u}^{( n - 1, 2)} - {\bf u}^{( n , 1)}\Big){\bf a}^{(2)} + (-{\bf a}^{(3)}) \cdot \Big({\bf u}^{(n , 2)} - {\bf u}^{( n , 1)}\Big)(-{\bf a}^{(3)}) \\ +i \frac{m \Omega \omega}{c} {\bf R} {\bf u}^{( n ,1)} - \frac{m G}{c} {\bf u}^{( n ,1)} - ( - {\bf a}^{(2)})\cdot {\bf u}^{( n , 1)}(- {\bf a}^{(2)}) - ( {\bf a}^{(3)})\cdot {\bf u}^{( n , 1)}({\bf a}^{(3)}) ,
\end{split}
\end{equation}

\begin{equation}\label{gove2abc}
\begin{split}
- \frac{m \omega^2}{c} {\bf u}^{( n , 2)} =  {\bf a}^{(1)} \cdot \Big({\bf u}^{( n+1 , 2)} - {\bf u}^{( n , 2)}\Big){\bf a}^{(1)} + (-{\bf a}^{(1)}) \cdot \Big({\bf u}^{(n - 1 , 2)} - {\bf u}^{( n , 2)}\Big)(-{\bf a}^{(1)}) \\ + {\bf a}^{(2)} \cdot \Big({\bf u}^{( n-1 , 3)} - {\bf u}^{(  n , 2)}\Big){\bf a}^{(2)} + (-{\bf a}^{(2)}) \cdot \Big({\bf u}^{( n+1 , 1)} - {\bf u}^{( n , 2)}\Big)(-{\bf a}^{(2)}) \\ + {\bf a}^{(3)} \cdot \Big({\bf u}^{( n , 1)} - {\bf u}^{( n , 2)}\Big){\bf a}^{(3)}  + (-{\bf a}^{(3)}) \cdot \Big({\bf u}^{( n , 3)} - {\bf u}^{( n , 2)}\Big)(-{\bf a}^{(3)})  - \frac{m G}{c} {\bf u}^{( n ,2)},
\end{split}
\end{equation}
and 
\begin{equation}
\begin{split}
- \frac{m \omega^2}{c} {\bf u}^{( n , 3)} =  {\bf a}^{(1)} \cdot \Big({\bf u}^{( n+1 , 3)} - {\bf u}^{( n , 3)}\Big){\bf a}^{(1)} + (-{\bf a}^{(1)}) \cdot \Big({\bf u}^{( n-1 , 3)} - {\bf u}^{( n , 3)}\Big)(-{\bf a}^{(1)}) \\ + (-{\bf a}^{(2)}) \cdot \Big({\bf u}^{( n+1 , 2)} - {\bf u}^{( n , 3)}\Big)(-{\bf a}^{(2)}) + {\bf a}^{(3)} \cdot \Big({\bf u}^{( n , 2)} - {\bf u}^{( n , 3)}\Big){\bf a}^{(3)}\\ - i \frac{m \Omega \omega}{c} {\bf R} {\bf u}^{( n ,3)} - \frac{m G}{c} {\bf u}^{( n ,3)} - ( {\bf a}^{(2)})\cdot {\bf u}^{( n , 3)}({\bf a}^{(2)}) - ( - {\bf a}^{(3)})\cdot {\bf u}^{( n , 3)}(-{\bf a}^{(3)}), \label{gove3abc}
\end{split}
\end{equation}
where $\Omega$ is the gyricity parameter which depends on the geometry of the spinners, $G=g/L$ is the gravity parameter with $L$ denoting the length of the pendulum arm, and ${\bf R}$ is the rotation matrix given by 
\begin{equation}
{\bf R} = \begin{pmatrix} 0 && 1 \\ -1 && 0\end{pmatrix}. \label{90rotation}
\end{equation}
In our analysis, the gyricities of the nodal points along the central chain ($j=2$) of the strip are chosen to be zero, while the gyricities in the lower ($j=1$) and upper ($j=3$) layers of nodal points are chosen to be the same in magnitude, but opposite in orientation. This is formulated in connection with the Coriolis force in the vicinity of the equator. In particular, the Coriolis force vanishes at the equator, and becomes increasingly significant for increasing latitudes.  In the following analysis, we investigate the combined effects of the gyricity and gravity parameters on the waveforms in the discrete structure as well as the dispersion characteristics.

\subsubsection{Elastic chiral gravitational Bloch-Floquet waves}\label{dirichletdispersion}
The amplitudes of the time-harmonic displacements of the nodal points within the periodic vortex-type gravitational Dirichlet strip satisfy the system of equations \eq{gove1abc}-\eq{gove3abc}, together with the quasi-periodicity of the system characterised by the Bloch-Floquet conditions: 
\begin{equation} \label{blockfa}
{\bf W}({\bf r} + n {\bf t}_{1}) = {\bf W}({\bf {r}}) e^{i k n l},
\end{equation}
where ${\bf W} = (u^{(1)}_{x}, u^{(1)}_{y},u^{(2)}_{x},u^{(2)}_{y},u^{(3)}_{x},u^{(3)}_{y})$ is the displacement vector in the $xy$-plane, ${\bf r} = (x,y)^T$ is the position vector and $k$ is the wavenumber. By applying \eq{blockfa} into the system \eq{gove1abc}-\eq{gove3abc}, we obtain the following system of equations in the matrix form:
\begin{equation}\label{disper123ab}
[{\bf C}_{D} - \omega^2 {\bf M} +  \omega {\bf A}]{\bf W} = {\bf 0},
\end{equation}
where ${\bf M}= m {\bf I}$ is the mass matrix (${\bf I}$ is the $6\times 6$ identity matrix),
\begin{equation}
 {\bf A} = i  \begin{pmatrix} 0 & - m \Omega & 0 & 0 & 0 & 0 \\  m \Omega & 0 & 0 & 0 & 0 & 0 \\ 0 & 0 & 0 & 0 & 0 & 0 \\ 0 & 0 & 0 & 0 & 0 & 0 \\ 0 & 0 & 0 & 0 & 0 &  m \Omega \\ 0 & 0 & 0 & 0 & -  m \Omega & 0\end{pmatrix}, \label{massspin1a}
\end{equation}
is the spinner matrix, and 
{\scriptsize
\begin{equation}
 {\bf C}_{D} = 
-\begin{pmatrix} 2c [\cos(k l) -\frac{3}{2}]- m G  & 0 & \frac{c}{4}(e^{-i k l} + 1) & \frac{c \sqrt{3} }{4}(-e^{-i k l} + 1) & 0 & 0 \\ 0 & -3 c - m G & \frac{c \sqrt{3}}{4}(- e^{- i k l}+1) & \frac{3 c}{4}(e^{- i k l}+1) & 0 & 0 \\ \frac{c}{4} (e^{i k l}+1) & \frac{c\sqrt{3}}{4} (-e^{i k l}+1) & 2c[\cos(k l) -\frac{3}{2}]  - m G & 0 & \frac{c}{4} (e^{- i k l} + 1) & \frac{c\sqrt{3}}{4} (-e^{- i k l} + 1) \\ \frac{c \sqrt{3}}{4}(-e^{i k l}+1) & \frac{3 c}{4} (e^{i k l}+1) & 0 & -3c - m G & \frac{c \sqrt{3}}{4}(-e^{- i k l} + 1) & \frac{3 c}{4} (e^{-i k l}+1) \\ 0 & 0 & \frac{c}{4} (e^{i k l} + 1) & \frac{c \sqrt{3}}{4} (-e^{i k l} + 1) & 2 c [\cos(k l) -\frac{3}{2}] - m G  & 0 \\ 0 & 0 & \frac{c \sqrt{3}}{4} (-e^{i k l}+1) & \frac{3 c}{4} (e^{i k l} + 1) & 0 & -3c - m G \end{pmatrix}, 
\end{equation}
}
is the gravity-stiffness matrix. Equation \eq{disper123ab} has non-trivial solutions if and only if the following condition is satisfied 
\begin{equation}\label{dispersionron1}
 \text{det}[{\bf C}_{D} - \omega^2 {\bf M} + \omega {\bf A}] =  \sigma_{D}^{(1)}(m, G, \Omega, k, \omega, c, l)  \sigma_{D}^{(2)}(m, G, \Omega, k, \omega, c, l) = 0,
\end{equation}
where
{\small
\begin{eqnarray}
\begin{gathered}
 \sigma_{D}^{(1)}(m, G, \Omega, k, \omega, c, l) = m^3\omega^6 + \Big(4c \cos( k l) - (\Omega^2 + 3G) m - 9 c \Big)m^2 \omega^4    \\   + \Big(4 c^2 \cos^2(k l) - \Big(\frac{47 c}{2} + 2 m ( \Omega^2 + 4 G ) \Big) c \cos(k l) + 26 c^2 + 3 ( \Omega^2 + 6 G) m c + G m^2 ( \Omega^2 + 3 G) \Big) m \omega^2 \\  - \frac{\sqrt{3}\Omega c^2 m \omega \sin(k l )}{2} - 4 G \cos^2(k l) c^2 m - \frac{21 \cos^2 (k l) c^3}{2} + 4 G^2 \cos(k l) c m^2 + \frac{47 G \cos(k l) c^2 m}{2} + 33 \cos(k l) c^3 \\   - G^3 m^3 - 9 G^2 c m^2 - 26 G c^2 m - 24 c^3,   \label{sigoddy}
\end{gathered}
\end{eqnarray}
}
and 
\begin{eqnarray}
\begin{gathered}
 \sigma_{D}^{(2)}(m, G, \Omega, k, \omega) = m^3 \omega^6 + \Big( 2 c \cos(k l) - (\Omega^2 + 3G) m - 9 c \Big)m^2 \omega^4 \\ + \Big(  G \Omega^2 m^2 + 3 G^2 m^2 - 4 G \cos(k l) c m + 3 \Omega^2 c m + 18 G c m - \frac{27 \cos(k l) c^2}{2} + 24 c^2 \Big) m \omega^2 \\ + \frac{3\sqrt{3}\Omega c^{2} m \omega\sin(k l)}{2} - \frac{9 \cos^2(k l) c^3}{2} + 2 G^2 \cos(k l) c m^2 + \frac{27 G \cos(k l) c^2 m}{2} + 18 \cos(k l) c^3 \\ -G^3 m^3 - 9 G^2 c m^2 - 24 G c^2 m - 18 c^3.  \label{sigeveny}
\end{gathered}
\end{eqnarray}
Equation \eq{dispersionron1} is referred to as the dispersion relation of the Dirichlet lattice strip. The quantities \eq{sigoddy} and \eq{sigeveny} define the frequencies and wavenumbers of the waveforms associated with only horizontal and only vertical motions of the central nodal points, respectively, while the upper and lower nodal points follow elliptical trajectories. The analytical forms of the trajectories are provided in the following section.

\subsection{Nodal trajectories of the Dirichlet strip}\label{akjshdfg}

This section includes auxiliary derivations related to the nodal point trajectories of the Dirichlet lattice strip of Section \ref{Dirichlet12a} in connection with the functions $\sigma_{D}^{(1)}(m, G, \Omega, k, \omega, c, l)$ and $\sigma_{D}^{(1)}(m, G, \Omega, k, \omega, c, l),$ which are given by \eq{sigoddy} and \eq{sigeveny}, respectively.

\subsubsection{Horizontal motions of the central nodal points }\label{centralnodalhorizontal1a}

The dispersion equation $\sigma_{D}^{(1)}(m, G, \Omega, k, \omega, c, l)=0$ for the Dirichlet strip (see also Section \ref{newdispersionproper1}), describes the connection between the frequency $\omega$ and wavenumber $k$ of the waveforms along the infinite chiral lattice strip. When $\sigma_{D}^{(1)}(m, G, \Omega, k, \omega, c, l)=0$ and equation \eq{dispersionron1} is satisfied, the nodal points in the central layer (denoted by $j=2$ in Fig. \ref{3dgyrops}(b) of Section \ref{dirichletdispersion}) move only in the horizontal direction, and the nodal points in the lower ($j=1$) and upper ($j=3$) layers follow elliptical trajectories of opposite orientations. In this case, the time-harmonic transverse displacements of the nodal points along the $j$-th horizontal strip, for $j=1,2,3,$ are given by (see Section \ref{Dirichlet12a})
\begin{equation}
{\bf u}^{(n)}_{j} = \text{Re}\Big( {\bf{U}}_{j} e^{i(k n l - \omega t)}\Big), \label{timeharmo1}
\end{equation}
where the displacement amplitude eigenvector ${\bf U}_{j}$ has the components  
\begin{equation}
{\bf U}_{1} =   \frac{e^{- i k l}}{{\frak{a}_{1} + i \frak{a}_{2}}} \Big( i \frak{A} , -  \frak{B} \Big)^{T}, ~~~~ {\bf U}_{2} =  (1, 0)^{T}, ~~~~ {\bf U}_{3} =  \frac{1}{\frak{a}_{1} + i \frak{a}_{2} }\Big(i \frak{A}, \frak{B} \Big)^{T},  \label{toge1as}
\end{equation}
 where
 \begin{eqnarray}
 \begin{gathered}
 \frak{A} = 2 m^2 \omega^4 -  4 m (G m - c \cos(k l) + 3 c )\omega^2 - 4 \Big( m G + \frac{21 c}{8} \Big) c \cos(k l) + 2 G^2 m^2 + 12 c G m + \frac{33 c^2}{2}, \nonumber \\ 
 \frak{B} = 2 m^2 \Omega \omega^3 + m \Omega (4 c \cos(k l) - 2 (G m + 3 c) ) \omega + \frac{\sqrt{3} c^2 \sin(k l)}{2}, \nonumber \\
 \frak{a}_{1} = -c m \sin(k l) \omega^2 +  \sqrt{3} m \Omega c ( \cos(k l) - 1) \omega + c (m G + 3 c )\sin(k l), \nonumber \\
 \frak{a}_{2} =  -mc ( \cos(k l) + 1) \omega^2 - c \sqrt{3} m \Omega \sin(k l)\omega + c ( m G + 3 c ) ( \cos(k l) + 1 ) . \nonumber
 \end{gathered}
 \end{eqnarray}
We note that the vertical displacement component of the central nodal points vanishes as shown by the representation of ${\bf U}_{2}$ in \eq{toge1as}. The components of the amplitude eigenvectors for the remaining nodal points are complex, resulting in elliptical trajectories of the nodal points in the lower ($j=1$) and upper ($j=3$) layers of the Dirichlet strip. These trajectories are identical, but are traced in opposite directions.
 
 The equations of the trajectories of the central nodal points are
 \begin{equation}
 x^{(n)}_{2} =  \cos(k n l - \omega t), ~~~ y^{(n)}_{2} = 0,
 \end{equation}
 and the trajectories of the nodal points in the lower ($j=1$) and upper ($j=3$) layers of the Dirichlet strip are given respectively by 
 \begin{equation}
 \Big(\frac{\sqrt{\frak{a}_{1}^2+\frak{a}_{2}^2}}{\frak{A}} x_{1}^{(n)} \Big)^2 + \Big( \frac{\sqrt{\frak{a}_{1}^2+\frak{a}_{2}^2}}{\frak{B}} y_{1}^{(n)}\Big)^2 = 1, ~~~~  \Big(\frac{\sqrt{\frak{a}_{1}^2+\frak{a}_{2}^2}}{\frak{A}} x_{3}^{(n)} \Big)^2 + \Big( \frac{\sqrt{\frak{a}_{1}^2+\frak{a}_{2}^2}}{\frak{B}} y_{3}^{(n)}\Big)^2 = 1, 
 \end{equation}
 where $x_{j}^{(n)}$ and $y_{j}^{(n)}$ are the local coordinates associated with the elliptical trajectory of the $n$-th node on the $j$-th layer, for $j=1$ and $j=3.$ The local origin corresponds to the equilibrium position of node $n$ for each layer. The $x_{j}^{(n)}$ and $y_{j}^{(n)}$ axes are parallel and perpendicular to the $j$-th layer, respectively. Thus, the trajectories of the nodal points of the Dirichlet lattice strip along $j=1$ and $j=3$ (see Fig. \ref{3dgyrops}(b)) are elliptical with axes aligned parallel and perpendicular to the $j$-th layer.

 \subsubsection{Vertical motions of the central nodal points  }\label{sektion1}

When the dispersion equation $\sigma_{D}^{(2)}(m, G, \Omega, k, \omega, c, l)=0$ is satisfied, the nodal points along the central layer ($j=2$) move only in the vertical direction, which differs from the horizontal motions discussed above, whereas the nodal points in the upper and lower layers follow elliptical trajectories. In this case, the vibrations of the Dirichlet strip can be expressed in a similar time-harmonic form as \eq{timeharmo1}, with the displacement amplitude eigenvector components represented as follows  
\begin{equation}
{\bf U}_{1} = \frac{e^{- i k l}}{{\frak{c}_{1} + i \frak{c}_{2}}} \Big(- i \frak{C}, \frak{D}  \Big)^{T}, ~~~~ {\bf U}_{2} =  (0, 1)^{T}, ~~~~ {\bf U}_{3} =  \frac{1}{{\frak{c}_{1} + i \frak{c}_{2}}} \Big(i \frak{C}, \frak{D} \Big)^{T}, 
\end{equation}
 where 
 \begin{eqnarray}
 \begin{gathered}
 \frak{C} = - 2 m^2 \Omega \omega^3 + 2 (G m + 3 c)\Omega m \omega + \frac{3 \sqrt{3} c^2 \sin(k l)}{2}, \nonumber \\ 
 \frak{D} = - 2 m^2 \omega^4 + 4 m ( G m -  c \cos(k l) + 3 c)\omega^2 + 4 \Big( G m + \frac{21c}{8} \Big) c \cos(k l) - 2 G^2 m^2 - 12 c G m - \frac{33 c^2}{2}, \nonumber \\
 \frak{c}_{1}  = 3 m c ( \cos(k l) + 1) \omega^2 - c \sqrt{3}m \Omega \sin(k l) \omega - 3 c \Big( -2 c \cos(k l)^2 + (G m + c) \cos(k l) + G m + 3 c \Big), \nonumber \\
 \frak{c}_{2} = - 3 c m \sin(k l) \omega^2 - \sqrt{3} m \Omega c ( \cos(k l) - 1)\omega + 3 c \Big( Gm - 2\cos(kl)c + 3c \Big) \sin(k l). \nonumber
 \end{gathered}
 \end{eqnarray}
Then, the central nodal point trajectories are described by 
 \begin{equation}
 x^{(n)}_{2} = 0, ~~~ y^{(n)}_{2} =   \cos(k n l - \omega t),
 \end{equation}
 while the equations governing the motion of the nodal points in the lower ($j=1$) and upper ($j=3$) layers of the Dirichlet strip are, respectively, given by 
 \begin{equation}
 \Big(\frac{\sqrt{\frak{c}_{1}^2+\frak{c}_{2}^2}}{\frak{C}} x_{1}^{(n)} \Big)^2 + \Big( \frac{\sqrt{\frak{c}_{1}^2+\frak{c}_{2}^2}}{\frak{D}} y_{1}^{(n)}\Big)^2 = 1, ~~~~  \Big(\frac{\sqrt{\frak{c}_{1}^2+\frak{c}_{2}^2}}{\frak{C}} x_{3}^{(n)} \Big)^2 + \Big( \frac{\sqrt{\frak{c}_{1}^2+\frak{c}_{2}^2}}{\frak{D}} y_{3}^{(n)}\Big)^2 = 1, 
 \end{equation}
 where $x_{j}^{(n)}$ and $y_{j}^{(n)}$ are the local coordinates used in Section \ref{centralnodalhorizontal1a}. Similar to the solutions presented in Section \ref{centralnodalhorizontal1a}, the trajectories of the nodal points along the $j=1$ and $j=3$ layers are also ellipses. These ellipses have axes aligned parallel and perpendicular to the $j$-th layer of the Dirichlet strip for $j=1$ and $j=3,$ while the central nodal points ($j=2$) exhibit only vertical motions. 
 
 At the crossing or touching points of the dispersion curves (see Section \ref{crossDirichlet}), the nodal points in the central layer ($j=2$) move along elliptical trajectories or linear paths that are generally not aligned with the horizontal or vertical directions. In this case, the above representations of the displacement components for the central nodal points do not apply. Moreover, at the degeneracy points, the displacements of the central nodal points can be written as a linear combination of eigenmodes consisting of horizontal and vertical components. At the degeneracy points, the nodal points along the $j=1$ and $j=3$ layers follow ellipses with different eccentricities and sizes. Illustrative examples of the nodal point trajectories related to the dispersion degeneracies are provided in Section \ref{crossDirichlet}.  At the crossing or touching points of the dispersion curves, the phase velocity can be used to describe the direction of motion of the waveforms.

\subsection{Dispersion properties of the gyro-elastic chiral strip with Dirichlet boundary conditions}\label{newdispersionproper1}

We analyse the dispersion properties of a discrete Dirichlet strip subjected to gravity, and demonstrate how the waveforms along the structure depend on the gravity and gyricity parameters. In the illustrative examples, we normalise the physical quantities by the natural units of the structure, which are assigned the unit values: $l=1,c=1$ and $m=1.$ In particular, the physical units of measurement will not be shown. In the following illustrative examples, we consider non-negative values of the gyricity parameter $\Omega.$ However, we also note that choosing $\Omega<0$ determines the preferential directions of the propagating and standing waves in a similar manner to $\Omega>0$; the shift in the dispersion curves for $\Omega<0$ is in the opposite direction compared to the case where $\Omega>0$.

\begin{table}[H]
        \setlength\tabcolsep{3pt}
    \expandafter\patchcmd\csname Gin@ii\endcsname   
      {\setkeys {Gin}{#1}}
      {\setkeys {Gin}
        {width=\dimexpr\linewidth-2\tabcolsep,      
         valign=c, margin=0pt 3pt 0pt 3pt,#1}       
      }
      {}{}
    \centering
\begin{tabular}{|p{1.4cm}|p{6cm}|p{6cm}|}
    \hline
    &  \begin{center}\textbf{ $G=0$}\end{center} & \begin{center}\textbf{ $G = 0.5$}\end{center}    \\
    \hline%
$\Omega=0$
                                 & \includegraphics{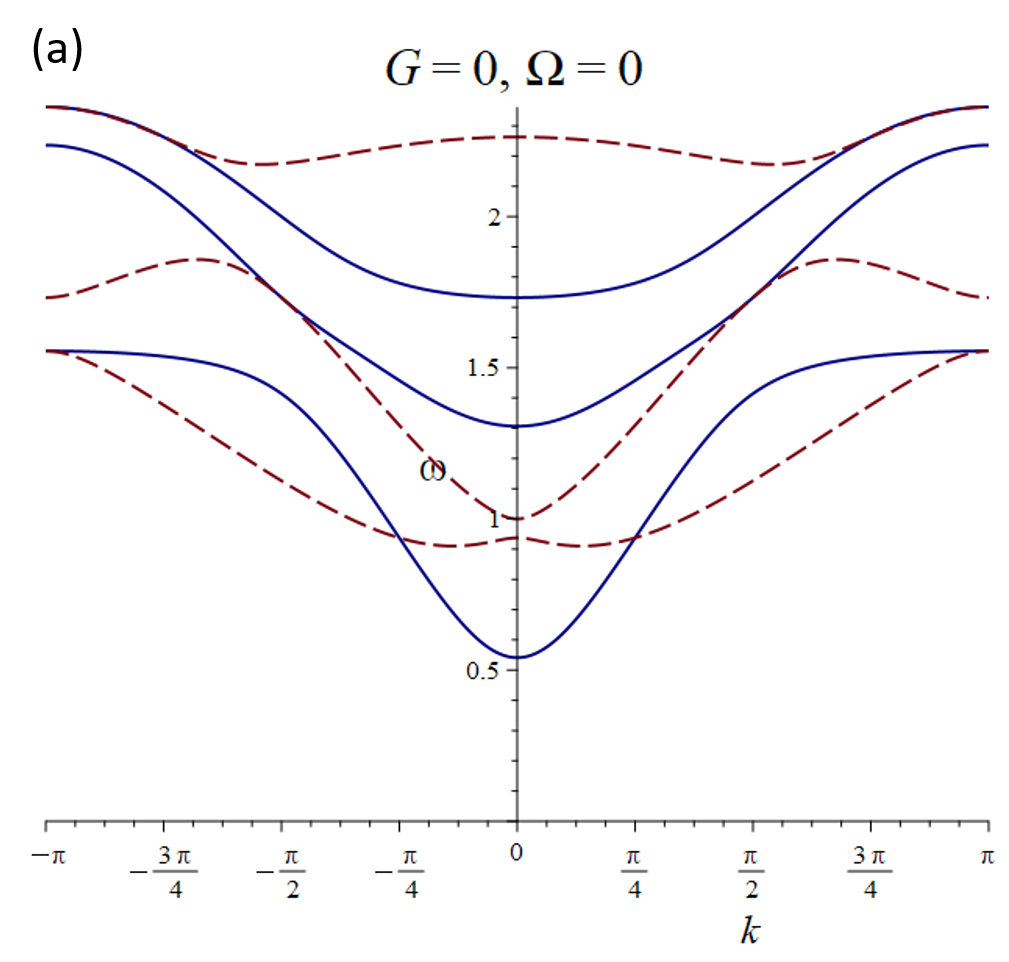}            & \includegraphics{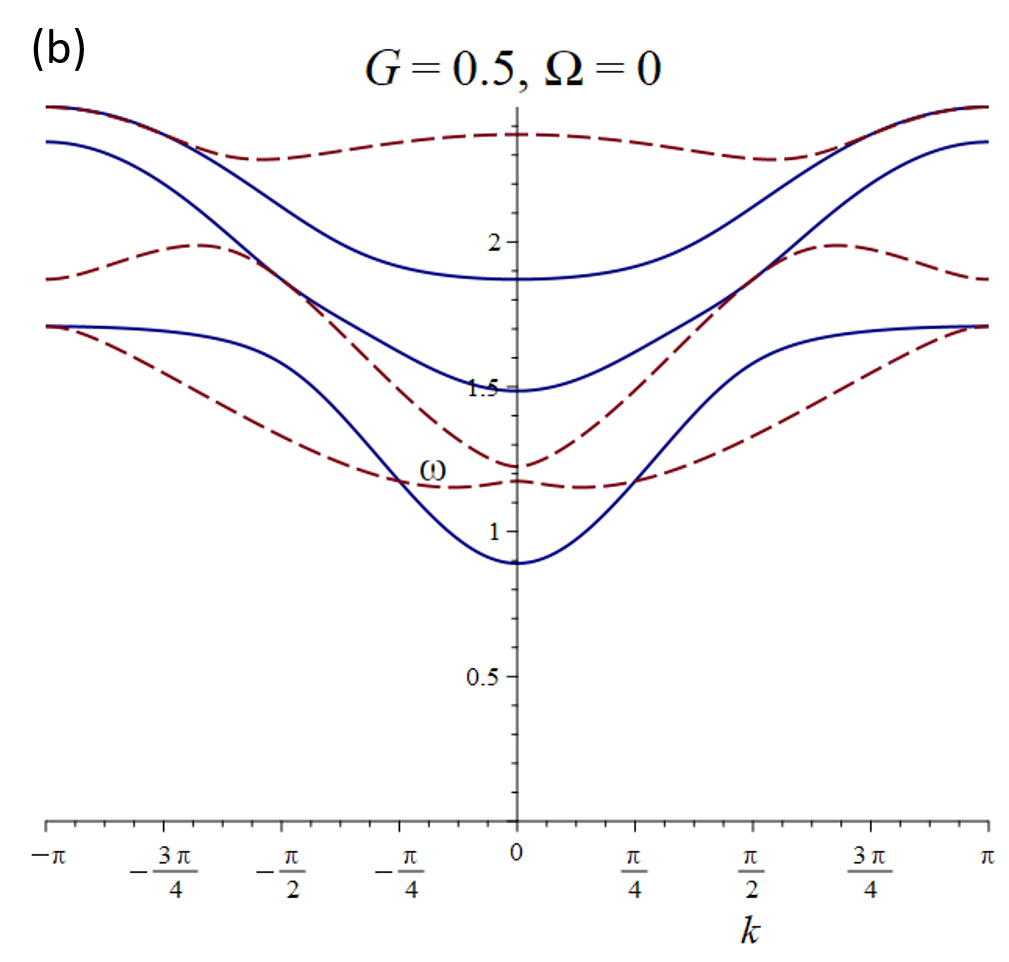}          \\
    \hline%
$\Omega = 0.8$
                                & \includegraphics{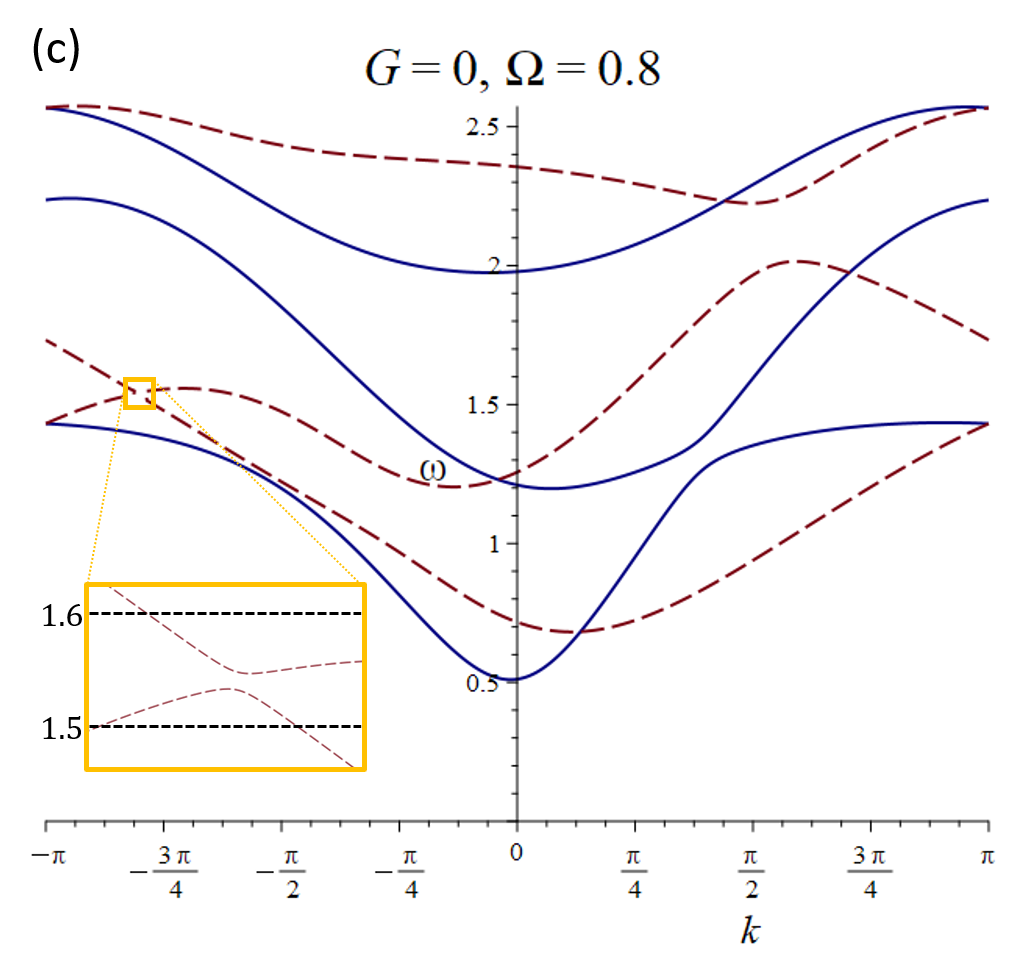}           &  \includegraphics{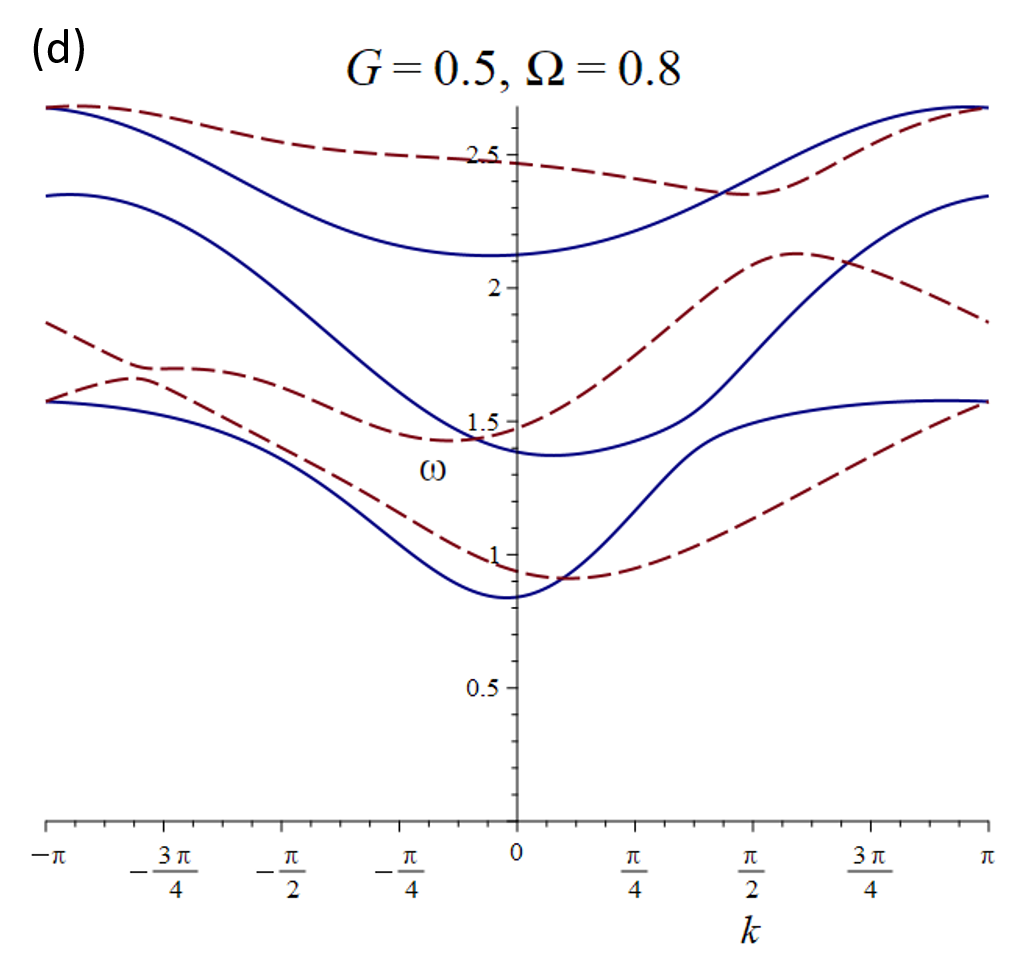}          \\
    \hline
    $\Omega = 1.5$
                                & \includegraphics{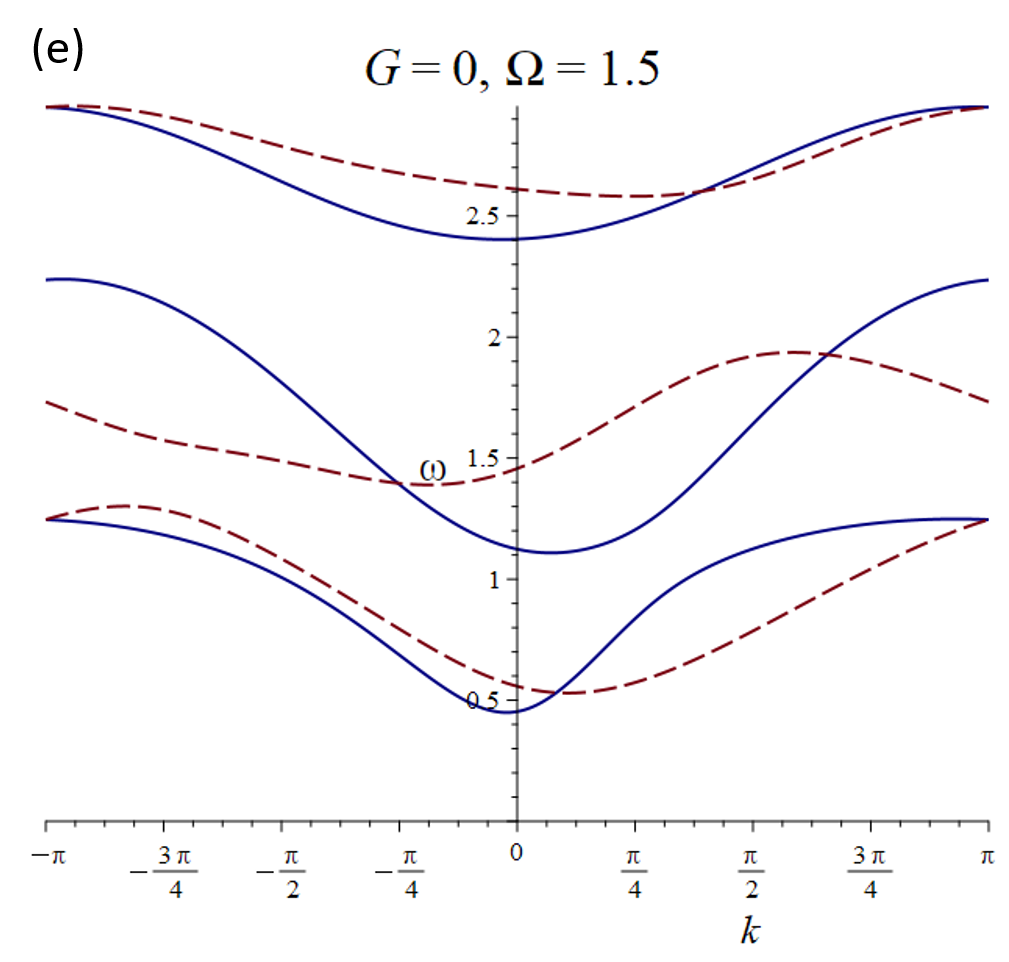}           &  \includegraphics{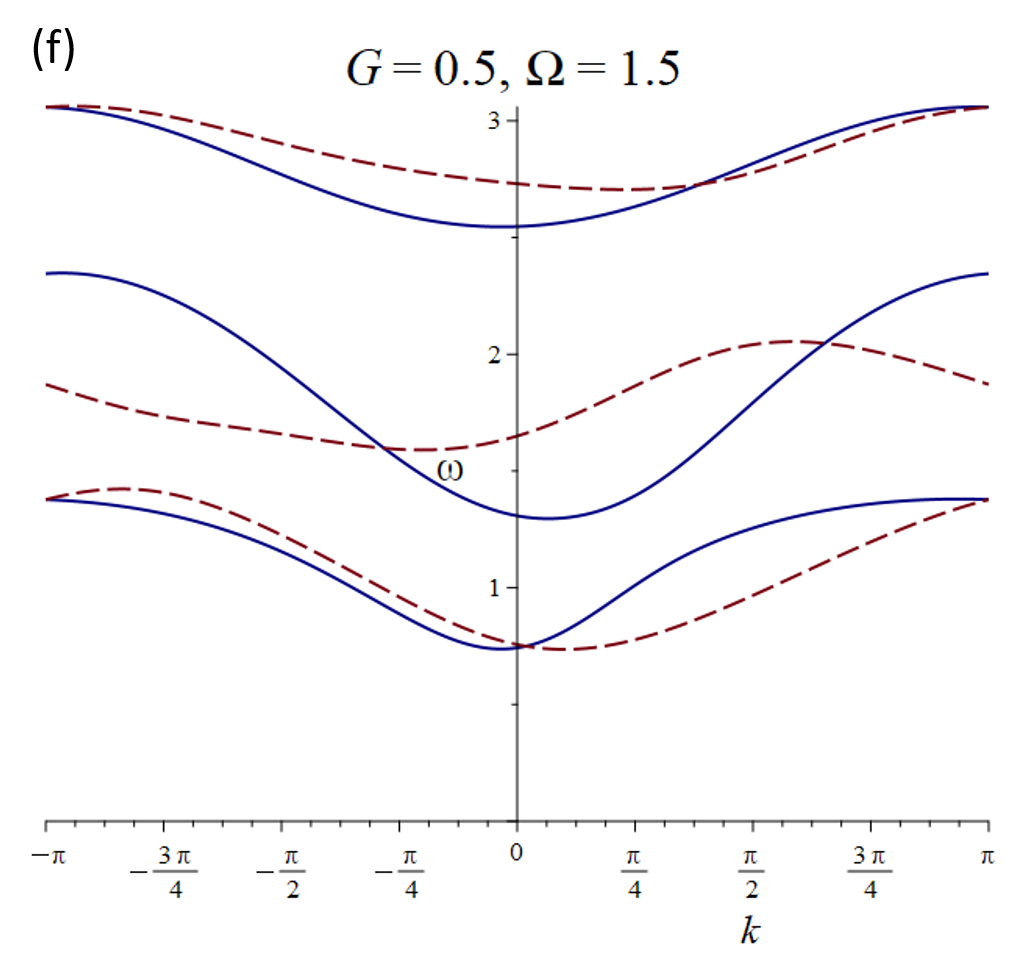}          \\
    \hline
\end{tabular}
 \captionof{figure}{\footnotesize Dispersion diagrams, illustrating the frequency $\omega$ against the wavenumber $k,$ of the lattice strip for a range of values of the gravity and gyricity parameters. The normalised parameter values of the Dirichlet strip are $c=1, l=1$ and $m=1.$ The solid and dashed curves correspond to the solutions of \eq{sigoddy} and \eq{sigeveny}, respectively, associated with the nodal point trajectories described in Section \ref{dirichletdispersion} and Section \ref{akjshdfg}.}
    \label{dirichlet1a}
    \end{table}
    
    Fig. \ref{dirichlet1a} shows the dispersion diagrams for different values of the gravity and gyricity parameters of the infinite lattice strip with Dirichlet boundary conditions. The dispersion diagrams can be used to select the different types of nodal trajectories of the Dirichlet lattice strip, in connection with travelling or standing waves. The solid dispersion curves correspond to zero vertical displacements of the central nodal points, while the dashed dispersion curves correspond to zero horizontal displacements of the central nodal points. The dispersion degeneracies linked to the crossing or touching points of the dispersion curves are discussed in Section \ref{crossDirichlet}. 
    
    When both gravity and gyricity are absent in the Dirichlet lattice strip, the dispersion diagram is shown in Fig. \ref{dirichlet1a}(a); there is a zero frequency stop band as well as a cut-off frequency above which no propagating waveforms exist.  It is noted that when $G=0$ and $\Omega=0,$ there are standing modes present at $k=0,$ at the boundaries $k= \pm \pi$ as well as points within the first Brillouin zone, and the dispersion curves are symmetric with respect to the frequency axis. There are six curves in each dispersion diagram, and the right- and left-travelling waves are determined by taking into account the locations of positive and negative slopes of the dispersion curves, respectively. In the presence of gravity and absence of gyricity, the dispersion diagram is shown in Fig. \ref{dirichlet1a}(b), where the size of the zero-frequency band gap increases, while the symmetry relative to the frequency axis of the dispersion curves remains as in the case with $G=0$ and $\Omega=0.$ Introducing the gyricity through the spinners breaks the symmetry of the dispersion curves relative to $k=0$ as demonstrated in Fig. \ref{dirichlet1a}, where the characteristics of mechanical chirality introduces a preferential directionality of the wave propagation. In addition when $\Omega \neq 0,$ the points on the dispersion curves associated with standing waveforms, shift away from $k=0,$ resulting in asymmetry in the dispersion curves; these points correspond to zero group velocities and non-zero phase velocities, which are investigated below. For non-trivial gyricities, the lattice strip also exhibits propagating waves with positive (negative) group velocities and negative (positive) phase velocities, that can affect the complex interaction between the motion of the wave and its individual components. Increasing the gyricity $\Omega$ leads to a larger separation between the two high-frequency and two low-frequency dispersion curves, as well as additional finite-frequency intervals associated with stop bands, as shown in Fig. \ref{dirichlet1a}(e) and Fig. \ref{dirichlet1a}(f). In the limit $\Omega\rightarrow \infty$, the two lower curves approach zero, while the two intermediate dispersion curves approach the curves described by 
    \begin{equation}
    \omega = \sqrt{\frac{ m G + 3 c }{m}}, ~~~ \omega = \sqrt{\frac{m G + 3c - 2 c \cos(k l)}{m}}.
    \end{equation}
We also note that the two upper curves increase for increasing $\Omega,$ with the frequencies being proportional to the gyricity parameter.

\subsubsection{Propagating waveforms along the chiral Dirichlet strip}\label{greoupscla1}
In this section, we present the typical trajectories of the nodal points in the Dirichlet strip. The motions of the nodal points are linked to the dispersion properties discussed in the previous section. We show that the displacements of the fixed points shown in Fig. \ref{3dgyrops} are zero, while the nodal points within the inner layers, i.e. for $j=1,2,3,$ follow linear or elliptical trajectories (see also Section \ref{akjshdfg}).

\begin{figure}[H]
  \centering
  \hspace{-3.5cm}
  \begin{minipage}[b]{0.55\textwidth}
    \includegraphics[width=1.25\textwidth]{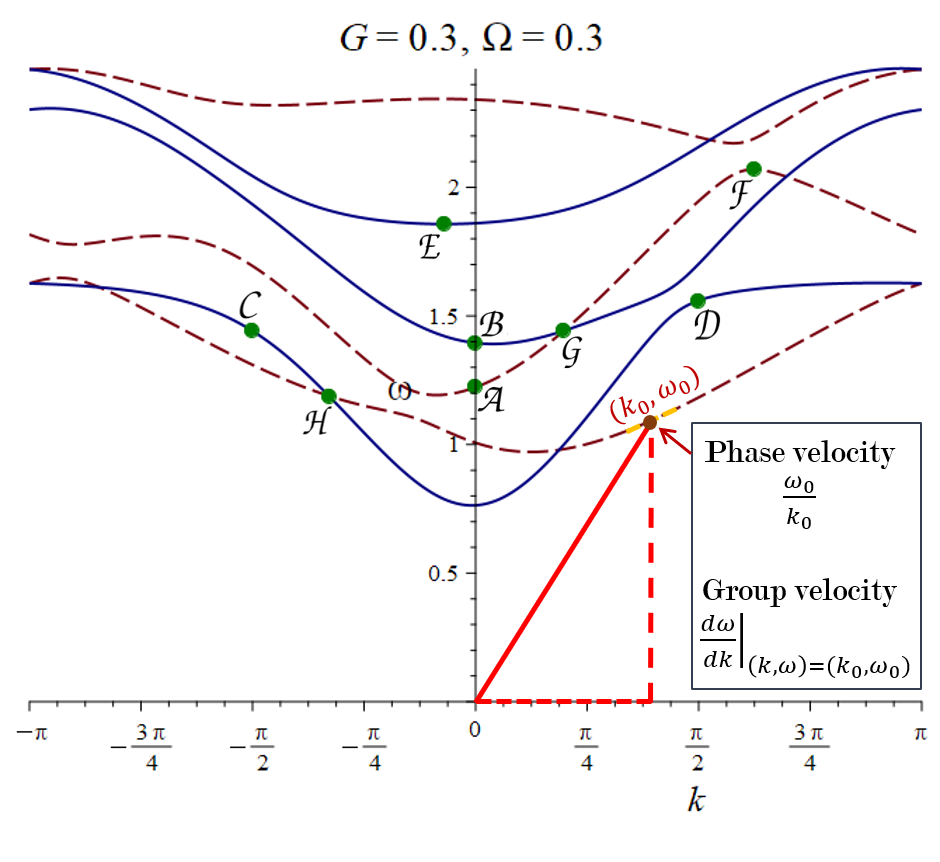}
  \end{minipage}
  \caption{\footnotesize Dispersion diagram for the chiral gravitational lattice ship shown in Fig. \ref{3dgyrops}, for $G=0.3$ and $\Omega=0.3.$ The remaining normalised physical quantities are $m=1,c=1$ and $l=1.$ The units of measurement will not be displayed. The definitions of the phase and group velocities at an arbitrary point $(k_{0}, \omega_{0})$ are also defined. The solid dispersion curves are associated with zero vertical displacements of the central nodal points, while the dashed dispersion curves are linked to zero horizontal displacements of the central nodal points. The nodal point trajectories of the Dirichlet strip linked to the points $\mathpzc{A},\mathpzc{B},\mathpzc{C},\ldots,\mathpzc{H},$ are discussed in the text. }
       \label{dirdispersion1}
\end{figure}

In Fig. \ref{dirdispersion1}, we present the dispersion diagram of the Dirichlet lattice strip for the parameter values $G=0.3$ and $\Omega=0.3.$ The points $\mathpzc{A},\mathpzc{B},\mathpzc{C},\ldots,\mathpzc{H},$ correspond to examples of nodal point trajectories which are detailed below. Fig. \ref{dirdispersion1} also shows the definitions of the group and phase velocities of the waves, where the former represents the rate at which the overall shape of the wave packet propagates in connection with the transmitted energy, while the latter describes the rate at which the phase of an individual wave propagates in the strip. For a stationary wave, characterised by a pattern that does not propagate in the spatial direction, there is zero net transfer of energy and thus, the group velocity vanishes; the points $\mathpzc{E}$ and $\mathpzc{F}$ on the dispersion diagram are associated with standing waves. The points $\mathpzc{A},\mathpzc{B},\mathpzc{C}$ and $\mathpzc{D}$ on Fig. \ref{dirdispersion1} correspond to four examples where the group velocity is non-zero, resulting in propagating waves along the infinite discrete chiral strip. The points of dispersion degeneracies, denoted by $\mathpzc{G}$ and $\mathpzc{H}$ correspond to two examples of crossing points of the dispersion curves. 

\begin{figure}[H]
  \centering
    \begin{minipage}[b]{0.49\textwidth}
    \hspace{-0.4cm}\includegraphics[width=1\linewidth]{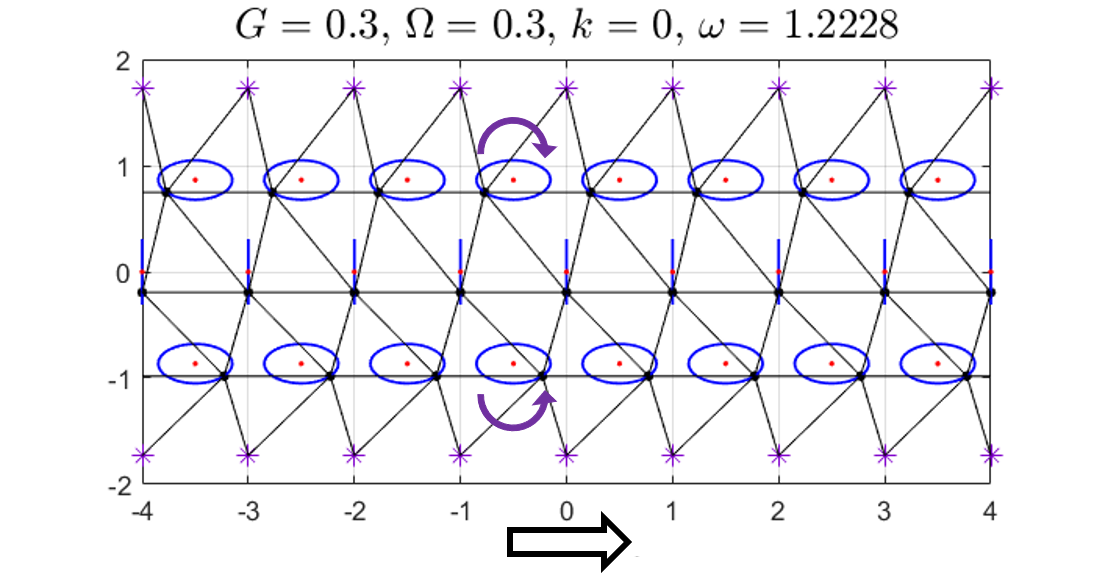}
    \centering\caption*{\footnotesize (a)}
  \end{minipage}
  \begin{minipage}[b]{0.49\textwidth}
    \hspace{-0.4cm}\includegraphics[width=1\linewidth]{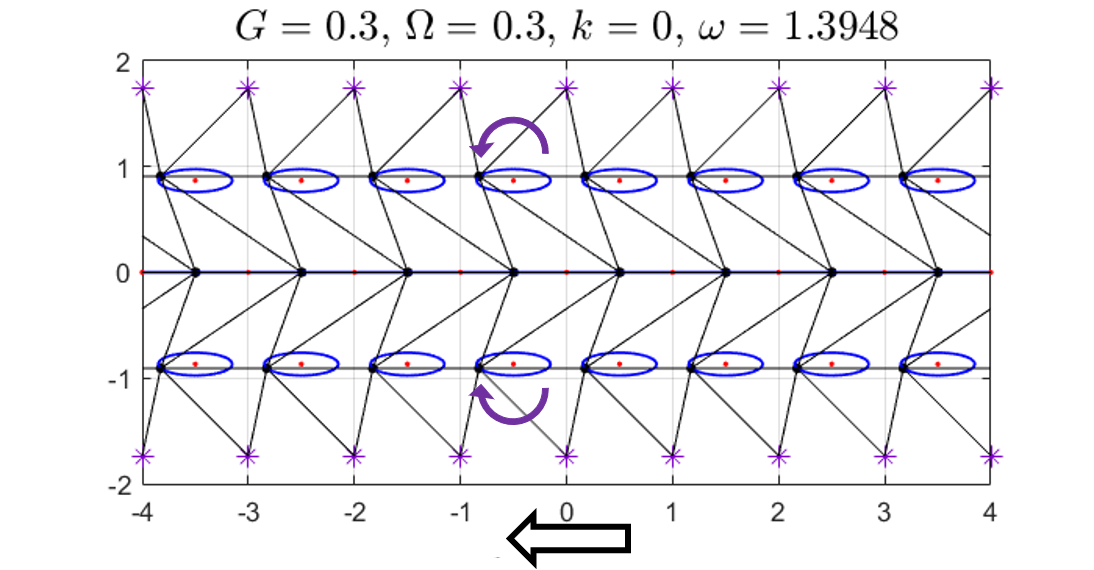}
    \centering\caption*{\footnotesize (b)}
  \end{minipage}
  \caption{\footnotesize Nodal point trajectories of the lattice strip for $k=0$ and the parameter values $G=0.3$ and $\Omega=0.3.$ The corresponding dispersion diagram is shown in Fig. \ref{dirdispersion1} with the points $\mathpzc{A}$ and $\mathpzc{B}$ linked to parts (a) and (b), respectively; (a) $(k,\omega)=(0, 1.2228)$ and (b) $(k,\omega)=(0, 1.3948).$ The arrows denote the direction of the propagating waveforms, and the circular arrows show the orientation of motion of the nodal points.}
  \label{nonactivedisper12}
\end{figure}

Illustrative trajectories of the nodal points at $k=0$ are shown in Fig. \ref{nonactivedisper12}, for two different frequency values. In Fig. \ref{nonactivedisper12}(a), the central nodal points move in a vertical motion, with no horizontal displacement, while in the example shown in Fig. \ref{nonactivedisper12}(b), the central nodal points are displaced only in the horizontal direction; these trajectories are associated with the points $\mathpzc{A}$ and $\mathpzc{B}$ in the dispersion diagram shown in Fig. \ref{dirdispersion1}. A characteristic feature of the Dirichlet strip geometry and the gyricity choices is the tendency of the nodal points in the upper ($j=3$) and lower ($j=1$) layers to move in opposite directions. In both the upper and lower layers of the strip, the nodal points follow ellipses with major axes aligned parallel to the direction of the wave motion. It is noted that although the propagating waveforms shown in Fig. \ref{nonactivedisper12} correspond to non-zero group velocities, the frequencies of the oscillations are near the standing mode frequencies of the structure, where the slope of the curves shown in Fig. \ref{dirdispersion1} is zero. This occurs due to the non-trivial gyricity value resulting in asymmetric dispersion diagrams.

\begin{figure}[H]
  \centering
  \begin{minipage}[b]{0.49\textwidth}
    \hspace{-0.6cm}\includegraphics[width=1\linewidth]{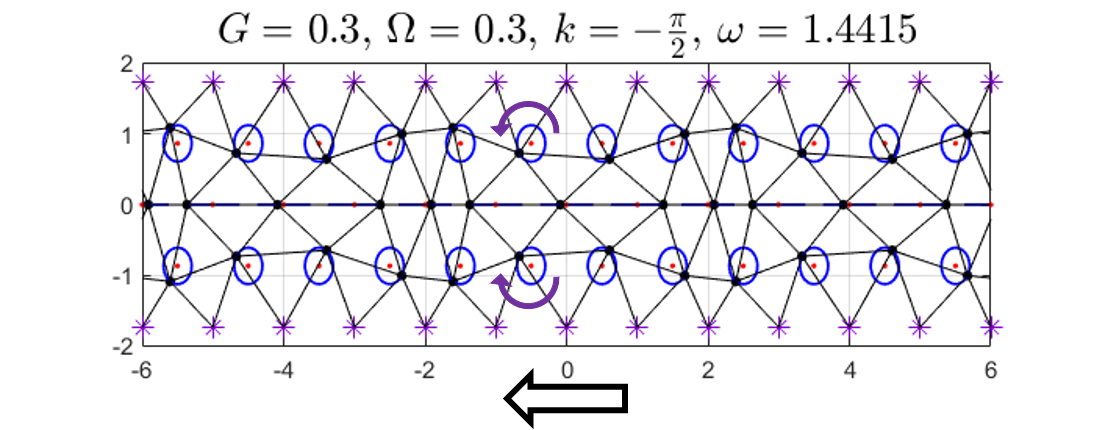}
    \centering\caption*{\footnotesize (a)}
  \end{minipage}
  \begin{minipage}[b]{0.49\textwidth}
    \hspace{-0.6cm}\includegraphics[width=1\linewidth]{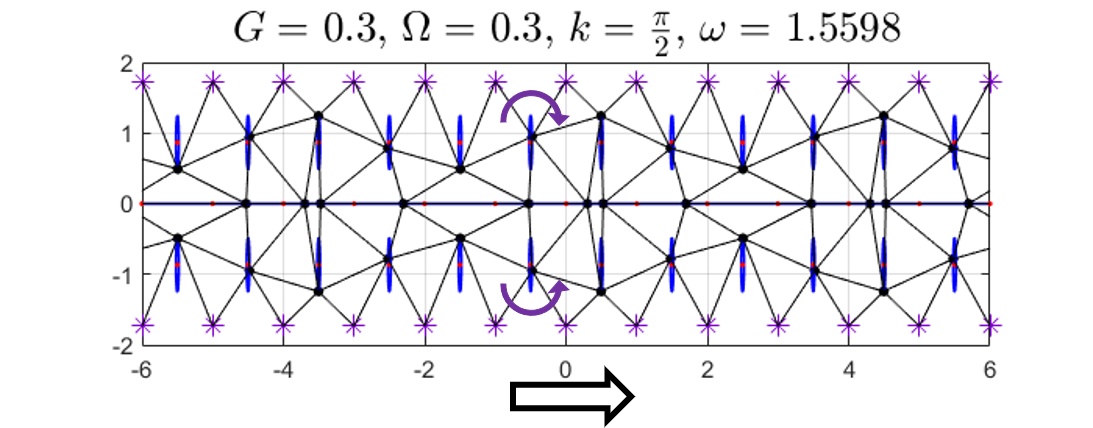}
    \centering\caption*{\footnotesize (b)}
  \end{minipage}
  \caption{ \footnotesize  Elliptical trajectories of the nodal points in connection with propagating waves. The associated dispersion diagram is shown in Fig. \ref{dirdispersion1} with the points $\mathpzc{C}$ and $\mathpzc{D}$ corresponding to parts (a) and (b), respectively; (a) $(k, \omega) = (-\pi/2, 1.4415)$ and (b) $(k,\omega)=(\pi/2, 1.5598).$   }
  \label{propgating_d}
\end{figure}

Additional examples of travelling wave solutions propagating in the negative and positive $x$-directions in the Dirichlet strip are shown in Fig. \ref{propgating_d}(a) and Fig. \ref{propgating_d}(b), respectively. The frequencies in the illustrative examples displayed in Fig. \ref{propgating_d} differ to those in Fig. \ref{nonactivedisper12}. In the trajectories shown in Fig. \ref{propgating_d}, the central nodal points move only horizontally, while the upper and lower nodal points follow elliptical paths with major axes aligned perpendicular to the direction of the propagating waveforms. The dispersion diagram associated with the examples is shown in Fig. \ref{dirdispersion1}, where the frequency and wavenumber values in Fig. \ref{propgating_d}(a) and Fig. \ref{propgating_d}(b) correspond to $\mathpzc{C}$ and $\mathpzc{D}$, respectively. Analogous features are also observed for the inertia-gravity waves in the equatorial region as discussed in Section \ref{rotatingshall1}.

Through the above illustrative examples, we observe that the frequencies alter the vibrational modes of the structure, impacting the trajectories of the nodal points, and thus, the oscillations and mode shapes of the Dirichlet lattice strip.

\subsubsection{Standing modes of the Dirichlet strip}\label{standingstrip1}
In this section we display the standing modes behaviour of the Dirichlet strip through illustrative examples. The related dispersion diagram is presented in Fig. \ref{dirdispersion1}, where the wavenumber and frequency values corresponding to standing waves are marked by $\mathpzc{E}$ and $\mathpzc{F}.$

\begin{figure}[H]
  \centering
  \begin{minipage}[b]{0.49\textwidth}
    \hspace{-0.6cm}\includegraphics[width=1\linewidth]{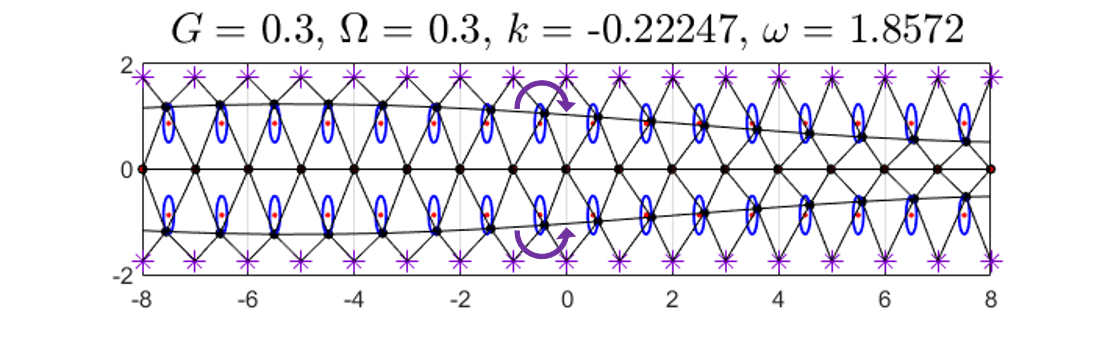}
    \centering\caption*{\footnotesize (a)}
  \end{minipage}
  \begin{minipage}[b]{0.49\textwidth}
    \hspace{-0.6cm}\includegraphics[width=1\linewidth]{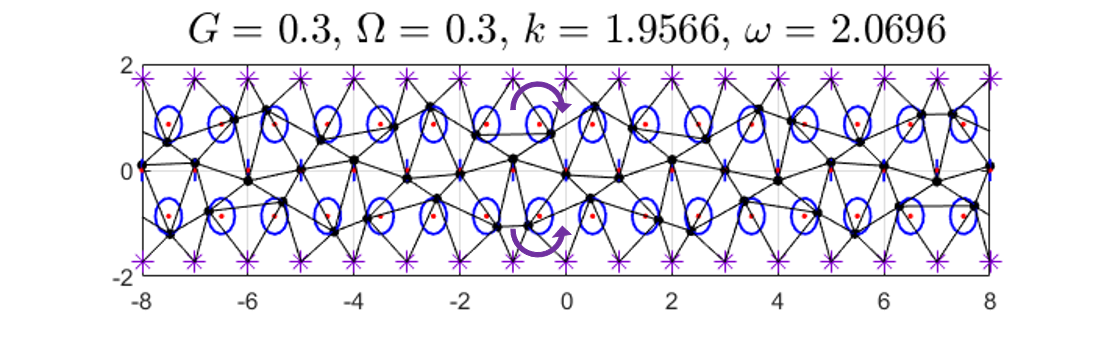}
    \centering\caption*{\footnotesize (b)}
  \end{minipage}
  \caption{ \footnotesize Trajectories of the lattice strip nodal points for the normalised parameters $c=1,l=1,m=1,G=0.3$ and $\Omega=0.3,$ corresponding to standing waves with two different pairs $(k,\omega).$ The associated dispersion diagram is shown in Fig. \ref{dirdispersion1}, where the points $\mathpzc{E}$ and $\mathpzc{F}$ are linked to parts (a) and (b), respectively; (a) $(k,\omega)= (-0.22247, 1.8572)$ and (b) $(k,\omega)= (1.9566, 2.0696).$}
  \label{easttowest1}
\end{figure}

The standing mode patterns of the lattice strip for $\omega=1.8572$ rad/s are shown in Fig. \ref{easttowest1}(a), with the central nodal points moving solely horizontally, whereas the upper and lower nodal points move in a clockwise and anticlockwise orientation, respectively. The standing wave with a higher frequency compared to the example presented in Fig. \ref{easttowest1}(a), is illustrated in Fig. \ref{easttowest1}(b) which corresponds to the vertically displaced central nodal points, while the upper and lower nodal point trajectories follow elliptical paths with major axes aligned perpendicular to their corresponding horizontal lattice strip layer. In both examples, the orientations and alignment of the nodal points in the lower ($j=1$) and upper ($j=3$) layers are similar, but the eccentricities of the ellipses differ as well as the motions of the central nodal points.

\subsubsection{Dispersion degeneracies}\label{crossDirichlet}
The dispersion degeneracies, linked to the crossing and touching points of the dispersion curves, are discussed in this section. At these points, there are repeated roots of the dispersion equation. The crossing points are identified by the intersection of two dispersion curves, while the touching points correspond to the tangents of two dispersion curves with equivalent group velocities at the touching point. This feature is inherent to the Dirichlet lattice strip, and does not depend on the gravitational and gyroscopic effects. However, the properties of the spinners and magnitude of the gravity parameter can be used to alter the behaviour of the band structure in the vicinity of the degeneracy points. At such points on the dispersion diagram, the oscillations of the nodal points in the strip do not exhibit symmetric modes relative to the central horizontal layer, and the central nodal points can display elliptical trajectories that were not observed in previous sections.  In the absence of gravity, the ability to control the locations of crossing points using gyroscopic spinners in a lattice system has been studied in \cite{garau2018interfacial}. The presence of crossing points is also related to the symmetries of the medium  \cite{hou2015hidden, he2015emergence}.

In the absence of gyroscopic forces, there are three touching points and two crossing points for both $k>0$ and $k<0.$ In this case, the points of degeneracy occur at the same $k$ values independently of the gravity parameter $G$ (see Fig. \ref{dirichlet1a}(a) and Fig. \ref{dirichlet1a}(b)). For $\Omega=0,$ the set of dispersion curves are symmetric with respect to the $\omega$-axis, so that if there is a crossing (touching) point at $(k_{*}, \omega_{*}),$ then there is also another distinct crossing (touching) point at $(-k_{*}, \omega_{*}).$ When $\Omega=0,$ the touching points occur at $k=\pi/2$ and $k=\pi,$ while the crossing points are observed at $k=\pi/4$ and $k=3 \pi/4.$ The introduction of gyricity leads to asymmetric dispersion curves relative to the frequency axis, and a change in the number of crossing and touching points - from ten such degeneracy points of the dispersion curves for $\Omega = 0$ to eight points for $\Omega \neq 0$ (see Fig. \ref{dirichlet1a}). Furthermore, there are no touching points when $\Omega\neq 0,$ and only crossing points appear as shown in Fig. \ref{dirichlet1a}.   

At the touching points of the dispersion curves with $\Omega=0,$ the trajectories of the nodal points can follow linear motions due to the absence of gyroscopic forces. The touching point at $kl=\pi/2$ occurs at the frequency $\omega=\sqrt{(m G + 3c)/m},$ while the two touching points at $kl=\pi$ correspond to the frequency values 
\begin{equation}
\omega= \sqrt{\frac{2 m G + 8 c - c \sqrt{10} }{2 m}}, ~~~~  \omega= \sqrt{\frac{2 m G + 8 c + c \sqrt{10} }{2m}}.
\end{equation}
The crossing points of the dispersion curves for the non-gyroscopic Dirichlet strip, for $k\geq 0,$ occur at $kl=\pi/4$ and $kl=3\pi/4$ with the frequencies, respectively, given by 
\begin{equation}
\omega= \sqrt{\frac{2 m G - 3 c \sqrt{2} + 6 c}{2 m}}, ~~~~  \omega=\sqrt{\frac{2 m G + 3 c \sqrt{2} + 6c}{2 m}}.
\end{equation}
The touching and crossing points of the dispersion curves for $k<0$ and $\Omega=0$ occur at the same frequencies shown above for the corresponding negative values of $k$.

\begin{figure}[H]
  \centering
  \begin{minipage}[b]{0.48\textwidth}
   \includegraphics[width=1\linewidth]{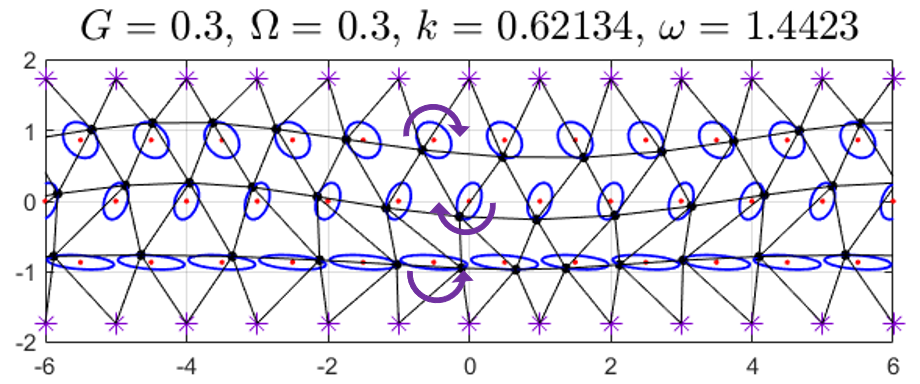}
    \centering \caption*{\footnotesize (a)}
  \end{minipage}
  \begin{minipage}[b]{0.48\textwidth}
   \hspace{0.8cm}
\includegraphics[width=1\linewidth]{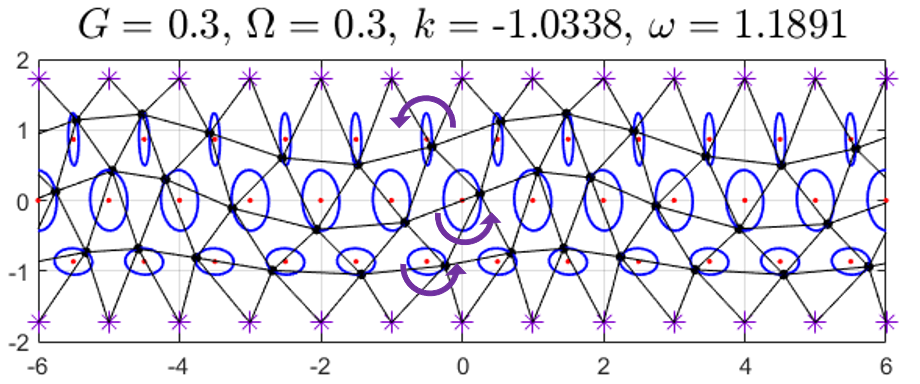}
    \centering\caption*{\footnotesize (b)}
  \end{minipage}
  \caption{\footnotesize Trajectories of the nodal points associated with the crossing points of the dispersion curves shown in Fig. \ref{dirdispersion1}. The normalised parameter values are $\Omega=0.3,$ $G=0.3,m=1,c=1$ and $l=1$; (a) $(k, \omega)=(0.62134, 1.4423)$ and (b) $(k, \omega)=(-1.0338, 1.1891).$ The trajectories of all nodal points are elliptical. The circular arrows show the orientation of motion of the nodal points. }
  \label{DiracPoints1a}
\end{figure}

In Fig. \ref{DiracPoints1a}, we present illustrative examples of the nodal trajectories in the chiral Dirichlet strip linked to the crossing points of the dispersion curves shown in Fig. \ref{dirdispersion1} (denoted by the points $\mathpzc{G}$ and $\mathpzc{H}$ for Fig. \ref{DiracPoints1a}(a) and Fig. \ref{DiracPoints1a}(b), respectively). In both examples, the central nodal points move in an elliptical motion, which differ from the horizontal or vertical motions described in Sections \ref{greoupscla1} and \ref{standingstrip1}. This motion of the central nodal points corresponds to a linear combination of two modes associated with a horizontal and vertical component. Fig. \ref{DiracPoints1a}(a) displays the elliptical trajectories of the nodal points with $\omega=1.4423$ rad/s, where the upper and central nodal points move in a clockwise orientation, while the lower nodal points move anticlockwise. The motion of the chiral Dirichlet strip for a different crossing point corresponding to the frequency $\omega=1.1891$ rad/s is shown in Fig. \ref{DiracPoints1a}(b). In this case, the elliptical trajectories of all nodal points follow an anticlockwise orientation. In both examples, the major axes of the ellipses traced by the nodal points are neither parallel nor perpendicular to the strip, and the eccentricities of the ellipses differ for each layer. We also note that the phase velocity is positive in the example presented in Fig. \ref{DiracPoints1a}(a) and negative in the example shown in Fig. \ref{DiracPoints1a}(b).

\subsubsection{Modes with zero vertical displacements of the central nodal points}\label{sygoddy}

In this section, we provide the analysis of the gyro-elastic Dirichlet strip modes which exhibit zero vertical displacements of the central nodal points (denoted by $j=2$ in Fig. \ref{3dgyrops}(b)). The conditions determining the motions of such nodal point trajectories are presented. In this case, we show that the horizontal displacements of the nodal points in the upper and lower layers of the Dirichlet strip follow the same direction, while their vertical displacements are in the opposite direction.

Applying the conditions $(u_{x}^{(1)}, u_{y}^{(1)})= (u_{x}^{(3)}, -u_{y}^{(3)})e^{- i k l}$ and $u^{(2)}_{y}=0$ (see Section \ref{dirichletdispersion}) and taking into account the analysis presented in Section \ref{dirichletdispersion}, results in a simplified dispersion equation, which is a sixth-order polynomial in $\omega,$ of the form 
\begin{equation}
\sigma_{D}^{(1)}(m, G, \Omega, k, \omega, c, l) = 0, \label{disperoddiny}
\end{equation}
where $\sigma_{D}^{(1)}$ is defined in \eq{sigoddy}. The group velocity of the waveforms is given by implicitly differentiating \eq{disperoddiny} with respect to $k,$ which yields the following:
\begin{equation}
\frac{d \omega}{d k} = \frac{\mathcal{F}_{1}}{\mathcal{G}_{1}}, \label{groupveloasioda}
\end{equation}
where 
\begin{eqnarray}
\begin{gathered}
\mathcal{F}_{1} = 2 c l \sin(k l) m^2 \omega^4 +  l \Big( 4 c m \cos(k l) - ( \Omega^2 + 4 G ) m^2 - \frac{47 c m}{4} \Big) \sin(k l) c \omega^2 \\ + \frac{l \sqrt{3} \cos(k l) \Omega c^2 m}{4} \omega +  l \Big( - \Big( 4 G m + \frac{21 c}{2} \Big) c \cos(k l) + 2 G^2 m^2 + \frac{47 c G m}{4} + \frac{33 c^2}{2} \Big) \sin(k l) c, \label{groupvelo1a}
\end{gathered}
\end{eqnarray}
and 
\begin{eqnarray}
\begin{gathered}
\mathcal{G}_{1} = 3 m^3 \omega^5 +  m \Big( 8 c m \cos(k l) - ( 2 \Omega^2 + 6 G ) m^2 - 18 c m \Big) \omega^3 \\ +  m \Big( 4 c^2 \cos^2(k l) - 8 \Big( \Big( \frac{\Omega^2}{4} + G \Big) m + \frac{47 c}{16} \Big) c \cos(k l) + G ( \Omega^2 + 3 G ) m^2 \\ +  3 (\Omega^2 + 6 G) c m + 26 c^2 \Big) \omega  - \frac{\sqrt{3} \Omega c^2 m \sin(k l)}{4}.
\end{gathered}
\end{eqnarray}
We note that when the group velocity \eq{groupveloasioda} vanishes, there are standing modes of the Dirichlet strip with no vertical translations of the central nodal points. Conversely, a positive group velocity indicates a forward motion of the propagating waves, while a negative group velocity signifies a backward motion of the waveforms. 
    \begin{figure}[H]
  \centering
  \begin{minipage}[b]{0.52\textwidth}
    \hspace{-0.4cm}\includegraphics[width=1\linewidth]{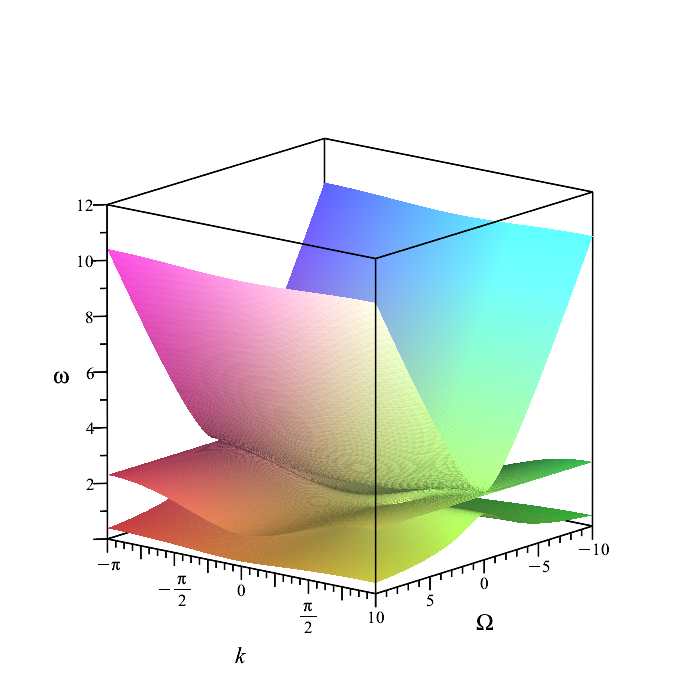}
    \centering\caption*{\footnotesize (a)}
  \end{minipage}
  \begin{minipage}[b]{0.45\textwidth}
    \hspace{-0.4cm}\includegraphics[width=1\linewidth]{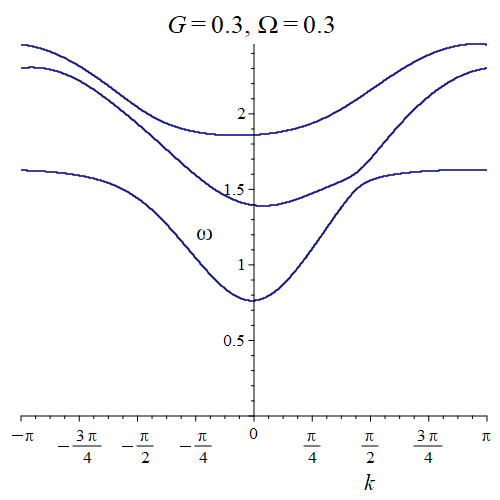}
    \centering\caption*{\footnotesize (b)}
  \end{minipage}
  \caption{\footnotesize (a) The graphs of $\omega$ as functions of $k$ and $\Omega$ for the normalised parameter values $G=0.3,$ $m=1, c=1$ and $l=1,$ satisfying the dispersion equation \eq{disperoddiny}. The graphs represent a collection of dispersion curves for varying values of the gyricity parameter $\Omega$. (b) Dispersion diagram for $\Omega=0.3$ corresponding to three dispersion curves obtained by taking the vertical cross section at $\Omega=0.3$ of the surfaces shown in part (a). The same three dispersion curves are shown by the solid curves in Fig. \ref{dirdispersion1}.}
  \label{surfaces}
\end{figure}

To illustrate the dynamic response of the chiral strip subjected to gravity for varying values of the gyricity parameter, we present the computations shown in Fig. \ref{surfaces}. In Fig. \ref{surfaces}(a), we have three surfaces corresponding to a set of three dispersion curves for varying values of $\Omega$. There is also a zero-frequency stop band due to the prescribed Dirichlet boundary conditions (see Section \ref{Dirichlet12a}). For $\Omega=0.3,$ there are three dispersion curves as shown in Fig. \ref{surfaces}(b), which correspond to the curves obtained by taking the vertical cross section of the surfaces presented in Fig. \ref{surfaces}(a) at $\Omega=0.3.$ For large values of $\Omega,$ the chiral Dirichlet strip is dominated by gyroscopic forces, with the lower and central dispersion curves approaching $\omega=0$ and the curve described by $\omega=\sqrt{(m G + 3c -2 c \cos(k l))/m},$ respectively, while the upper dispersion curve increases with $\Omega.$ This phenomenon is referred to as gyroscopic rigidity, where the gyroscopic effects become very dominant rendering the structure rigid. Additionally, we note that the dispersion equation \eq{disperoddiny} is also dependent on the sign of the gyricity parameter $\Omega,$ and thus the surfaces presented in Fig. \ref{surfaces}(a) are not symmetric relative to the $\Omega=0$ plane - changing the sign of $\Omega$ is linked to the change in the orientation of motion of the nodal points. In the illustrative examples we consider the case of $\Omega\geq 0$ since the dispersion curves for $\{\Omega \geq 0, k \geq 0\}$ and $\{\Omega \geq 0, k \leq 0\}$ are equivalent to those for $\{\Omega \leq 0, k \leq 0\}$ and $\{\Omega\leq 0, k \geq 0\},$ respectively.

\begin{table}[H]
        \setlength\tabcolsep{3pt}
    \expandafter\patchcmd\csname Gin@ii\endcsname   
      {\setkeys {Gin}{#1}}
      {\setkeys {Gin}
        {width=0.94\dimexpr\linewidth-2\tabcolsep,      
         valign=c, margin=0pt 3pt 0pt 3pt,#1}       
      }
      {}{}
    \centering
\begin{tabular}{|p{1cm}|p{6.5cm}|p{6.5cm}|}
    \hline
    &  \begin{center}\textbf{ $G=0$}\end{center} & \begin{center}\textbf{ $G = 1$}\end{center}    \\
    \hline%
$\Omega=0$
                                & \raisebox{2.8cm}{\text{(a)}}\includegraphics{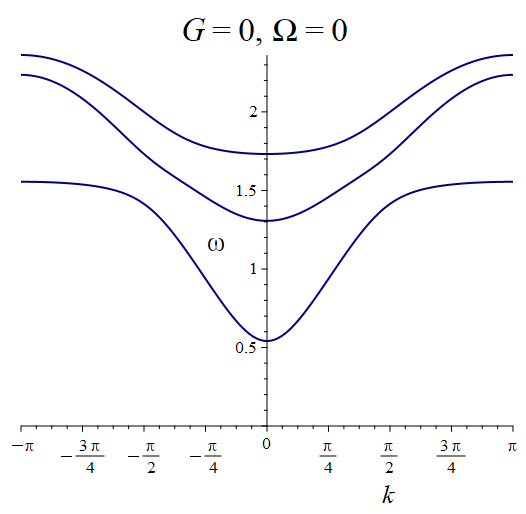}            & \raisebox{2.8cm}{\text{(b)}}\includegraphics{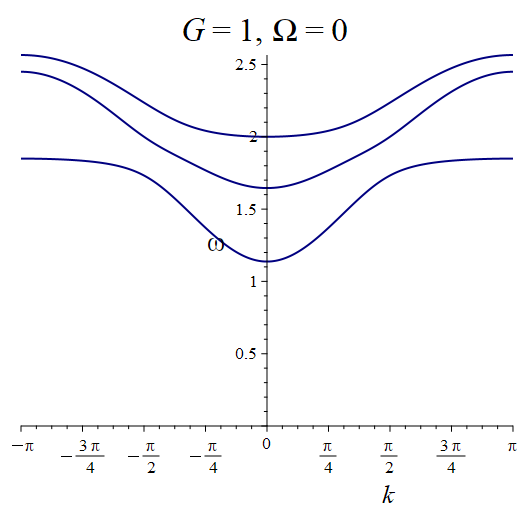}          \\
    \hline%
$\Omega = 1$
                                & \raisebox{2.8cm}{\text{(c)}}\includegraphics{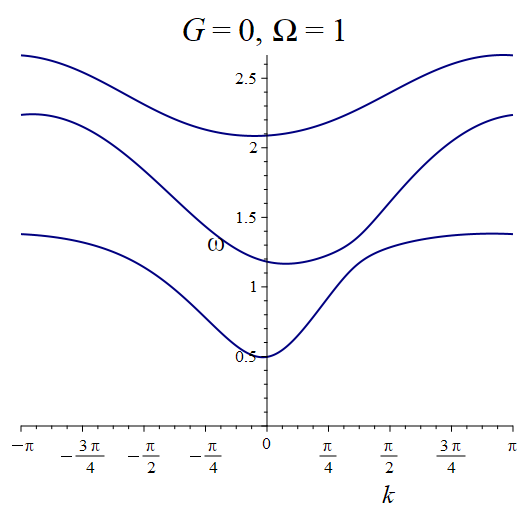}           &  \raisebox{2.8cm}{\text{(d)}}\includegraphics{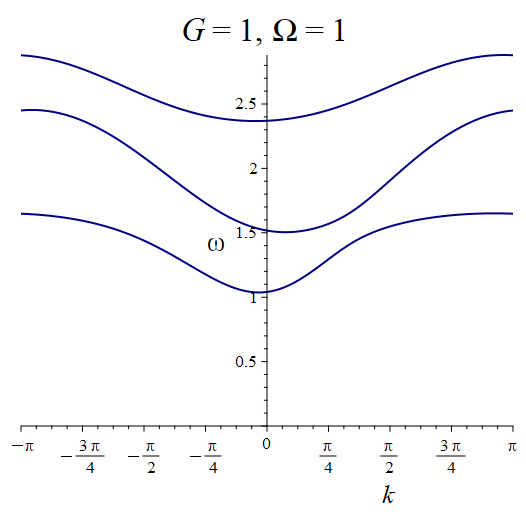}          \\
    \hline
\end{tabular}
 \captionof{figure}{\footnotesize Dispersion diagrams defined by the relation \eq{disperoddiny}, illustrating the frequency $\omega$ against the wavenumber $k,$ of the Dirichlet strip for a range of values of the gravity and gyricity parameters with $c=1, l=1$ and $m=1$.}
    \label{dirichletsym1}
    \end{table}
    
    The set of three dispersion curves for varying values of the parameters $G$ and $\Omega$ are shown in Fig. \ref{dirichletsym1} for the symmetry-constrained system with no vertical motions of the central nodal points. In the absence of gyricity, i.e. $\Omega=0$, the dispersion curves shown in Fig. \ref{dirichletsym1}(a) and Fig. \ref{dirichletsym1}(b) are symmetric relative to the $\omega$-axis, and the signs of the group and phase velocities are the same. In contrast, additional wave characteristics are present when gyricity is introduced in the model as detailed in Section \ref{newdispersionproper1} (see Fig. \ref{dirichletsym1}(c), (d)). Standing mode vibrations of the discrete lattice strip are also present for $\Omega>0$, but unlike the $\Omega=0$ case, they do not occur at $k=0$ or $k=\pm \pi.$ We also note that the dispersion diagram shown in Fig. \ref{dirichletsym1}(a) with $(G, \Omega)=(0,0)$ coincides with the three solid dispersion curves in Fig. \ref{dirichlet1a}. The solid dispersion curves are linked to a particular vibrational mode of the Dirichlet strip with zero vertical displacements of the central nodal points. In particular, the three dashed curves in the dispersion diagrams illustrated in Fig. \ref{dirichlet1a} correspond to the vibrations of the strip with zero horizontal displacements of the central nodal points, and are analysed in the next section.

\begin{table}[H]
        \setlength\tabcolsep{3pt}
    \expandafter\patchcmd\csname Gin@ii\endcsname   
      {\setkeys {Gin}{#1}}
      {\setkeys {Gin}
        {width=0.95\dimexpr\linewidth-2\tabcolsep,      
         valign=c, margin=0pt 3pt 0pt 3pt,#1}       
      }
      {}{}
    \centering
\begin{tabular}{|p{1.6cm}|p{7.3cm}|p{7.3cm}|}
    \hline
    &  \begin{center}\textbf{ $\Omega=0$}\end{center} & \begin{center}\textbf{ $\Omega = 1$}\end{center}    \\
    \hline%
$k=-0.2$
                                &  \raisebox{1.5cm}{\text{(a)}}\includegraphics{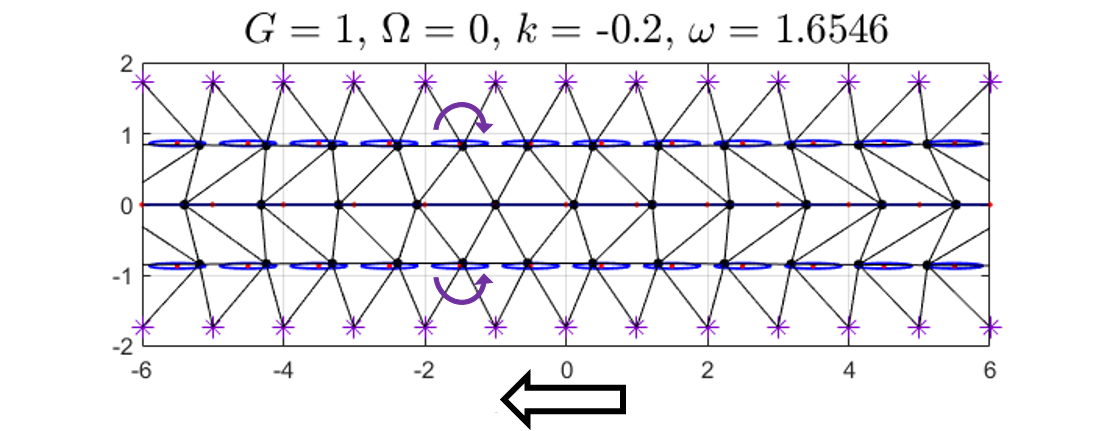}            & \raisebox{1.5cm}{\text{(b)}}\includegraphics{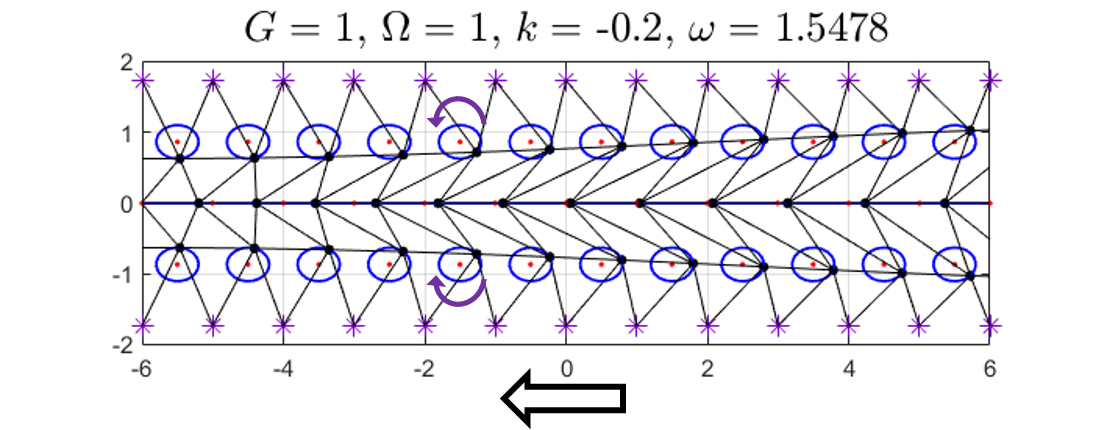}          \\
    \hline%
$k = 0.2$
                                & \raisebox{1.5cm}{\text{(c)}}\includegraphics{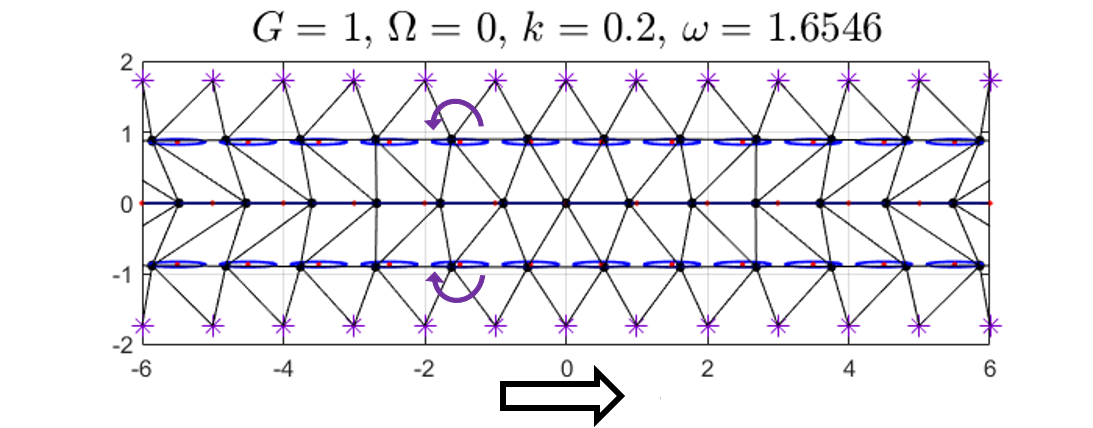}           &  \raisebox{1.5cm}{\text{(d)}}\includegraphics{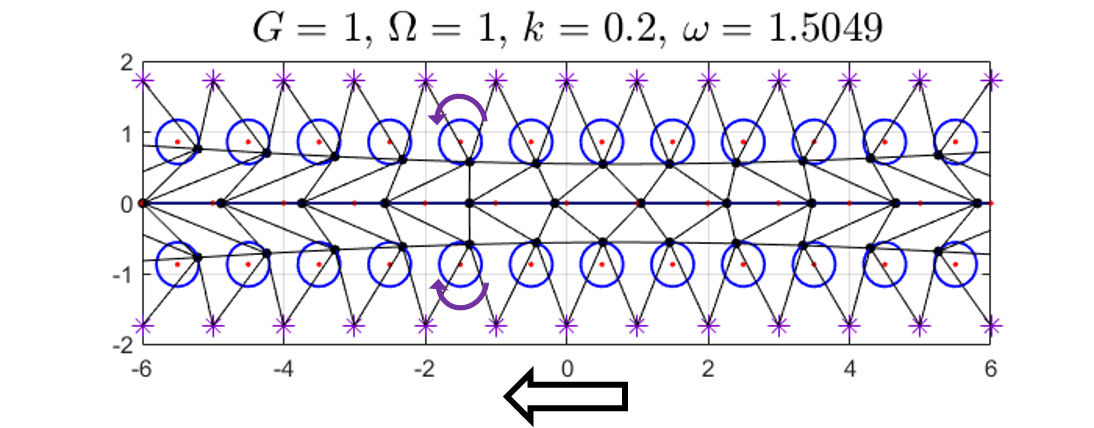}          \\
    \hline
\end{tabular}
 \captionof{figure}{\footnotesize Nodal point trajectories for the elastic Dirichlet strip for $\Omega=0$ and $\Omega=1$, with zero vertical displacements of the central nodal points. The normalised geometrical and physical parameter values of the model are $c=1, l=1$ and $m=1$. The associated dispersion diagrams are shown in Fig. \ref{dirichletsym1}(b), (d). }
    \label{symnewsys1}
    \end{table}
    
    Examples of nodal point trajectories for the elastic chiral Dirichlet lattice strip with zero vertical displacements of the central nodal points, are presented in Fig. \ref{symnewsys1} for $G=1$ and two different values of $\Omega.$ In the absence of gyricity, with $\Omega=0,$ the choices $k=-0.2$ and $k=0.2$ result in elastic waves that propagate along the negative and positive $x$-direction, respectively, and for each case, the group and phase velocities have the same sign. The related dispersion diagrams are shown in Fig. \ref{dirichletsym1}(b) and Fig. \ref{dirichletsym1}(d), which illustrate the symmetric and asymmetric dispersion curves for $\Omega=0$ and $\Omega \neq 0,$ respectively.  In particular, in Fig. \ref{symnewsys1}(b) and Fig. \ref{symnewsys1}(d) we also show the examples of the nodal point trajectories for $\Omega=1$ with $k=-0.2$ and $k=0.2,$ where the group velocities are negative in both cases, while the phase velocities are negative when $k=-0.2$ and positive when $k=0.2.$ This characteristic of chiral wave phenomena can be exhibited in metamaterial structures, with applications to dynamic shielding, cloaking devices and vibration damping \cite{brun2012vortex}.

\subsubsection{Modes with zero horizontal displacements of the central nodal points}\label{sigeveny1}

In this section we analyse vibrations of the chiral Dirichlet lattice strip where the horizontal displacements of the central nodal points are zero, which differ from the motions of central nodal points with zero vertical displacements presented in Section \ref{sygoddy}. We show that the horizontal displacements of the nodal points in the upper and lower layers of the strip are in opposite directions, while their vertical motions are in the same direction. 

In this case, the motions of the nodal points satisfy the conditions  $(u_{x}^{(1)}, u_{y}^{(1)})= (- u_{x}^{(3)},  u_{y}^{(3)})e^{- i k l}$ and $u^{(2)}_x=0,$ associated with the dispersion equation 
\begin{equation}
\sigma_{D}^{(2)}(m, G, \Omega, k, \omega, c, l) = 0, \label{anotherdisper1ion}
\end{equation}
where $\sigma_{D}^{(2)}$ is defined by \eq{sigeveny}. In a similar manner to Section \ref{sygoddy}, we can also implicitly differentiate \eq{anotherdisper1ion} with respect to $k$ and define the group velocity as follows
\begin{equation}
\frac{d \omega}{d k} = \frac{\mathcal{F}_{2}}{\mathcal{G}_{2}}, \label{groupg1}
\end{equation}
where 
\begin{eqnarray}
\begin{gathered}
{\footnotesize \mathcal{F}_{2} = c \sin(k l) m^2 l \omega^4 - c \Big( 2 G m^2 + \frac{27 c m}{4} \Big) \sin(k l) l \omega^2 - \frac{3 c^2 \cos(k l) \sqrt{3} \Omega m l}{4} \omega } \\  { \footnotesize + c \Big( - \frac{9 \cos(k l) c^2}{2} + G^2 m^2 + \frac{27 c G m}{4} + 9 c^2 \Big) \sin(k l) l,} \label{groupvelo2a}
\end{gathered}
\end{eqnarray}
and 
\begin{eqnarray}
\begin{gathered}
{\footnotesize{\mathcal{G}_{2} = 3 m^3 \omega^5 +  m \Big( 4 c m \cos(k l) - ( 2 \Omega^2 + 6 G ) m^2 - 18 c m \Big) \omega^3 }} \\  {\footnotesize{+  m \Big( - 4 \Big( G m + \frac{27 c}{8}\Big) c \cos(k l) + G \Big( \Omega^2 + 3 G \Big) m^2 + 3 c ( \Omega^2 + 6 G ) m + 24 c^2 \Big) \omega} } \\ {\footnotesize + \frac{3 \sqrt{3} \Omega c^2 m \sin(k l)}{4}.}  \label{groupvelo2b}
\end{gathered}
\end{eqnarray}

   \begin{figure}[H]
  \centering
  \hspace{-2cm}
  \begin{minipage}[b]{0.45\textwidth}
    \includegraphics[width=1.25\textwidth]{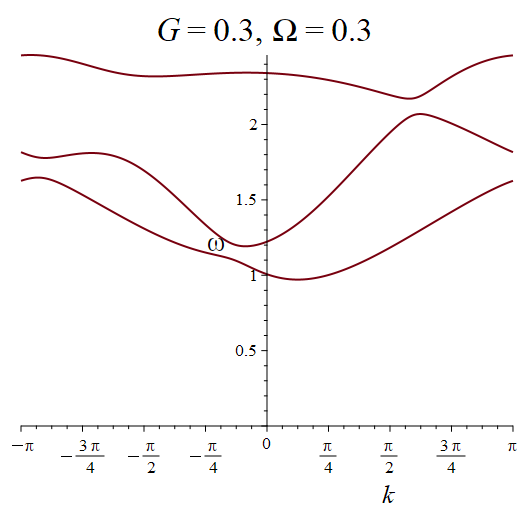}
  \end{minipage}
  \caption{\footnotesize Dispersion diagram of the Dirichlet strip for $G=0.3$ and $\Omega=0.3,$ defined by \eq{anotherdisper1ion}, corresponding to the strip vibrations with zero horizontal displacements of the central nodal points. The same three dispersion curves coincide with the dashed dispersion curves in Fig. \ref{dirdispersion1}. }
       \label{evenydisper1}
\end{figure}

The  dispersion diagram illustrated in Fig. \ref{evenydisper1} corresponds to the chiral Dirichlet strip modes with solely vertical motions of the central nodal points; the chosen normalised parameter values are the same as in the example presented in Fig. \ref{surfaces}(b). We note that in the limit when $\Omega\rightarrow\infty$, the two lower dispersion curves shown in Fig. \ref{evenydisper1} approach the curves described by $\omega=0$ and $\omega=\sqrt{(m G + 3 c)/m}.$ Accordingly, the upper dispersion curve increases for increasing $\Omega$. 

\begin{figure}[H]
  \centering
  \begin{minipage}[b]{0.49\textwidth}
    \hspace{-0.4cm}\includegraphics[width=1\linewidth]{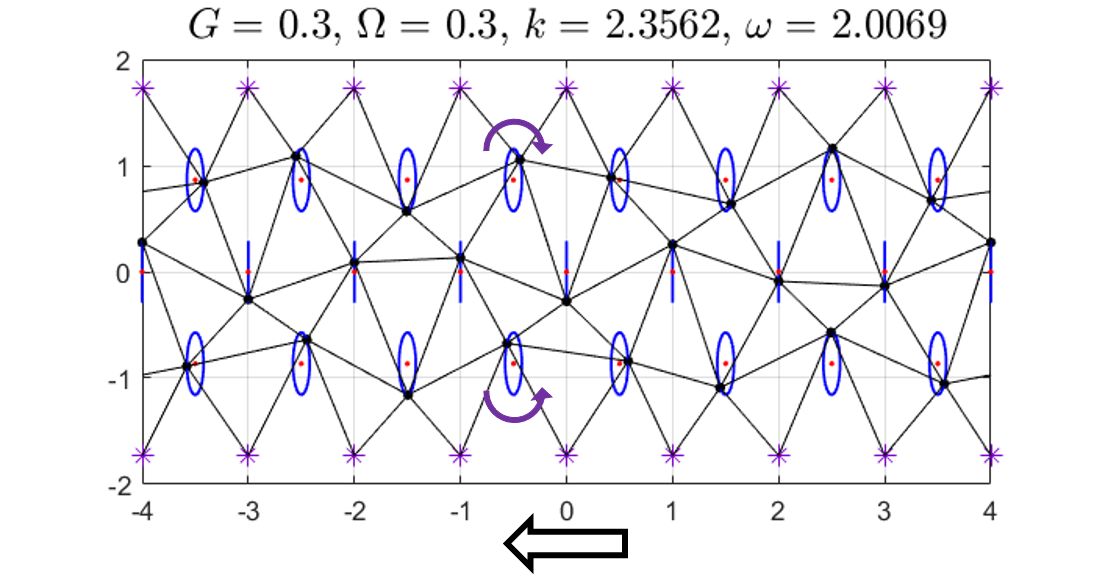}
    \centering\caption*{\footnotesize (a) Wave propagation in the negative $x$-direction}
  \end{minipage}
  \begin{minipage}[b]{0.49\textwidth}
    \hspace{-0.4cm}\includegraphics[width=1\linewidth]{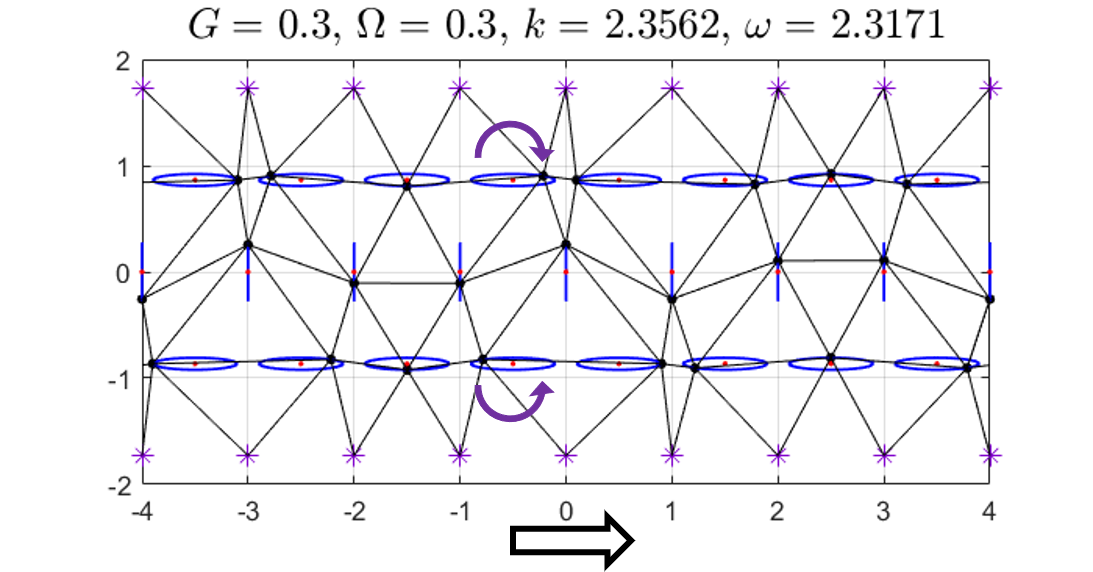}
    \centering\caption*{\footnotesize (b) Wave propagation in the positive $x$-direction}
  \end{minipage}
    \begin{minipage}[b]{0.7\textwidth}
    \hspace{-0.4cm}\includegraphics[width=1\linewidth]{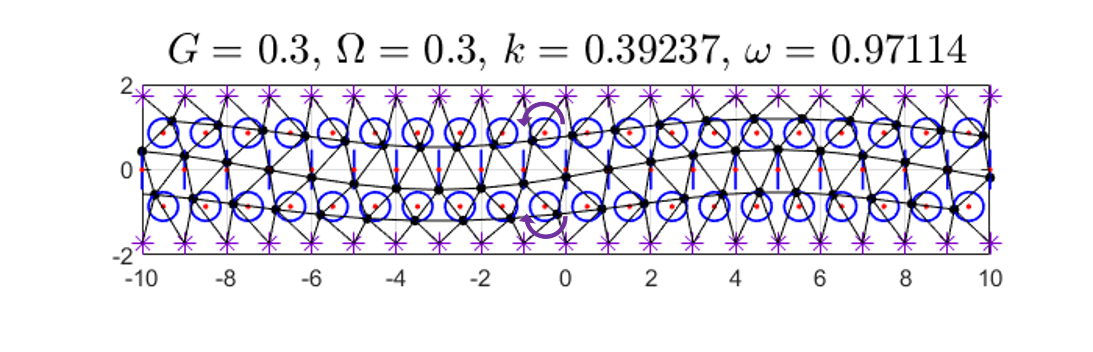}
    \centering\caption*{\footnotesize (c) Standing wave}
  \end{minipage}
  \caption{\footnotesize Nodal point trajectories of the chiral strip with Dirichlet boundary conditions for $\Omega=0.3,$ $\Omega=0.3$ and varying values of $\omega$ and $k$; (a) wave propagating in the negative $x$-direction with $(k,\omega) = (2.3562, 2.0069)$, (b) wave propagating in the positive $x$-direction with $(k,\omega) = (2.3562, 2.3171)$ and (c) standing wave with $(k,\omega) = (0.39237, 0.97114).$ For each example, the horizontal displacements of the central nodal points are zero.}
  \label{verticalcentralmodes1}
\end{figure}

Typical nodal point trajectories of the Dirichlet strip with zero horizontal displacements of the central nodal points are shown in Fig. \ref{verticalcentralmodes1} (see also Section \ref{greoupscla1}). The presence of a gyroscopic action for the nodal points in the upper and lower layers of the strip result in a coupling of the transverse displacement components, leading to elliptical trajectories of such masses. The group velocity (see \eq{groupg1}) is negative for the example shown in Fig. \ref{verticalcentralmodes1}(a) and positive for the oscillations displayed in Fig. \ref{verticalcentralmodes1}(b), resulting in waveforms propagating, respectively, in the negative and positive $x$-directions along the strip. Conversely, a representative example of the nodal trajectories for a standing mode is shown in Fig. \ref{verticalcentralmodes1}(c). For each example, the pair $(k,\omega)$ corresponds to a point on the dispersion diagram shown in Fig. \ref{evenydisper1}. 

\subsection{Active two-dimensional chiral strip with Dirichlet-type boundary conditions}\label{sdoublepend}

In this section, we investigate the dynamics of a lattice strip with Dirichlet boundary conditions (see Section \ref{Dirichlet12a}) for an active system. The term active refers to the gyricity being proportional to the frequency of the vibrations, where the gyricity is adjusted according to the ambient system. Mathematically, we enforce an active system by choosing $\Omega= \alpha \omega,$ where $\alpha$ is the gyricity control parameter that ensures the compatibility of the gyroscopic spinner with the time-harmonic strip vibrations \cite{brun2012vortex}. The latter choice leads to a dispersion equation similar to that of \eq{dispersionron1}, but each polynomial in $\omega$ can now be written as a product of two cubic polynomials in $\omega^2$ as detailed below. The characteristic motions of the central nodal point trajectories for the active Dirichlet strip are similar to the passive (non-active) system discussed in Section \ref{Dirichlet12a}, but the dispersion properties change according to the choice of $\alpha.$

For the active system, the dispersion relation for the Dirichlet strip takes the form
\begin{equation}
\text{det}[{\bf C}_{D} - \omega^2 ( {\bf M} - {\bf A}_{\alpha})] = 0, \label{activedisp}
\end{equation}
where the matrices ${\bf C}_{D}$ and ${\bf M}$ are defined in Section \ref{dirichletdispersion} and
\begin{equation}
 {\bf A}_{\alpha} = i  \begin{pmatrix} 0 & - m \alpha & 0 & 0 & 0 & 0 \\  m \alpha & 0 & 0 & 0 & 0 & 0 \\ 0 & 0 & 0 & 0 & 0 & 0 \\ 0 & 0 & 0 & 0 & 0 & 0 \\ 0 & 0 & 0 & 0 & 0 &  m \alpha \\ 0 & 0 & 0 & 0 & -  m \alpha & 0\end{pmatrix}.
\end{equation}
The dispersion relation \eq{activedisp} can be written as follows  (see \eq{sigoddy} and \eq{sigeveny}):
\begin{equation}\label{activf1a}
 \sigma_{D}^{(1)}(m, G, \Omega, k, \omega, c, l) \sigma_{D}^{(2)}(m, G, \Omega, k, \omega, c, l) \Big|_{\Omega=\alpha \omega} = 0,
\end{equation}
where
{\small
\begin{eqnarray}
\begin{gathered}
 \sigma_{D}^{(1)}(m, G, \Omega, k, \omega, c, l)\Big|_{\Omega=\alpha \omega} =  m^3 (1-\alpha^2) \omega^6 + m^2 ( m G \alpha^2  - 2c\cos(k l)\alpha^2 + 3c\alpha^2 - 3mG + 4c\cos(kl) - 9c) \omega^4 \\ 
+ \frac{m}{2}(-\sqrt{3}\sin(k l)\alpha c^2 + 6 G^2 m^2 - 16 G c m \cos(k l) + 8 c^2 \cos(k l)^2 + 36 c m G - 47 \cos(k l) c^2 + 52 c^2) \omega^2\\
-4mG\cos(k l)^2 c^2 - \frac{21}{2}\cos(k l)^2 c^3 + 4 G^2 \cos(kl)c m^2 + \frac{47}{2}G\cos(k l) c^2 m + 33\cos(k l) c^3\\  - G^3 m^3 - 9 G^2 c m^2 - 26 G c^2 m - 24 c^3,  \label{Actuvea1a}
\end{gathered}
\end{eqnarray}
}
and 
{\small
\begin{eqnarray}
\begin{gathered}
 \sigma_{D}^{(2)}(m, G, \Omega, k, \omega, c, l)\Big|_{\Omega=\alpha \omega} = m^3 (1-\alpha^2) \omega^6 + m^2 (m G \alpha^2 + 3c \alpha^2 - 3mG + 2c\cos(k l) - 9 c)\omega^4 \\ 
+ \frac{m}{2}(3\sqrt{3}\sin(k l)\alpha c^2 + 6 G^2 m^2 - 8 G\cos(k l) c m + 36 G c m - 27\cos(k l) c^2 + 48 c^2)\omega^2 \\ 
-\frac{9}{2}\cos(k l)^2 c^3 + 2 G^2\cos(k l) c m^2 + \frac{27}{2} G\cos(k l) c^2 m + 18\cos(k l) c^3 - G^3 m^3 - 9 G^2 c m^2 - 24 G c^2 m - 18 c^3.  \label{Actuvea2a}
\end{gathered}
\end{eqnarray}
}
To simplify our analysis, we take the normalised physical quantities $c=1, l=1$ and $m=1.$ Then, the leading coefficient of the dispersion relation \eq{activf1a} is $(1-\alpha^2)^2,$ i.e. the coefficient of $\omega^{12},$ which allows for a description of the dispersion properties by comparing the spinner constant $\alpha$ against the unit value. When $\alpha<1,$ the equation \eq{activedisp} yields six solutions for $\omega,$ resulting in six dispersion curves. Conversely, when $\alpha\geq 1,$ the dispersion equation degenerates, leading to four positive solutions for $\omega.$ The dispersion curves provide the frequencies and wavenumbers corresponding to the elliptical trajectories of the nodal points in the upper and lower layers of the strip, and to the horizontal or vertical motions of the central nodal points. 

\begin{figure}[H]
\centering
\subfloat[$\alpha=0$]{\label{fig:a}\includegraphics[width=0.45\linewidth]{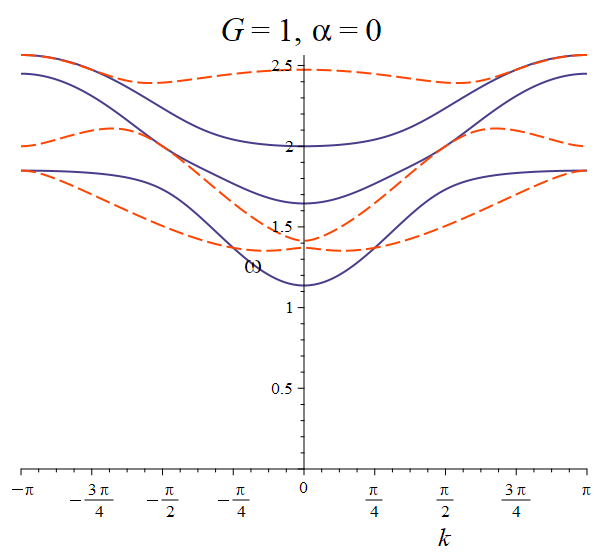}}\qquad
\subfloat[$\alpha=0.5$]{\label{fig:b}\includegraphics[width=0.45\linewidth]{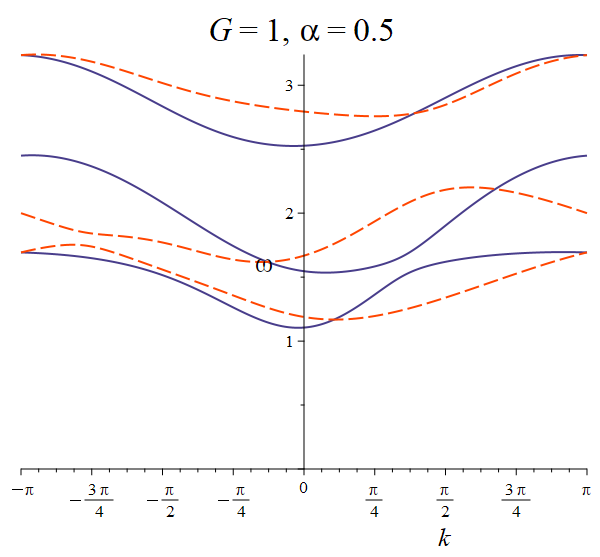}}\\
\subfloat[$\alpha=1$]{\label{fig:c}\includegraphics[width=0.45\textwidth]{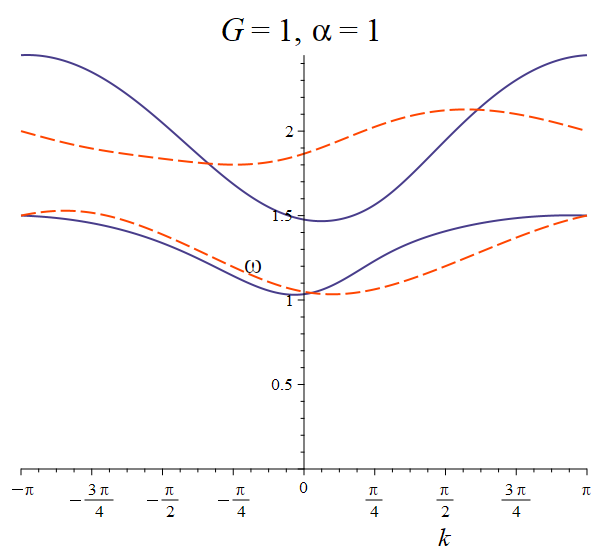}}\qquad%
\subfloat[$\alpha=2.5$]{\label{fig:d}\includegraphics[width=0.45\textwidth]{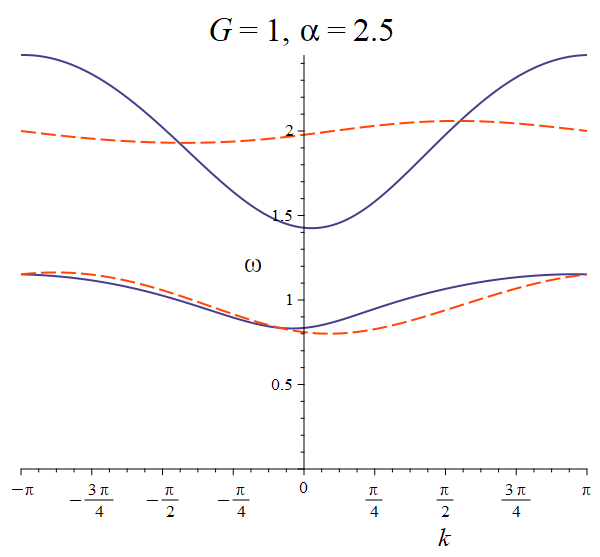}}%
\caption{\footnotesize Dispersion diagrams for the active Dirichlet lattice strip for various values of the gyricity control parameter $\alpha$. The gravity parameter is chosen as $G=1,$ and the physical parameter values, which are normalised by the system's natural units, are chosen as $c=1, l=1$ and $m=1$ (see Section \ref{Dirichlet12a}). The solid and dashed dispersion curves (defined by the roots of \eq{Actuvea1a} and \eq{Actuvea2a}) correspond to the trajectories of the central nodal points with zero vertical and zero horizontal displacements, respectively; (a) $\alpha=0,$ (b) $\alpha=0.5,$ (c) $\alpha=1$ and (d) $\alpha=2.5.$}
       \label{activesystem1a}
\end{figure}

Illustrative examples of the dispersion diagrams for varying values of the spinner constant $\alpha$, linked to the active chiral system, are presented in Fig. \ref{activesystem1a} for the gravity parameter $G=1$. We note that, similarly to the passive system discussed in Section \ref{governos1z}, the gravity parameter can be used to modulate the zero-frequency band gap of the active Dirichlet strip, while the presence of gyricity breaks the symmetry of the dispersion curves relative to the frequency axis. When $\alpha=0$ (no spinners are attached to the strip), there are generally six frequency values $\omega$ for a fixed wavenumber $k$, which may degenerate to frequencies of a higher multiplicity corresponding to crossing or touching points of the dispersion curves (see also Section \ref{crossDirichlet}). In this case, the locations of the degeneracy points coincide with the degeneracy points of the dispersion curves for the passive Dirichlet strip with zero gyricity. The dispersion degeneracies of the passive Dirichlet lattice strip are discussed in Section \ref{crossDirichlet}, which also includes examples of the corresponding nodal point trajectories. 
When $0< \alpha <1,$ the dispersion curves are asymmetric relative to the $\omega$-axis, as shown in Fig. \ref{activesystem1a}(b). When $\alpha \geq 1,$ there are four dispersion curves as illustrated in Fig. \ref{activesystem1a}(c) and Fig. \ref{activesystem1a}(d).  In the limit as $\alpha\rightarrow\infty$, the two lower dispersion curves tend towards $\omega=0$ for any wavenumber $k,$ while the two upper dispersion curves approach the curves described by 
\begin{equation}
\omega = \sqrt{\frac{m G + 3c }{m}}, ~~~ \omega = \sqrt{\frac{m G + 3c  - 2c\cos(k l)}{m}}.
\end{equation} 
Increasing $\Omega$ results in an additional finite-width stop band and a change in the number of crossing and touching points compared to $\alpha=0.$

The trajectories of the nodal points are comparable in behaviour between the active and passive Dirichlet lattice strips, while the dispersion characteristics differ as detailed above. In particular, for $\alpha <1,$ the solid and dashed dispersion curves (see Fig. \ref{activesystem1a}) for the active Dirichlet strip are associated with zero vertical and zero horizontal displacements of the central nodal points, respectively.

\subsection{Chiral gravitational lattice strip with Neumann boundary conditions}
\label{class}
In this section, we present a chiral lattice strip system with prescribed Neumann boundary conditions, subjected to gravity, for which the equations of motion are derived. The dispersion properties of the structure are analysed, and we show that, in a similar manner to the chiral Dirichlet lattice strip presented in Section \ref{Dirichlet12a}, waveforms with preferential directionality can also be exhibited in this type of lattice strip.

   \begin{figure}[H]
  \centering
  \begin{minipage}[b]{0.8\textwidth}
    \hspace{-0.6cm}\includegraphics[width=1\linewidth]{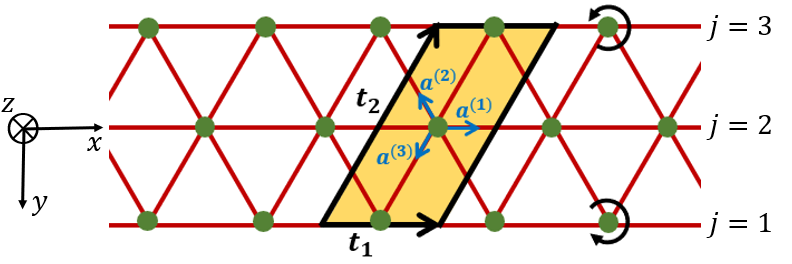}
  \end{minipage}
  \caption{\footnotesize Top view of the Neumann lattice strip consisting of gyropendulums, with nodal points connected by massless springs. The arrows indicate the choice of the lattice vectors ${\bf t}_1,$ ${\bf t}_2,$ and unit vectors ${\bf a}^{(j)},$ $j=1,2,3,$ (see Section \ref{governos1}) together with the shaded region demonstrating the choice of the unit cell of the lattice. The circular arrows represent the positive direction of spin of the gyroscopic spinners. The gravitational force acts in the positive $z$-direction, which points into the page.}
       \label{tstrip}
\end{figure}

\subsubsection{Equations of motion}\label{governos1}
We consider an infinite discrete lattice strip comprising of gyropendulums connected by springs with Neumann boundary conditions applied on the outer horizontal boundaries of the strip, as illustrated in Fig. \ref{tstrip}. The lengths and stiffnesses of the springs, as well as the nodal point masses, are the same as in the model analysed in Section \ref{Dirichlet12a}; only the boundary conditions and the geometries of the structure differ between the Dirichlet and Neumann strips. The problem assumptions and general parameter definitions used here are the same as in Section \ref{Dirichlet12a}.

The elementary cell of the periodic structure is shown in Fig. \ref{tstrip}, whose in-plane displacements are denoted by ${\bf u}^{( p, j)}$, for $j=1,2,3,$ where $p$ is the integer index. Assuming time-harmonic displacements with the radian frequency $\omega,$ it follows that the equations of motion of the three nodal points are given by 
\begin{equation}\label{gove1cc}
\begin{split}
- \frac{m \omega^2}{c} {\bf u}^{(p, 1)} =  {\bf a}^{(1)} \cdot \Big({\bf u}^{(p + 1, 1)} - {\bf u}^{(p , 1)}\Big){\bf a}^{(1)} + (-{\bf a}^{(1)}) \cdot \Big({\bf u}^{(p - 1, 1)} - {\bf u}^{(p , 1)}\Big)(-{\bf a}^{(1)}) \\ + {\bf a}^{(2)} \cdot \Big({\bf u}^{( p - 1, 2)} - {\bf u}^{( p , 1)}\Big){\bf a}^{(2)} + (-{\bf a}^{(3)}) \cdot \Big({\bf u}^{( p , 2)} - {\bf u}^{( p , 1)}\Big)(-{\bf a}^{(3)})+i \frac{m \Omega \omega}{c} {\bf R} {\bf u}^{( p ,1)} - \frac{m G}{c} {\bf u}^{( p ,1)},
\end{split}
\end{equation}

\begin{equation}\label{gove2cc}
\begin{split}
- \frac{m \omega^2}{c} {\bf u}^{(p , 2)} =  {\bf a}^{(1)} \cdot \Big({\bf u}^{( p + 1, 2)} - {\bf u}^{( p , 2)}\Big){\bf a}^{(1)} + (-{\bf a}^{(1)}) \cdot \Big({\bf u}^{( p - 1, 2)} - {\bf u}^{( p , 2)}\Big)(-{\bf a}^{(1)}) \\ + {\bf a}^{(2)} \cdot \Big({\bf u}^{( p - 1, 3)} - {\bf u}^{( p , 2)}\Big){\bf a}^{(2)} + (-{\bf a}^{(2)}) \cdot \Big({\bf u}^{( p + 1 , 1)} - {\bf u}^{( p , 2)}\Big)(-{\bf a}^{(2)}) \\ + {\bf a}^{(3)} \cdot \Big({\bf u}^{( p , 1)} - {\bf u}^{( p , 2)}\Big){\bf a}^{(3)}  + (-{\bf a}^{(3)}) \cdot \Big({\bf u}^{( p , 3)} - {\bf u}^{( p , 2)}\Big)(-{\bf a}^{(3)})  - \frac{ m G}{c} {\bf u}^{( p ,2)},
\end{split}
\end{equation}
and 
\begin{equation}
\begin{split}
- \frac{m \omega^2}{c} {\bf u}^{( p , 3)} =  {\bf a}^{(1)} \cdot \Big({\bf u}^{( p + 1 , 3)} - {\bf u}^{( p , 3)}\Big){\bf a}^{(1)} + (-{\bf a}^{(1)}) \cdot \Big({\bf u}^{( p - 1, 3)} - {\bf u}^{( p , 3)}\Big)(-{\bf a}^{(1)}) \\ + (-{\bf a}^{(2)}) \cdot \Big({\bf u}^{( p + 1, 2)} - {\bf u}^{( p , 3)}\Big)(-{\bf a}^{(2)}) + {\bf a}^{(3)} \cdot \Big({\bf u}^{( p , 2)} - {\bf u}^{( p , 3)}\Big){\bf a}^{(3)}- i \frac{ m \Omega \omega}{c} {\bf R} {\bf u}^{( p ,3)} - \frac{m G }{c} {\bf u}^{( p ,3)}, \label{gove3cc}
\end{split}
\end{equation}
where the parameters $m, c, \Omega$ and $G$ are defined in Section \ref{governos1z}. The vectors ${\bf a}^{(j)}$ for $j=1,2,3,$ are given in \eq{unitoa2}, and the rotation matrix ${\bf R}$ is defined by \eq{90rotation}. The zero gyricity parameter is chosen for the nodal points along the central layer ($j=2$ in Fig. \ref{tstrip}) and the gravity parameter acts uniformly at each nodal point. We also assume that the gyricities of the lower ($j=1$) and upper ($j=3$) layers of nodal points have the opposite sign, but are the same in magnitude. 

Similarly to the method provided in Section \ref{governos1z}, by applying the Bloch-Floquet conditions (see \eq{blockfa}) into the system \eq{gove1cc}-\eq{gove3cc}, we obtain the following system
\begin{equation}\label{disper123acc}
[{\bf C}_{N} - \omega^2 {\bf M} +  \omega {\bf A}]{\bf W} = {\bf 0},
\end{equation}
where 
{ \scriptsize
\begin{equation}
 {\bf C}_{N} = 
-\begin{pmatrix} 2c [\cos(k l) -\frac{5}{4}]- G m  & 0 & \frac{c}{4}(e^{-i k l} + 1) & \frac{c \sqrt{3} }{4}(-e^{-i k l} + 1) & 0 & 0 \\ 0 & -\frac{3 c}{2} - G m & \frac{c \sqrt{3}}{4}(- e^{- i k l}+1) & \frac{3 c}{4}(e^{- i k l}+1) & 0 & 0 \\ \frac{c}{4} (e^{i k l}+1) & \frac{c\sqrt{3}}{4} (-e^{i k l}+1) & 2c[\cos(k l) -\frac{3}{2}]  - G m & 0 & \frac{c}{4} (e^{- i k l} + 1) & \frac{c\sqrt{3}}{4} (-e^{- i k l} + 1) \\ \frac{c \sqrt{3}}{4}(-e^{i k l}+1) & \frac{3 c}{4} (e^{i k l}+1) & 0 & -3c - G m & \frac{c \sqrt{3}}{4}(-e^{- i k l} + 1) & \frac{3 c}{4} (e^{-i k l}+1) \\ 0 & 0 & \frac{c}{4} (e^{i k l} + 1) & \frac{c \sqrt{3}}{4} (-e^{i k l} + 1) & 2 c [\cos(k l) -\frac{5}{4}] - G m  & 0 \\ 0 & 0 & \frac{c \sqrt{3}}{4} (-e^{i k l}+1) & \frac{3 c}{4} (e^{i k l} + 1) & 0 & -\frac{3c}{2} - G m \end{pmatrix} \label{newstiffy1} 
\end{equation}
}
and ${\bf M}$ and ${\bf A}$ are given in Section \ref{dirichletdispersion}. The change in the prescribed conditions on the outer boundaries of the chiral lattice strip, from the Dirichlet boundary conditions used in Section \ref{Dirichlet12a} to the Neumann boundary conditions used here, results in different representations of the gravity-stiffness matrices. In the following analysis, we demonstrate how this change in boundary conditions affects the dispersion properties. In particular, we show that zero-frequency band gaps occur only due to the presence of gravitational forces when Neumann boundary conditions are prescribed. Conversely, zero-frequency stop bands exist independently of gravitational effects when Dirichlet boundary conditions are applied as discussed in Section \ref{newdispersionproper1}. However, in both cases, the gravity parameter can be used to control the widths of the stop bands. We also show that the presence of gyricity leads to asymmetric dispersion diagrams for the chiral Neumann lattice strip, similar to those observed for the Dirichlet strip. This asymmetric feature of the dispersion curves is also present for the high-frequency inertia-gravity waves in the equatorial region as detailed in Section \ref{rotatingshall1}.

\subsubsection{Dispersion characteristics of the discrete Neumann strip}\label{combinedf1a}

To obtain non-trivial solutions of the system \eq{disper123acc}, the matrices ${\bf C}_{N}, {\bf M}$ and ${\bf A}$ must satisfy the condition 

\begin{equation}
\text{det}[{\bf C}_{N} - \omega^2 {\bf M} + \omega {\bf A}] =  \sigma_{N}^{(1)} (\Omega, G, m, \omega, k) \sigma_{N}^{(2)}(\Omega, G, m, \omega, k) = 0, \label{neumangrav1a}
\end{equation}
where
\begin{eqnarray}
\begin{gathered}
{\footnotesize \sigma_{N}^{(1)}(m, G, \Omega, k, \omega) = m^3\omega^6 - \Big( - 4 c \cos(k l) + ( \Omega^2 + 3 G )m + 7 c \Big) m^2 \omega^4  }  \\  {\footnotesize{  + \Big( G \Omega^2 m^2 - 2 c \cos(k l) m \Omega^2 + 3 G^2 m^2 - 8 c \cos(k l) G m + 4 c^2 \cos(k l)^2 + 3 m c \Omega^2 + 14 G c m }} \\  {\footnotesize - \frac{33 c^2 \cos(k l)}{2} + \frac{59 c^2}{4} \Big) m \omega^2   - \frac{\sqrt{3}\Omega c^2 m \omega \sin(k l)}{2} -4 G \cos(k l)^2 c^2 m - \frac{9 \cos^2(k l) c^3}{2} } \\ {\footnotesize + 4 G^2 \cos(k l) c m^2 + \frac{33 G \cos(k l) c^2 m}{2}   + \frac{27 \cos(k l) c^3}{2}  -G^3 m^3 - 7 G^2 c m^2 - \frac{59 G c^2 m}{4} - 9 c^3}, \label{Nsigoddy11}
\end{gathered}
\end{eqnarray}
and 
\begin{eqnarray}
\begin{gathered}
\sigma_{N}^{(2)}(m, G, \Omega, k, \omega) = m^3\omega^6 - \Big(  - 2 c \cos(k l) + (\Omega^2  + 3 G)m + 7 c \Big) m^2 \omega^4   \\ + \Big(  G \Omega^2 m^2 + 3 G^2 m^2 - 4 c \cos(k l) m G + 3 m c \Omega^2 + 14 G c m - \frac{ 21 c^2 \cos(k l) }{2} + \frac{51 c^2}{4} \Big) m \omega^2\\ + \frac{3 \sqrt{3}\Omega c^2 m \omega \sin(k l)}{2} - \frac{9 \cos^2(k l) c^3}{2} + 2 G^2 \cos(k l) c m^2 + \frac{21 G \cos(k l) c^2 m}{2} \\ + 9 \cos(k l) c^3 - G^3 m^3 - 7 G^2 c m^2 - \frac{51 G c^2 m}{4} - \frac{9c^3}{2}. \label{Nsigevenyzz}
\end{gathered}
\end{eqnarray}
Similarly to the dispersion relation obtained for the Dirichlet lattice strip (see \eq{dirichletdispersion}), the dispersion equation for the Neumann lattice strip can also be written as a product of two sixth-order polynomials in $\omega$ as shown above.  In particular, the dispersion curves for $\Omega<0$ are equivalent to the dispersion curves for $\Omega>0$ with a reflection in the vertical $\omega$-axis and thus, without loss of generality, we only consider the latter case.
 The polynomial \eq{Nsigoddy11} is associated with the zero vertical displacements of the central nodal points (the $j=2$ layer in Fig. \ref{tstrip}), while the roots of the polynomial \eq{Nsigevenyzz} correspond to the strip modes with zero horizontal displacements of the central nodal points. An analogous characteristic of the nodal point motions, in connection with the dispersion relation, also holds for the chiral Dirichlet strip. Furthermore, the nodal points in the lower ($j=1$) and upper ($j=3$) layers of the chiral Neumann strip move in opposite orientations.

\begin{figure}[H]
  \centering
  \begin{minipage}[b]{0.45\textwidth}
    \hspace{-0.4cm}\includegraphics[width=0.8\linewidth]{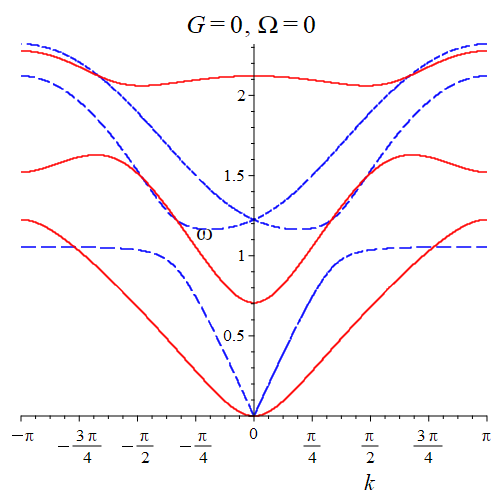}
    \centering\caption*{\footnotesize (a)}
  \end{minipage}
  \begin{minipage}[b]{0.45\textwidth}
    \hspace{-0.4cm}\includegraphics[width=0.8\linewidth]{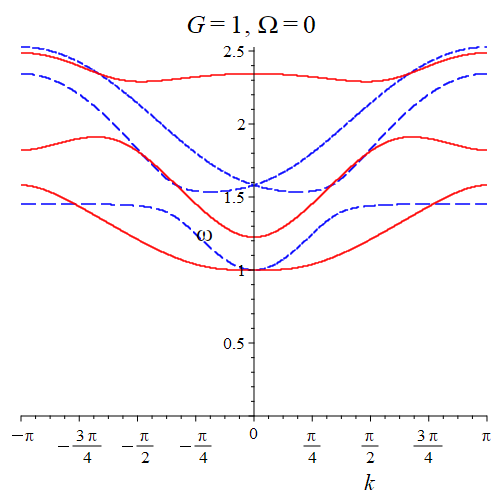}
    \centering\caption*{\footnotesize (b)}
  \end{minipage}
  \begin{minipage}[b]{0.45\textwidth}
    \hspace{-0.4cm}\includegraphics[width=0.8\linewidth]{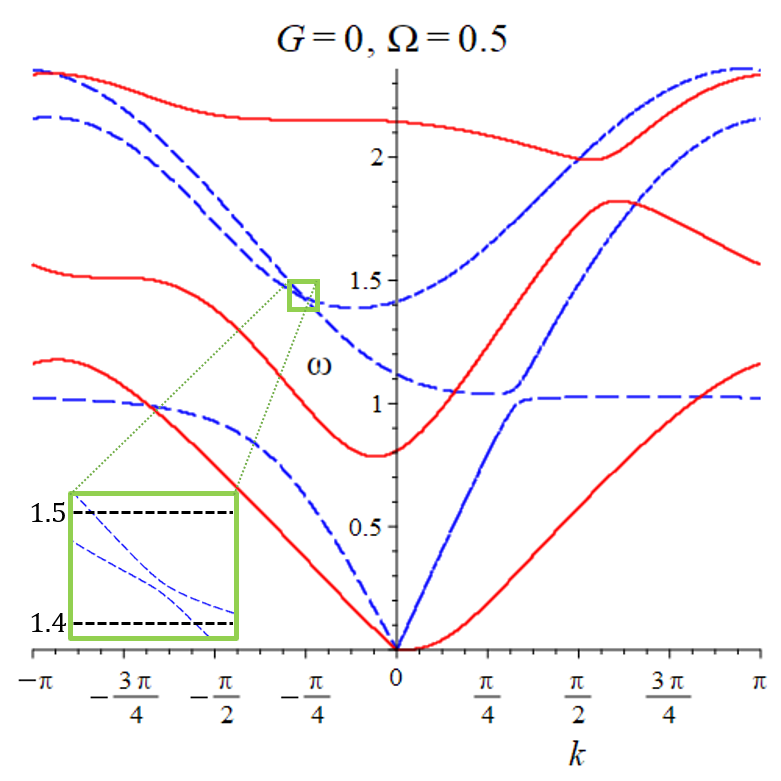}
    \centering\caption*{\footnotesize (c)}
  \end{minipage}
  \begin{minipage}[b]{0.45\textwidth}
    \hspace{-0.4cm}\includegraphics[width=0.8\linewidth]{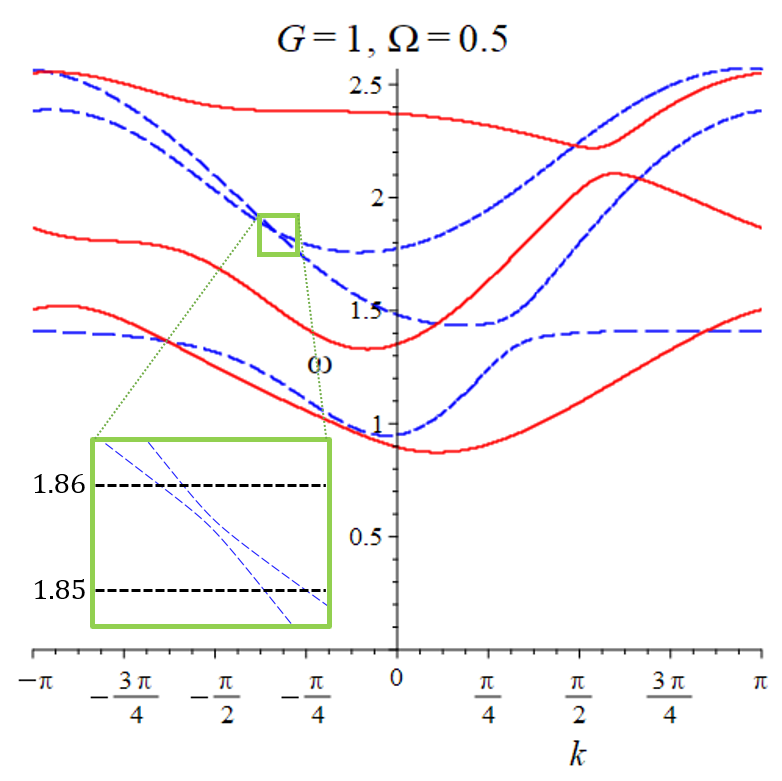}
    \centering\caption*{\footnotesize (d)}
  \end{minipage}
  \caption{\footnotesize Dispersion diagrams for the chiral gravitational lattice strip with Neumann boundary conditions; (a) $(G,\Omega)=(0,0),$ (b) $(G, \Omega)=(1, 0),$ (c) $(G, \Omega)=(0, 0.5)$ and (d) $(G, \Omega)=(1, 0.5).$}
  \label{nonactivedisper}
\end{figure}

\begin{figure}[H]
  \centering
    \begin{minipage}[b]{0.6\textwidth}
    \hspace{-0.25cm}\includegraphics[width=1\linewidth]{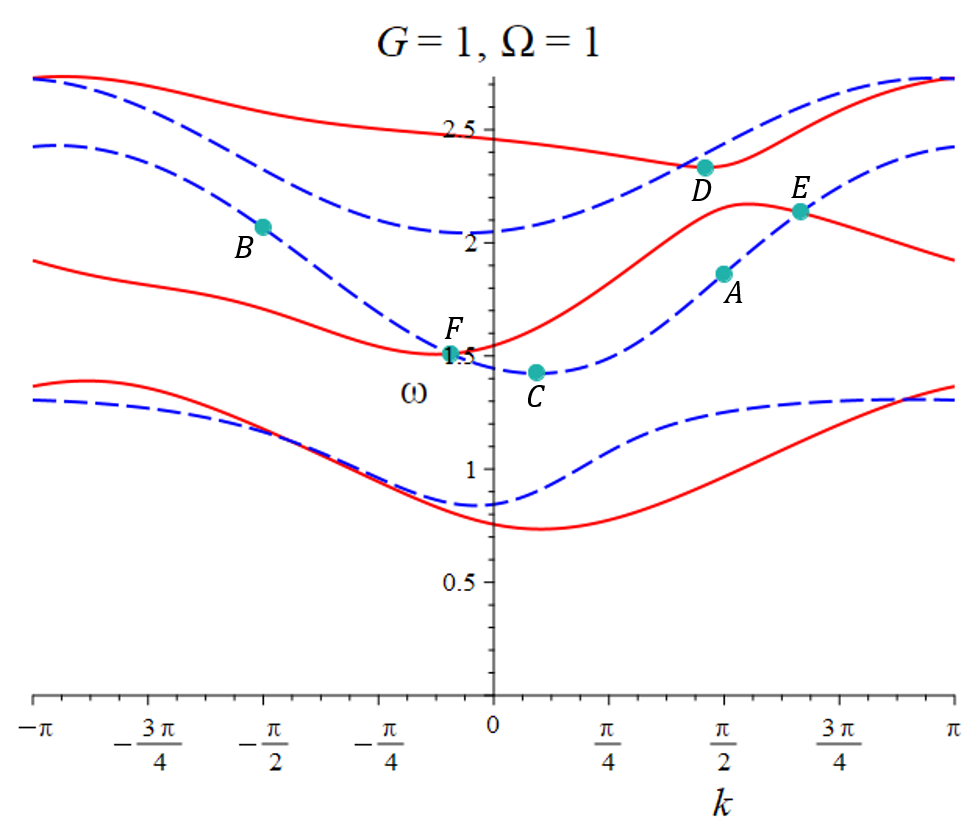}
  \end{minipage}
  \caption{\footnotesize Dispersion diagram for the Neumann strip with $G=1$ and $\Omega=1.$ The nodal trajectories associated with the points $A,B,C,D,E$ and $F$ are discussed below.}
  \label{newpassivesystema}
\end{figure}

 In the illustrative examples, we take the normalised physical quantities $l=1,$ $c = 1$ and $m=1.$ Similarly to Section \ref{newdispersionproper1}, the units of measurement will not be shown. Fig. \ref{nonactivedisper} and Fig. \ref{newpassivesystema} illustrate the dispersion curves of the Neumann strip for different values of the gyricity and gravity parameters. There are six dispersion curves; the three dashed curves correspond to the roots of \eq{Nsigoddy11}, while the three solid curves are characterised by the zeros of \eq{Nsigevenyzz}. It is apparent that there are intersection points of the dispersion curves, which correspond to the degeneracy points of the dispersion equation. These points also occur for the Dirichlet lattice strip as discussed in Section \ref{crossDirichlet}. In both the Dirichlet and Neumann lattice strips, the gyroscopic effect introduced by the spinners allows for the chiral strip to support vortex-type waves. The nodal point trajectories of the Neumann strip corresponding to the points $A,B,C,D,E$ and $F$ in Fig. \ref{newpassivesystema} are presented in the next section.

  When gravitational forces are absent in the system ($G=0$), there are no zero-frequency band gaps as shown in Fig. \ref{nonactivedisper}(a) and Fig. \ref{nonactivedisper}(c).  When $G>0,$ a zero-frequency stop band is formed (see Figs. \ref{nonactivedisper}(b), (d)). A similar feature was also observed in \cite{kandiah2023effect}, where a chiral chain of gyropendulums subjected to gravity was considered. In addition, for $\Omega=0,$ the dispersion curves are symmetric relative to the $\omega$-axis, whereas for $\Omega > 0,$ they exhibit asymmetry. An analogous effect of the gyroscopic action was noted for the dispersion diagrams of the Dirichlet strip as detailed in Section \ref{newdispersionproper1}. However, the presence of gravity leads to different properties of the dispersion curves due to the different boundary conditions on the lattice strip, especially in the low-frequency regime. Gyroscopic spinners also affect band gap widths as well as the locations of the degeneracy points as illustrated in Fig. \ref{nonactivedisper}(c), Fig. \ref{nonactivedisper}(d) and Fig. \ref{newpassivesystema}, where the gyricity parameter of the spinners are non-zero. In the limit when $\Omega\rightarrow \infty,$ the two low-frequency dispersion curves tend to zero and the two intermediate dispersion curves approach the curves described by 
  \begin{equation}
  \omega = \sqrt{\frac{m G + 3c}{m}}, ~~~ \omega = \sqrt{\frac{mG + 3 c - 2 c \cos(k l)}{m}}.
  \end{equation}
  Additionally, the two high-frequency dispersion curves increase for increasing $\Omega.$
 The above frequencies, linked to gyroscopic rigidity, are the same as the limiting frequencies of the chiral Dirichlet strip (see Section \ref{newdispersionproper1}). This shows that gyroscopic rigidity is a property intrinsic to the rotating spinners.

\subsubsection{Dynamics of the chiral gravitational Neumann strip}\label{neumanstaois}

In this section, we analyse the modes of an infinite chiral Neumann strip with the normalised parameter values $G=1, \Omega=1, c=1,l=1$ and $m=1,$ in connection with the propagating and standing waveforms along the structure.
\begin{figure}[H]
  \centering
  \begin{minipage}[b]{0.49\textwidth}
    \hspace{-0.6cm}\includegraphics[width=1\linewidth]{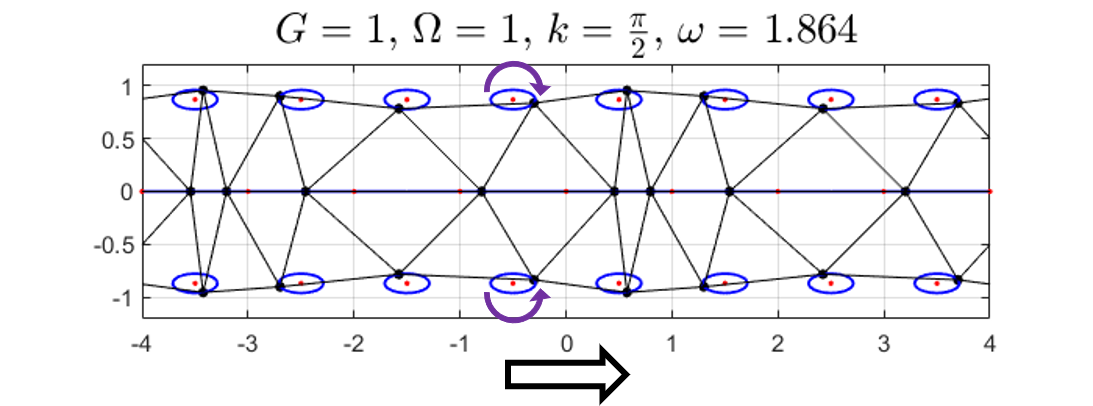}
    \centering\caption*{\footnotesize (a)}
  \end{minipage}
  \begin{minipage}[b]{0.49\textwidth}
    \hspace{-0.6cm}\includegraphics[width=1\linewidth]{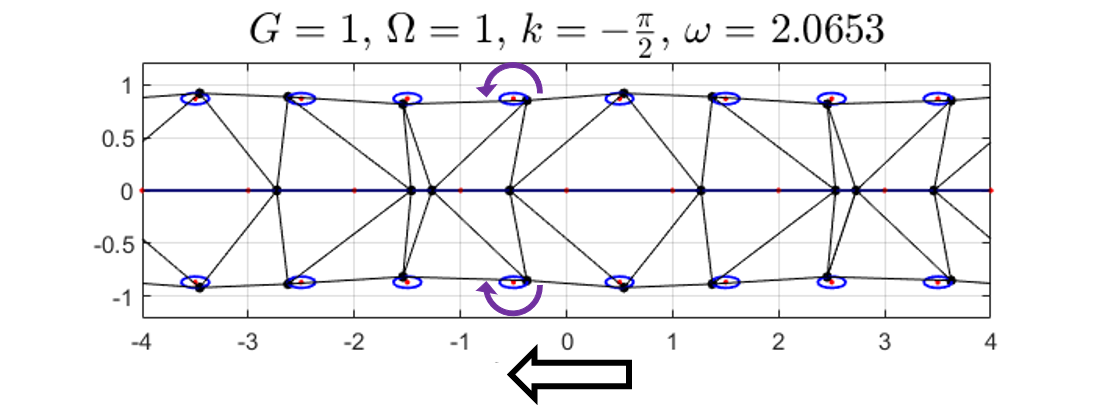}
    \centering\caption*{\footnotesize (b)}
  \end{minipage}
  \caption{\footnotesize  Nodal point trajectories of the chiral gravitational lattice strip with prescribed Neumann boundary conditions. The circular arrows show the orientation of motion of the nodal points. The motions are associated with the points on the dispersion diagram shown in Fig. \ref{newpassivesystema}, denoted by $A$ and $B$ for parts (a) and (b), respectively; (a) wave propagating in the positive $x$-direction with $(k,\omega)= (\pi/2, 1.864)$ and (b) wave propagating in the negative $x$-direction with $(k,\omega)=(-\pi/2, 2.0653).$ }
  \label{inertiogravity}
\end{figure}

Fig. \ref{inertiogravity} shows the nodal points of the Neumann lattice strip for two different frequency values corresponding to the propagating waveforms. The nodal points in the upper and lower layers of the strip follow elliptical paths with the major axis aligned parallel to the strip, while the nodal points along the central layer move horizontally with no vertical components. Similar motions were also observed for the nodal points of the Dirichlet strip in Section \ref{newdispersionproper1}. Additionally, the group velocity is positive for the illustrative example shown in Fig. \ref{inertiogravity}(a), resulting in waveforms propagating in the positive $x$-direction, whereas Fig. \ref{inertiogravity}(b) displays waves travelling in the negative $x$-direction, where the group velocity is negative. We note that the vibration frequency is higher in Fig. \ref{inertiogravity}(b) compared to Fig. \ref{inertiogravity}(a) due to the asymmetry in the dispersion curves shown in Fig. \ref{newpassivesystema}. The examples in Fig. \ref{inertiogravity}(a) and Fig. \ref{inertiogravity}(b) are linked to the points $A$ and $B$ in Fig. \ref{newpassivesystema}.

\begin{figure}[H]
  \centering
  \begin{minipage}[b]{0.49\textwidth}
    \hspace{-0.6cm}\includegraphics[width=1\linewidth]{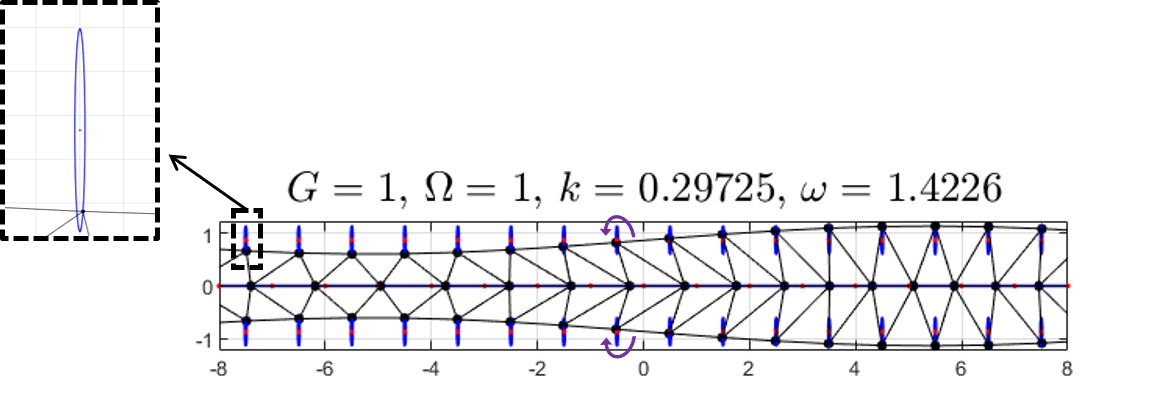}
    \centering\caption*{\footnotesize (a)}
  \end{minipage}
  \begin{minipage}[b]{0.49\textwidth}
    \hspace{-0.6cm}\includegraphics[width=1\linewidth]{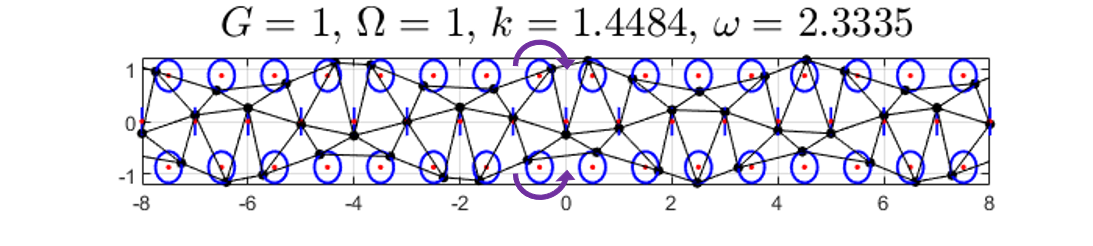}
    \centering\caption*{\footnotesize (b)}
  \end{minipage}
  \caption{\footnotesize Standing modes of the Neumann strip for the normalised parameter values $G=1,\Omega=1, m=1, c=1$ and $l=1$ (see Section \ref{governos1}); (a) $(k, \omega)= (0.29725, 1.4226)$ and (b) $(k, \omega) = (1.4484, 2.3335).$ }
  \label{standing modes}
\end{figure}

Examples of trajectories for the chiral Neumann lattice strip corresponding to standing modes are shown in Fig. \ref{standing modes}; these are associated with zero group velocities and zero net energy transfer along the strip. In both examples, the positive phase velocity corresponds to the phase of wave moving in the positive $x$-direction. In Fig. \ref{standing modes}, the trajectories of the nodal points in the upper and lower layers trace ellipses, while the central nodal points move either vertically or horizontally depending on the chosen vibration frequency. The frequencies and wavenumbers of the nodal point motions in Fig. \ref{standing modes}(a) and Fig. \ref{standing modes}(b) correspond to the points $C$ and $D$ in the dispersion diagram shown in Fig. \ref{newpassivesystema}. We also note that the eigenmodes of the finite three-dimensional belt shown in Fig. \ref{eigenmode12}, exhibit similar motions to the standing modes of the Neumann strip illustrated in Fig. \ref{standing modes} for the same gravity and gyricity parameters. In particular, the nodal points on the central ring of the chiral belt also oscillate either horizontally or vertically, while the nodal points on the upper and lower rings trace elliptical trajectories with opposite orientations. Moreover, changes in frequency values affect the magnitudes and eccentricities of the ellipses, as well as the orientations of motion of the nodal points.

\begin{figure}[H]
  \centering
  \begin{minipage}[b]{0.49\textwidth}
    \hspace{-0.6cm}\includegraphics[width=1\linewidth]{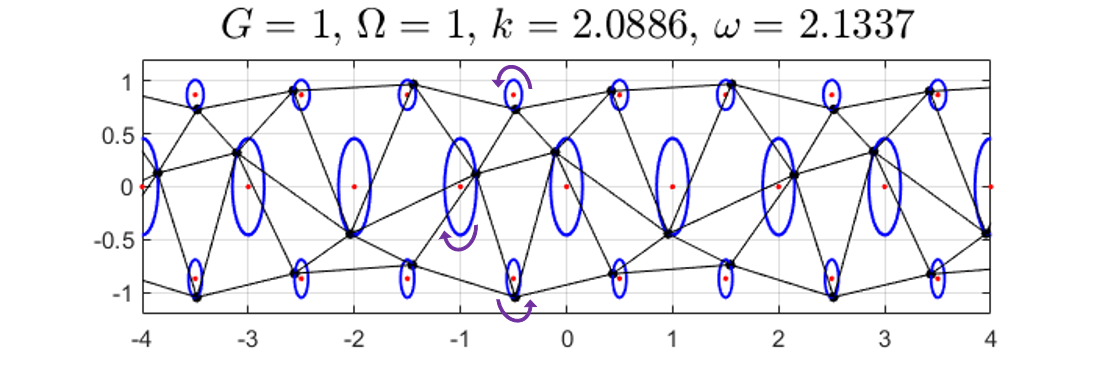}
    \centering\caption*{\footnotesize (a)}
  \end{minipage}
  \begin{minipage}[b]{0.49\textwidth}
    \hspace{-0.6cm}\includegraphics[width=1\linewidth]{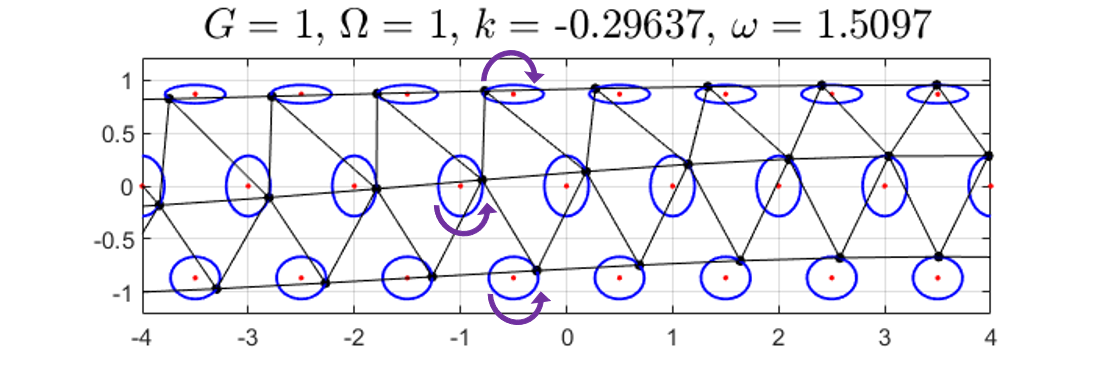}
    \centering\caption*{\footnotesize (b)}
  \end{minipage}
  \caption{\footnotesize Nodal point trajectories linked to the crossing points of dispersion curves. The corresponding dispersion diagram is shown in Fig. \ref{newpassivesystema}, and the examples are associated with the points $E$ and $F$ for parts (a) and (b), respectively. (a) $(k, \omega) = (2.0886, 2.1337)$ and (b) $(k, \omega) = (-0.29637, 1.5097).$}
  \label{crossing_modes1}
\end{figure}

In Fig. \ref{crossing_modes1}, we present two examples of nodal point trajectories of the Neumann strip, with the frequencies and wavenumbers corresponding to the crossing points of the dispersion curves shown in Fig. \ref{newpassivesystema}. In the examples shown in Fig. \ref{crossing_modes1}, the trajectories of the nodal points are ellipses of varying sizes and orientations. Additionally, compared to the examples presented in Fig. \ref{inertiogravity} and  Fig. \ref{standing modes}, there is no symmetry between the upper and lower nodal point trajectories shown in Fig. \ref{crossing_modes1}. This asymmetry was also observed in the nodal point trajectories of the Dirichlet lattice strip corresponding to the dispersion degeneracy points (see Fig. \ref{DiracPoints1a}). It is noted that although the central nodal points are characterised by zero gyricity, their trajectories are elliptical as displayed in Fig. \ref{crossing_modes1} in contrast with the linear motions observed in Fig. \ref{inertiogravity} and Fig. \ref{standing modes}. 

The examples presented in this section show a novel approach to modelling chiral elastic waveforms in a discrete strip with Neumann boundary conditions without perturbing the system, while taking into account the combined actions of gravitational and gyroscopic forces. The choices of frequency and wavenumber are important in describing the trajectories of the nodal points, influencing wave properties such as propagation speeds and the preferential directionality of propagation of the waveforms. The dispersion diagrams presented in Fig. \ref{nonactivedisper} and Fig. \ref{newpassivesystema} determine how waveforms of different frequencies propagate through the discrete Neumann strip. Typical trajectories of the non-chiral Neumann strip are discussed in Section \ref{nochiralitysec}, with an emphasis on standing and propagating waveforms.

\subsection{Non-chiral discrete Neumann lattice strip}\label{nochiralitysec}
In this section, we examine the modes of a lattice strip subjected to gravity with Neumann boundary conditions in the absence of gyroscopic forces, corresponding to $\Omega=0.$ In this case, the dispersion equation can be written as product of two cubic polynomials in $\omega^2$ as follows (see also \eq{neumangrav1a}):
\begin{equation}
F_{1}(G, k, \omega) F_{2}(G, k, \omega) = 0,  \label{nodisper1}
\end{equation}
where
{\small
\begin{eqnarray}
\begin{gathered}
F_{1}(G, k, \omega) = \sigma_{N}^{(1)}\Big|_{\Omega=0} = m^3 \omega^6 + (4  c \cos(k l) - 3 G m - 7 c) m^2 \omega^4 \\ + \Big( 4 c^2  \cos^2(k l) - 4 \Big( 2 G m + \frac{33}{8} c  \Big) c \cos(k l) + 3 G^2 m^2 + 14 m c G + \frac{59 c^2}{4} \Big) m \omega^2 \\ - 4 \Big(G m + \frac{9c}{8}\Big) c^2 \cos^2(k l) + 4 \Big( G^2 m^2 + \frac{33}{8} G c m + \frac{27}{8} c^2 \Big) c \cos(k l) - G^3 m^3 - 7 G^2 c m^2 - \frac{59 G c^2 m}{4} - 9 c^3,  \vspace{0.4cm} \\ 
F_{2}(G, k, \omega) = \sigma_{N}^{(2)}\Big|_{\Omega=0} = m^3 \omega^6 + (2 m^2 c \cos(k l) - 3 G m^3 - 7 m^2 c) \omega^4 \\ + \Big(2 \Big(-2 G m^2 - \frac{21}{4} c m \Big) c \cos(k l) + 3 G^2 m^3 + 14 m^2 c G + \frac{51 c^2 m}{4} \Big)\omega^2 \\ - \frac{9 \cos^2(k l) c^3}{2} + 2 \Big( G^2 m^2 + \frac{21}{4} G c m + \frac{9}{2} c^2 \Big) c \cos(k l) - G^3 m^3 - 7 G^2 c m^2 - \frac{51 G c^2 m}{4} - \frac{ 9 c^3}{2},
\end{gathered}
\end{eqnarray}
}
where $\sigma_{N}^{(1)}$ and $\sigma_{N}^{(2)}$ are defined by \eq{Nsigoddy11} and \eq{Nsigevenyzz}.

The absence of spinners limits the coupling between the transverse displacement components of the nodal points; in this case the nodal point trajectories are determined by their spring connections with neighbouring nodal points and gravitational forces.  For $\Omega=0$, the lowest frequency mode occurs at $\omega=\sqrt{G}$ when $k=0,$ while the highest frequency of the vibrating non-chiral Neumann lattice strip occurs at $\omega=\sqrt{(4mG +13 c +\sqrt{73}c)/m}/2$ when $k l=\pm \pi.$ In the following sections, we explore the dynamic features of the non-chiral Neumann lattice strip. 

\subsubsection{Standing modes}\label{standing1a}
In this section, we demonstrate the typical standing modes of the Neumann lattice strip subjected to gravity. We assume that gravity acts uniformly along the strip with $G=0.5,$ and that the gyricity is zero for each spinner ($\Omega=0$). For simplicity, all remaining normalised parameter values of the structure are set to unity ($c=1,l=1$ and $m=1$). 

\begin{figure}[H]
  \centering
  \begin{minipage}[b]{0.49\textwidth}
 \hspace{-3.7cm}\includegraphics[width=1.9\linewidth]{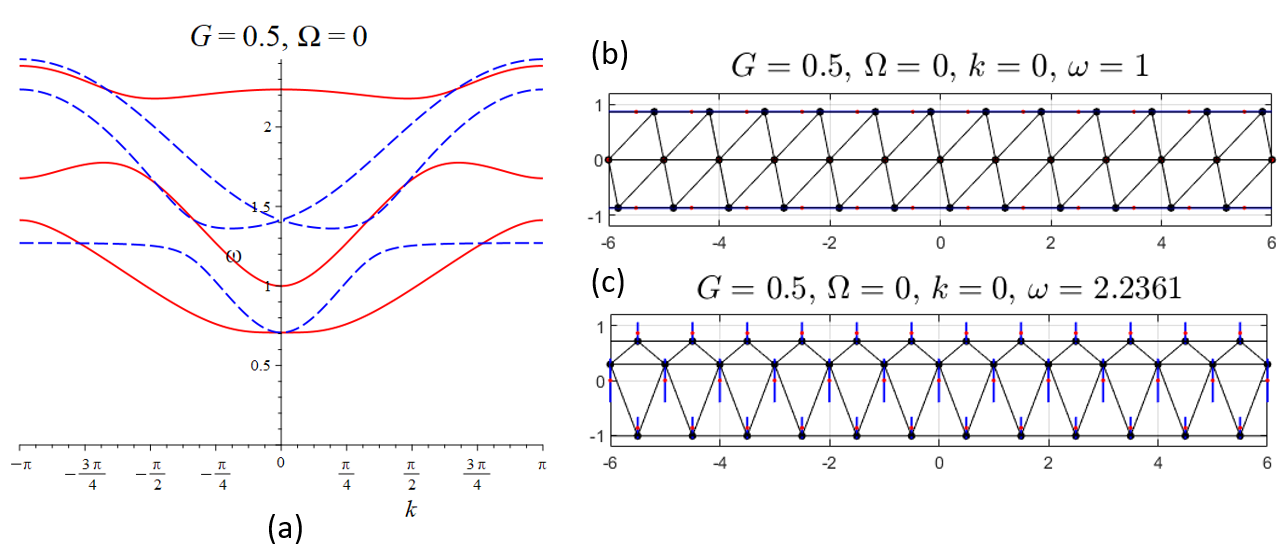}
  \end{minipage}
  \caption{\footnotesize (a) Dispersion diagram, showing $\omega$ versus $k,$ for an infinite non-chiral Neumann lattice strip for $\Omega=0$ and $G=0.5.$ The trajectories of the strip for $k=0$ and two values of $\omega$ corresponding to the standing modes are illustrated in parts (b) and (c), where $\omega=1$ and $\omega=2.2361,$ respectively. }
  \label{kzerostanding}
\end{figure}

\begin{figure}[H]
  \begin{minipage}[b]{0.49\textwidth}
   \hspace{1.7cm}\includegraphics[width=1.65\linewidth]{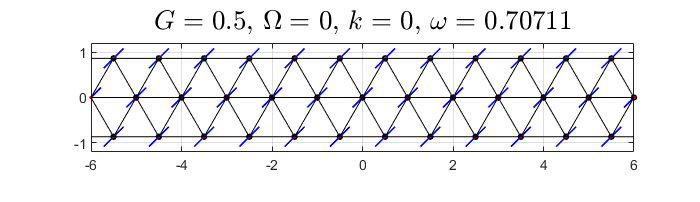}
  \end{minipage}
  \caption{\footnotesize Standing wave pattern in the non-chiral Neumann lattice strip for $k=0.$ The wavenumber and frequency of the oscillations at $(k, \omega)= (0, 0.70711)$ correspond to the touching point of the dispersion curves shown in Fig. \ref{kzerostanding}(a).}
  \label{kequalspimodesax}
\end{figure}

\begin{figure}[H]
  \centering
  \begin{minipage}[b]{0.49\textwidth}
    \hspace{-0.85cm}\includegraphics[width=1.1\linewidth]{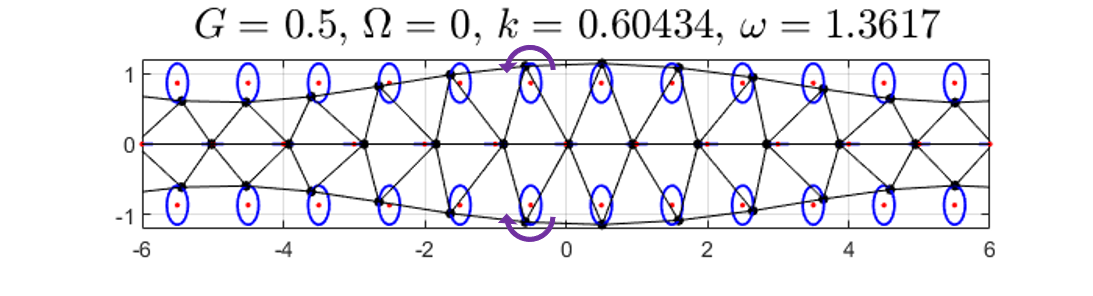}
    \centering\caption*{\footnotesize (a)}
  \end{minipage}
    \begin{minipage}[b]{0.49\textwidth}
    \hspace{-0.85cm}\includegraphics[width=1.1\linewidth]{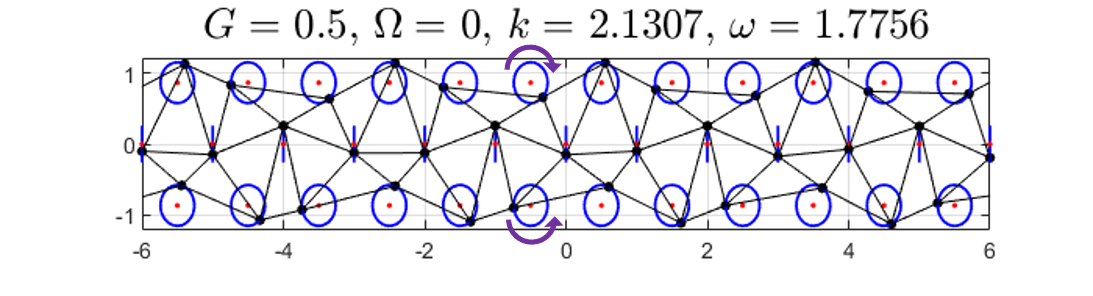}
    \centering\caption*{\footnotesize (b)}
  \end{minipage}
    \begin{minipage}[b]{0.49\textwidth}
    \hspace{-0.9cm}\includegraphics[width=1.1\linewidth]{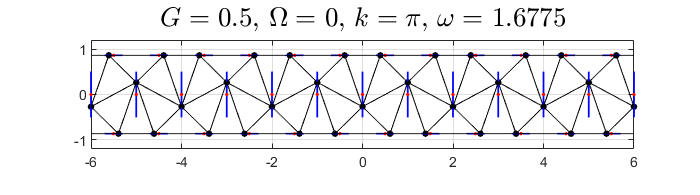}
    \centering\caption*{\footnotesize (c)}
  \end{minipage}
  \caption{ \footnotesize Trajectories of the Neumann lattice strip in the absence of gyricity corresponding to standing waves with $k\neq 0.$ The associated dispersion diagram is provided in Fig. \ref{kzerostanding}(a): (a) $(k, \omega)=(0.60434,1.3617),$ (b) $(k, \omega) = (2.1307, 1.7756)$ and (c)  $(k, \omega)=(\pi, 1.6775)$. The circular arrows indicate the orientation of motion of the nodal points.}
  \label{kequalspimodes}
\end{figure}

In Fig. \ref{kzerostanding}(a), we present the dispersion diagram for a non-gyroscopic Neumann lattice strip with identical masses connected by massless unit-length springs. A non-zero gravity parameter creates a low-frequency band gap (see also Section \ref{combinedf1a}) in the range $0\leq \omega < \sqrt{G}~~  (\approx 0.7071).$ The cut-off frequency is $\omega=\sqrt{4G + 13 +\sqrt{73}}/2 ~~ (\approx 2.4261),$ beyond which there are no propagating modes. In contrast, propagating waveforms occur for frequency values in the range $0.7071 \leq \omega \leq 2.4261.$ Fig. \ref{kzerostanding}(b) and Fig. \ref{kzerostanding}(c) show the standing modes of the non-chiral strip for two different frequency values at $k=0$, leading to oscillation patterns with zero net energy transfer. In Fig. \ref{kzerostanding}(b), the nodal points in the upper and lower layers of the strip oscillate horizontally, resembling the harmonic motion of a simple pendulum, while the central nodal points exhibit no vibrations. Conversely, in the example shown in  Fig. \ref{kzerostanding}(c), the nodal points in the upper and lower layers move in opposite directions compared to the central nodal points, each displaying vertical movements.  
The touching point between the solid and dashed dispersion curves for $k=0$ occurs at $(k,\omega)=(0, 0.70711)$ (see also Section \ref{combinedf1a}). This corresponds to a standing mode where the central nodal points are displaced in both horizontal and vertical directions as shown in Fig.  \ref{kequalspimodesax}. The degeneracy point of the two high-frequency dashed curves shown in Fig. \ref{kzerostanding}(a), along the frequency axis, corresponds to the vibration frequency $\omega = \sqrt{2} ~~ (\approx 1.4142).$ In this case, the structure can exhibit standing wave patterns resulting from the superposition of travelling waves.
Standing waves for the non-chiral Neumann strip are also observed for non-zero wavenumbers as shown in Fig. \ref{kequalspimodes}. The trajectories of the nodal points in the upper and lower layers presented in Fig. \ref{kequalspimodes}(a) and Fig. \ref{kequalspimodes}(b) consist of ellipses with their major axes aligned perpendicular to the strip. However, the eccentricities of the ellipses and the orientations of motion differ between the two examples. In addition, the central nodal points move horizontally in Fig. \ref{kequalspimodes}(a) and vertically in Fig. \ref{kequalspimodes}(b). An illustrative example of the standing mode at $k=\pi$ is displayed in Fig. \ref{kequalspimodes}(c), where the central nodal points move vertically, while the nodal points in the lower and upper layers move horizontally, differing from the elliptical paths shown in Figs. \ref{kequalspimodes}(a), (b). In the presented examples of standing modes with $k>0,$ the central nodal points are displaced linearly and the phase velocities are positive. Although there are no gyroscopic effects, the spring connections and gravitational forces result in elliptical motions of the nodal points. 
\subsubsection{Travelling waveforms}
We turn our attention to the travelling waves along the gravitational Neumann strip in the absence of gyricity. In this section, we provide two illustrative examples of the nodal point trajectories associated with propagating waves along the non-chiral strip. The parameter values used in the examples are identical to those in Section \ref{standing1a}, while the frequencies and wavenumbers correspond to the eigenmodes exhibiting travelling waves instead of standing waves. The corresponding dispersion diagram is shown in Fig. \ref{kzerostanding}(a).

\begin{figure}[H]
  \centering
  \begin{minipage}[b]{0.49\textwidth}
    \hspace{-0.85cm}\includegraphics[width=1.1\linewidth]{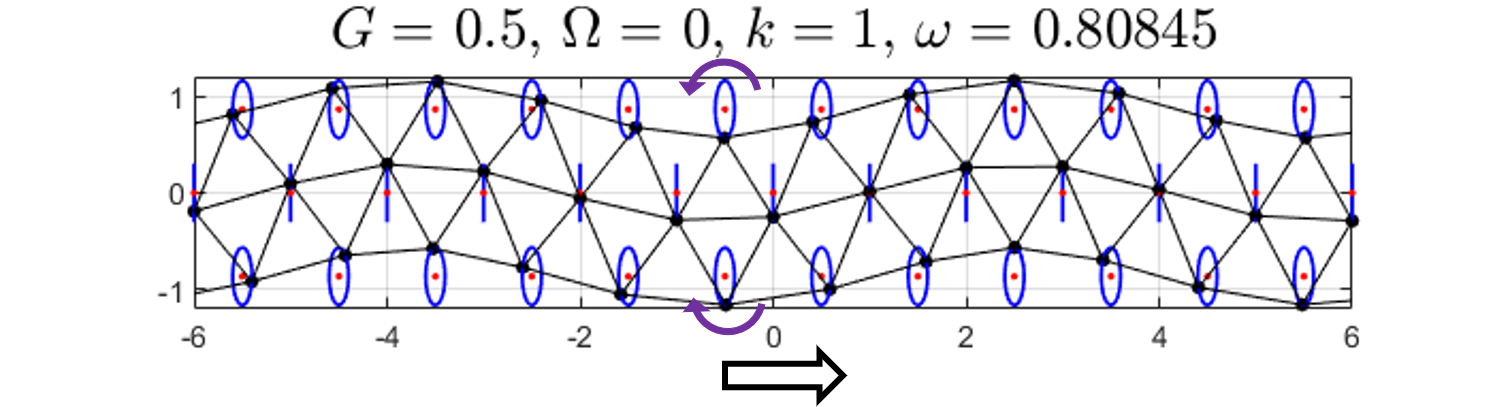}
    \centering\caption*{\footnotesize (a)}
  \end{minipage}
  \begin{minipage}[b]{0.49\textwidth}
    \hspace{-0.85cm}\includegraphics[width=1.1\linewidth]{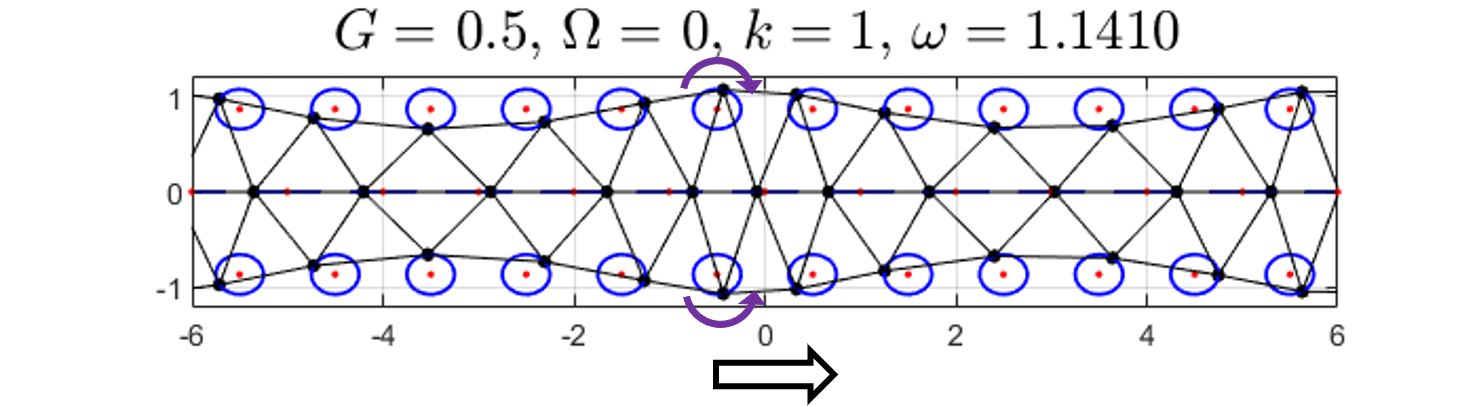}
    \centering\caption*{\footnotesize (b)}
  \end{minipage}
  \caption{\footnotesize Nodal trajectories of the non-chiral Neumann lattice strip for the parameter values $G=0.5$ and $\Omega=0,$ corresponding to waves travelling in the positive $x$-direction along the strip: (a) $(k, \omega) = (1, 0.80845)$ and (b) $(k, \omega) = (1, 1.1410).$ }
  \label{travellingkmodes}
\end{figure}

Fig. \ref{travellingkmodes} illustrates the nodal point trajectories of the non-chiral Neumann strip corresponding to propagating waves with a positive group velocity. The trajectories of the nodal points on the upper and lower layers are elliptical in both Fig. \ref{travellingkmodes}(a) and Fig. \ref{travellingkmodes}(b), while the central nodal points move vertically in Fig. \ref{travellingkmodes}(a) and horizontally in Fig. \ref{travellingkmodes}(b) (see also Section \ref{combinedf1a}). It is also observed that since the group velocity is positive, the energy is transported along the direction of the wave propagation, that is in the positive $x$-direction. Due to the symmetry of the dispersion diagram shown in Fig. \ref{kzerostanding}(a) relative to the frequency axis, a wave propagating in the negative $x$-direction can also be obtained by choosing the corresponding negative value of the wavenumber $k$ compared to the examples presented in Fig. \ref{travellingkmodes}.

\section{Continuum asymptotic models in equatorial and polar regions}\label{intro1a}\label{rotatingshall1}

In this section we discuss the connection between the dynamics of chiral gravitational waves in discrete structures and atmospheric and oceanic waves in the vicinity of the equator. Equatorial wave dynamics have been extensively explored using models based on the shallow water equations \cite{wheeler2000large, matsuno1966quasi}. We construct analytical asymptotic solutions of a model in a narrow equatorial band and use appropriate boundedness conditions at the boundaries of the band in the meridional direction. 
The analysis is followed by illustrative examples of the types of waves present in such bands: Kelvin waves, inertia-gravity waves and Rossby waves. We show that an equatorial fluid layer, subjected to a variable Coriolis force, can act as a chiral waveguide for propagating inertia-gravity waves and equatorial Rossby waves.

The full set of equations governing the time evolution of three-dimensional large-scale atmospheric flow are termed the primitive equations. These include the momentum equations for the $x$- and $y$-components, the hydrodynamic and thermodynamic laws, continuity equation and equation of state \cite{lorenz1967nature, eckart2013hydrodynamics}. We provide an analogy between the Coriolis term in the continuum system and the gyroscopic action in the discrete lattice strip. It is noted that both Coriolis and gyroscopic forces couple the velocity components of a moving object.  In our analysis, we consider the linearised shallow water equations, in its simplest form, governed by the continuity and momentum equations, respectively, given as follows: 
\begin{eqnarray}
\frac{\partial u}{\partial t} - 2\Omega \sin(\phi) v = - g\frac{\partial h}{\partial x}, \label{whe1} \\
\frac{\partial v}{\partial t} + 2\Omega \sin(\phi) u = -g\frac{\partial h}{\partial y}, \label{whe2} \\
\frac{\partial h}{\partial t} + H \Big(\frac{\partial u}{\partial x} + \frac{\partial v}{\partial y}\Big) = 0, \label{whe3}
\end{eqnarray} 
where ${\bf u} = (u ,v)^{T}$ with $u=u(x,y,t)$ and $v=v(x,y,t)$ denoting the velocities in the $x$- and $y$-directions (or zonal and meridional velocities) respectively, $h=h(x,y,t)$ is the height deviation of the fluid surface, $t$ measures time, $H$ is the depth of the undisturbed fluid, $\Omega$ is the angular speed of the rotating body, $\phi$ is the latitude and $g$ is the gravity acceleration. The zonal and meridional velocity and height field variables are functions of the horizontal fluid position $(x,y)$ and time $t$ only. The system \eq{whe1}-\eq{whe3} provides an approximate description of a layer of incompressible fluid of constant density in the presence of rotation.
In our study, we consider a geostrophic approximation, neglect viscous forces, and assume that the surface height deviation is small compared to the depth, that is $h \ll H.$

\subsection{Problem formulation of equatorial waves}\label{shallowwater1}\label{rotatvec1}
Equatorial waves, which consist of eastward- and westward-moving disturbances, are oceanic and atmospheric waves confined near the equator. These waves can propagate in both zonal and meridional directions. Equatorial waves can be separated into subclasses based on their dynamics, taking into account their periods, speeds and directions of propagation. The well-known waves in the equatorial atmosphere are Kelvin, Rossby and inertia-gravity waves. In this section, we provide the full analytical asymptotic solution of the linearised shallow water equations in a narrow band linked to atmospheric flows near the equatorial region. We also investigate the dispersion properties of equatorial waves by considering the combined effects of the rotating system in the presence of gravitational forces.

Given our focus on wave motions near the equator, we consider the linearised shallow water equations given by \eq{whe1}-\eq{whe3} and prescribe Dirichlet boundary conditions in the meridional direction for the velocity component $v$ at $y=\pm a,$ where $a$ is an appropriate positive quantity. This condition is important for maintaining the validity of the equatorial approximation in our model. Thus, the following boundary conditions are prescribed on the boundaries of the equatorial region:
\begin{equation}
v(x,y,t)\Big|_{y=\pm a} = 0. \label{boundarycondequa}
\end{equation}

 In meteorology and oceanography, the spatial variation of the Coriolis parameter is of importance.  The Coriolis parameter, given by $f = 2\Omega\sin(\phi),$ depends on the angular speed $\Omega$ and latitude $\phi.$  We approximate the Coriolis parameter with the latitude being linearly proportional to the distance from the equator, that is $\phi=y/R,$ where $R$ denotes the radius of the Earth.

We look for time-harmonic solutions in the form of zonally propagating waves, given by 
\begin{equation}
(u,v,h) = \text{Re}\{(A(y),B(y), C(y))e^{i (k x + \omega t)}\}, \label{harmonc1}
\end{equation}
where $k$ is the (zonal) wavenumber, $\omega$ is the radian frequency and $A(y), B(y)$ and $C(y)$ are functions describing the horizontal and transverse velocity components and height field component of the fluid. 
Substituting the above assumptions and given expressions into the system \eq{whe1}-\eq{whe3} results in a set of ordinary differential equations for $A$, $B$, and $C$:
\begin{eqnarray}
i\omega A(y) - 2\Omega \sin( \frac{y}{R} ) B(y) = - i g k C(y), \label{pete1} \\
i\omega B(y) + 2\Omega \sin(\frac{y}{R} ) A(y) = - g \frac{d C}{d y},  \label{pete2} \\
i \omega C(y) + H \Big(i k A(y) + \frac{d B}{d y}\Big) = 0, \label{pete3}
\end{eqnarray}
subject to the Dirichlet boundary conditions (see \eq{boundarycondequa}): 
\begin{equation}
B(y)\Big|_{y=\pm a} = 0. \label{boundarycondequa1a}
\end{equation}

The above problem formulation in a finite-width equatorial strip differs from the analysis of the Weber-type equations in an infinite domain discussed in  \cite{matsuno1966quasi, cane1976forced1, cane1977forced2}.
The solutions of the shallow water equations related to the Kelvin waves, with the dispersion equation $\omega^2 = g H k^2$, are discussed in Section \ref{nonrotatus}. To obtain other solutions of the shallow water equations we consider the case when $\omega^2 \neq g H k^2.$ In this case, waves of two different types are obtained as solutions, which are referred to as the inertia-gravity wave and the Rossby wave. Re-arranging \eq{pete3} for $C(y)$ and substituting into \eq{pete1} yields the representations for the height field and zonal velocity components, respectively, in the form:
\begin{equation}
 C(y) = i \frac{H (i k A(y) + \frac{\partial B}{\partial y} )}{\omega}, ~~~~ A(y) = -i \frac{(2\Omega \omega \sin\Big(\frac{y}{R}\Big)B(y) + gkH\frac{\partial B}{\partial y})}{\omega^2 - g H k^2}. \label{CAexpre}
\end{equation}
Upon substituting \eq{CAexpre} into \eq{pete2}, we obtain the following second-order meridional velocity component equation together with the associated boundary conditions:
\begin{equation}
gH\omega \frac{d^2 B}{d y^2} + \Big[\omega^3 - g H k^2 \omega -4\Omega^2 \omega \sin^2\Big(\frac{y}{R}\Big) + \frac{2 g k H \Omega}{R}\cos\Big(\frac{y}{R}\Big) \Big]B(y) = 0, ~~~~ B(\pm a) = 0. \label{Bdiffeqa}
\end{equation} 
The above problem is an eigenvalue problem with the general solution consisting of a linear combination of confluent Heun functions \cite{decarreau1978formes, ronveaux1995heun, abramowitz1968handbook}. Here, we consider the shallow water equations within a narrow horizontal layer of fluid, where the flows are confined near the equator, i.e. within $-a \leq y \leq a.$ Thus, we neglect higher-order terms in the expansions of $\sin(y/R)$ and $\cos(y/R).$ In particular, we present the asymptotic solutions for the eigenvalues and eigenfunctions of the equatorial waves for two different analytic approximations: first, by neglecting terms of order $O((y/R)^2),$ and second, by neglecting the terms of order $O((y/R)^3)$ in the meridional velocity equation in \eq{Bdiffeqa}. It is shown that the dispersion properties of inertia-gravity and Rossby waves are captured by both approximations, while the second approximation yields additional asymptotic terms in the solutions for the equatorial waves.

 In the following analysis it is convenient to introduce the dimensionless variables
  \begin{equation}
\tilde{Y} = \frac{y}{R}, ~~ \tilde{\omega} = \omega \sqrt{\frac{R}{g}}, ~~ \tilde{k} = k R, ~~ \tilde{\Omega} = 2  \Omega \sqrt{\frac{R}{g}}, ~~ \tilde{H} = \frac{H}{R}, \label{normalos1a}
 \end{equation}
   where the quantities with the symbol ``$\sim$'' are dimensionless. Hence, we consider the non-dimensional form of the meridional velocity problem \eq{Bdiffeqa}, given by (the symbol  ``$\sim$'' has been dropped for convenience)
   \begin{equation}
   \frac{d^2 \hat{B}}{d Y^2} + \Big[\frac{\omega^2}{H} - k^2 - \frac{\Omega^2}{H} \sin^2(Y) + \frac{k \Omega}{\omega}\cos(Y) \Big]\hat{B}(Y) = 0, ~~~~ \hat{B}(\pm \epsilon)=0, \label{newmark1}
   \end{equation} 
   where $\hat{B}(Y) = B(RY)$ and $\epsilon= a/R.$  It is noted that $Y=\pm \epsilon$ correspond to the upper and lower boundaries of the equatorial strip.

   \subsection{ Kelvin waves}\label{nonrotatus}
In this section we analyse the non-dispersive Kelvin waves, which are a type of equatorial waves that occur in rotating fluid systems. Kelvin waves propagate parallel to the equator or coastlines \cite{holton1968note, philander1989nino}, with no meridional velocity variations, and thus we take $v(x,y,t)=0$ (see Section \ref{rotatingshall1}). In the description of Kelvin waves, the conditions at the boundaries of the narrow equatorial strip are important since the wave amplitudes can increase away from the boundaries. Such waves can also propagate along interfaces between regions of different fluid densities. Then, setting $v=0$ in the equations \eq{whe1} and \eq{whe3}, and eliminating one of the variables yields the following dispersion relation: 
\begin{equation}
\omega^2 = g H k^2. \label{geigen1}
\end{equation}
The above dispersion equation for Kelvin waves also represent the shallow water gravity waves dispersion relation \cite{eckart1195219}. By taking into account \eq{geigen1}, we write the phase speed of the Kelvin waves in the zonal $x$-direction, as follows: 
\begin{equation}
c = \frac{\omega }{k} = \pm \sqrt{g H}. \label{phaseps}
\end{equation}
The quantity \eq{phaseps} refers to the velocity at which the phase of the waves propagates in the spatial direction. Additionally, the group velocities of the Kelvin waves, which represent the velocities of the oscillating wave packets in the medium, are given by $\pm \sqrt{gH}$. The eastward-propagating waves are known as the Kelvin waves, while the westward-propagating waves are referred to in the literature as the anti-Kelvin waves \cite{philander1989nino, cane1979forced}. They are distinguished by their amplitudes away from the equator as well as their direction of propagation.

By combining equations \eq{whe1} and \eq{whe3} of Section \ref{intro1a} with the additional condition that $v=0$, we obtain the following wave equations: 
\begin{equation}
 \frac{\partial^2 u }{\partial t^2} = g H \frac{\partial^2 u}{\partial x^2}, ~~~~  \frac{\partial^2 h}{\partial t^2} = g H \frac{\partial^2 h}{\partial x^2}.
\end{equation}
Thus, the time-harmonic variations of $u$ and $h$ in the domain $(x,t)$ are given by $e^{ ik (x + c t)}.$ Finally, equation \eq{whe2} is used to determine the $y$-dependence for the quantities $u$ and $h,$ yielding the time-harmonic representations for the zonal and meridional velocities and height field, respectively, as follows
\begin{equation}
u = \text{Re}\Big\{-\frac{g}{c} h_{0} e^{-\frac{2\Omega R}{c } \cos( \frac{y}{R})} e^{ ik (x + c t)} \Big\}, ~~~ v = 0, ~~~  h = \text{Re}\Big\{h_{0} e^{-\frac{2\Omega R}{c } \cos( \frac{y}{R})} e^{ ik (x + c t)}\Big\}, \label{Kelvindance}
\end{equation}
where $h_{0}$ is an arbitrary constant. The sign of $c$ (see \eq{phaseps}) determines the direction of propagation for the travelling Kelvin waves. For $\Omega>0,$ if $c>0,$ the amplitude of the waveform decreases as $y$ increases, whereas if $c<0,$ the wave amplitude increases as $y$ increases.

Applying the normalisations \eq{normalos1a} of Section \ref{rotatvec1} together with $x=R \tilde{X}$ and $t=\sqrt{R/g} \tilde{T},$ where the quantities with the symbol ``$\sim$'' are dimensionless, to the dimensional representations \eq{Kelvindance}, leads to the following non-dimensional forms for the zonal and meridional velocities and height field of the Kelvin modes (the symbol ``$\sim$'' has been dropped for convenience)
\begin{eqnarray}
\begin{gathered}
\hat{u} = -\frac{{h}_{0}}{\hat{c}} e^{-\frac{\Omega}{\hat{c}}\cos(Y) }\cos( k (X+ \hat{c} T) ), \\
\hat{v} =0,  \label{nondiKel}\\
\hat{h} = {h}_{0} e^{-\frac{\Omega }{\hat{c}}\cos(Y)} \cos( k (X+ \hat{c} T) ),
\end{gathered}
\end{eqnarray}
where { \footnotesize $ {\footnotesize \hat{u}(X,Y,T)=\sqrt{\frac{R}{g}}u\Big(RX, RY, \sqrt{\frac{R}{g}} T\Big), \hat{v}(X,Y,T)=\sqrt{\frac{R}{g}}v\Big(RX, RY, \sqrt{\frac{R}{g}} T\Big), \hat{h}(X,Y,T) = h\Big(RX,RY,\sqrt{\frac{R}{g}} T\Big)} $ } and the normalised velocity is given by $\hat{c}=\pm \sqrt{H}$ for the Kelvin and anti-Kelvin waves, respectively. The non-dimensional dispersion relation for the non-dispersive Kelvin waves is given by
\begin{equation}
\omega = \hat{c} k. \label{kelviaw11a}
\end{equation}
Thus, the normalised phase speeds and group velocities of the Kelvin waves are given by $\pm \sqrt{H}.$

\begin{figure}[H]
  \centering
  \begin{minipage}[b]{0.45\textwidth}
    \hspace{-0.5cm}\includegraphics[width=1\linewidth]{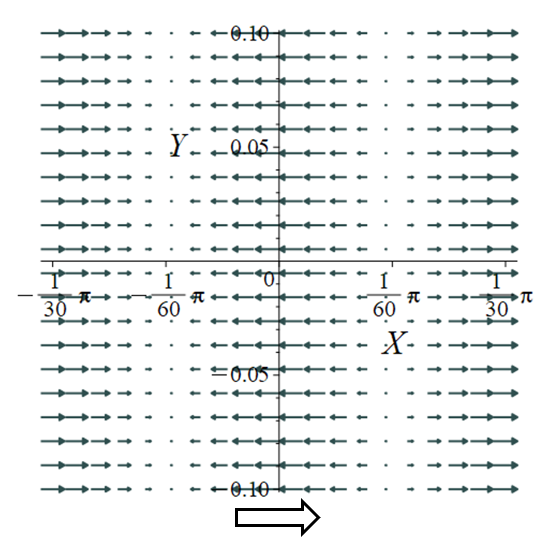}
  \end{minipage}
  \caption{\footnotesize Eastward-propagating Kelvin mode for $(k,\omega)=(30, 6.7082)$ with the velocity components defined by \eq{nondiKel} for $h_0=1$ and the dimensionless parameter values $\Omega=0.25,$ $H=0.05$ and $\epsilon=0.1.$ The non-dimensional propagation speed is $\hat{c}=0.2236.$}
  \label{eigenmodes3c}
\end{figure}

A typical example of an eastward-propagating Kelvin wave is shown in Fig. \ref{eigenmodes3c}, where the zonal velocities are non-zero and the meridional velocities vanish, for the parameter values $H=0.05, \Omega=0.25$ and $\epsilon=0.1$. The non-dimensional eigenfunctions of the Kelvin mode displayed in Fig. \ref{eigenmodes3c} are defined by \eq{nondiKel} with $h_0=1.$ These waves are non-dispersive and propagate either eastward or westward along the narrow equatorial band, depending on the sign of the group velocity. We note that the phase and group velocities of the Kelvin waves are either both positive or both negative. Compared to Rossby and inertia-gravity waves, Kelvin waves typically do not exhibit standing modes. Moreover, Kelvin waves do not display vortices, unlike Rossby modes (see Section \ref{rotatingshallow2a}), and are characterised by their horizontal motion as shown in Fig. \ref{eigenmodes3c}. Kelvin, Rossby and inertia-gravity waves show distinct dispersion and propagation properties, resulting in different spatial and temporal patterns of the equatorial fluid motions.

\subsection{Asymptotic models of the meridional velocity equation}\label{newase22a}

\subsubsection{Case $1$: Dirichlet problem for the Weber-type differential equation}\label{webs1a}

 When third-order terms in the expansions of $\sin(Y)$ and $\cos(Y)$ are neglected in \eq{newmark1}, we obtain the following boundary value problem:  
 \begin{eqnarray}
 \frac{d^2 \hat{B}}{d Y^2} + \Big[\Lambda - \Phi Y^2 \Big]\hat{B}(Y) = 0, ~~~~ \hat{B}(\pm \epsilon) = 0, \label{Bdiffeqanew1}
 \end{eqnarray}
where 
 \begin{equation}
 \Lambda = \frac{\omega^2}{H} - k^2 + \frac{k \Omega}{\omega} , ~~~ \Phi = \frac{\Omega^2}{H} + \frac{k \Omega}{2 \omega}. \label{newdefin1}
 \end{equation}
The differential equation shown in \eq{Bdiffeqanew1} seems to have first appeared in the paper \cite{weber1869ueber}. The above non-dimensional coefficients are fully consistent with the asymptotic expansions to the order $O(Y^2).$ The coefficient of $Y^2$ in \eq{Bdiffeqanew1}, denoted by $\Phi,$ depends on the non-dimensional frequency $\omega$ and the non-dimensional wavenumber $k.$ This coefficient is different from that in \cite{matsuno1966quasi}, where the order $O(Y^2)$ terms were missing due to the fact that the cosine term was approximated as a constant, which was an omission. When $\Lambda=0,$ the solutions are represented by the modified Bessel functions \cite{abramowitz1968handbook}, whereas when $\Phi=0,$ the solutions can be written as a linear combination of sinusoidal functions. The boundary conditions at $Y=\pm \epsilon$ also capture the meridional velocity conditions for the waveforms in the narrow band, which are also consistent with the problem assumptions. The analysis of the above meridional velocity equation is presented in Section \ref{movasymp1}.

 \subsubsection{Case $2$: harmonic oscillator model}
 
 If the terms of order $O(Y^2)$ are neglected in \eq{newmark1}, we obtain a non-dimensional equation with constant coefficients resembling the well-known linear harmonic oscillator: 
  \begin{eqnarray}
 \frac{d^2 \hat{B}}{d Y^2} + \Lambda \hat{B}(Y) = 0, ~~~~ \hat{B}(\pm \epsilon) = 0, \label{Bdiffeqanew1a1}
 \end{eqnarray}
 where the non-dimensional quantities $\Lambda$ and $\epsilon$ are defined in \eq{newdefin1} and Section \ref{rotatvec1}, respectively.
  In this case, the solution for $\hat{B}(Y)$ consists of sinusoidal terms; an exact non-trivial solution which satisfies the boundary conditions at $Y=\pm \epsilon$ can be obtained, as detailed in Section \ref{harmosh1a}. Quantitative comparisons between the dispersion properties of the harmonic waves in a bounded equatorial channel and those of the trapped equatorial waves in an unbounded channel, derived in \cite{matsuno1966quasi}, are presented in Section \ref{Matsuno1ac}.

\subsection{Harmonic waves in a narrow equatorial channel}\label{harmosh1a}
In this section, the full analytical solution of the shallow water equations is presented by neglecting terms of order $O(Y^2).$ Accordingly, Dirichlet boundary conditions are imposed on the narrow equatorial channel. The corresponding eigenvalues and eigenfunctions of the equatorial waveforms are also analysed.  
 
\begin{figure}[H]
  \centering
  \begin{minipage}[b]{0.95\textwidth}
    \includegraphics[width=1\linewidth]{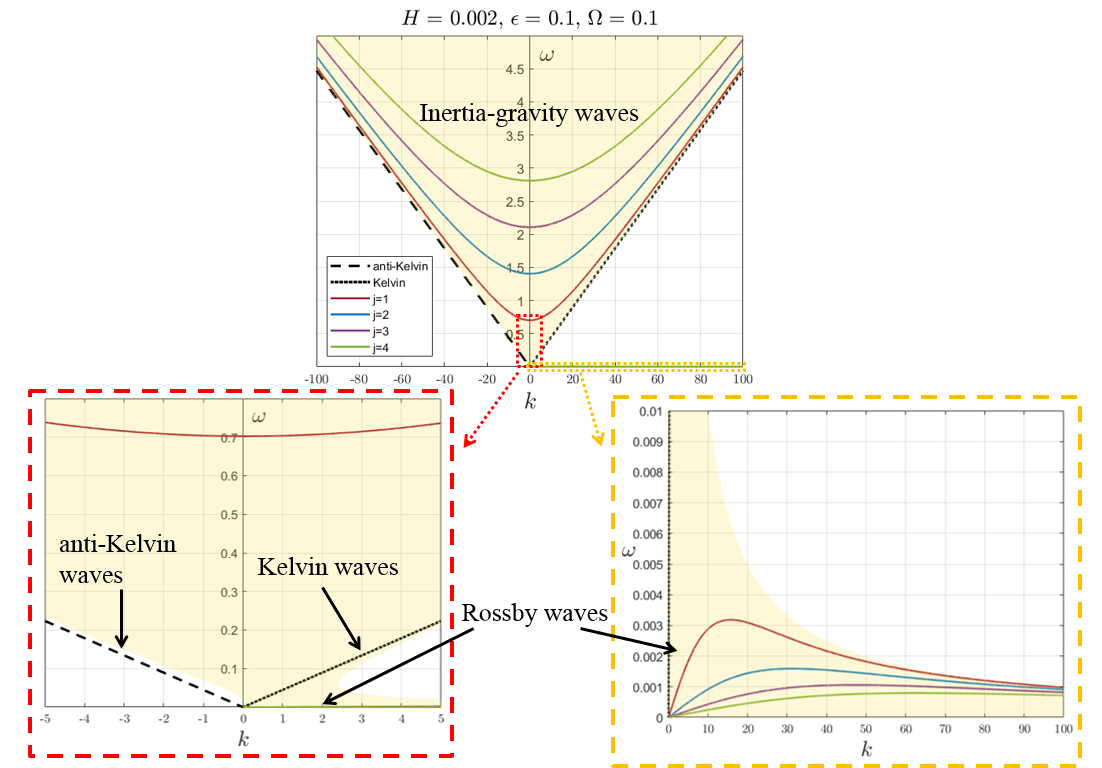}
  \end{minipage}
  \caption{\footnotesize Dispersion diagram for equatorial waves trapped in a narrow strip as a  function of the non-dimensional frequency $\omega,$ and non-dimensional zonal wavenumber $k$. The non-dimensional parameter values are $H=0.002, \epsilon=0.1$ and $\Omega=0.1.$ The anti-Kelvin and Kelvin modes are described in Section \ref{nonrotatus} and Section \ref{movasymp1}. The remaining dispersion curves correspond to the solutions of \eq{dispsim1}. Inertia-gravity waves are represented by the high-frequency curves for the values $j=1,2,3,4,$ resulting in westward- or eastward-propagating waveforms, while for the same values of $j$, the low-frequency curves denote the Rossby waves. 
  The shaded region corresponds to $\Lambda > 0.$
   }
  \label{new1a}
\end{figure}

  In the vicinity of the equator, by neglecting terms of order $O(Y^2)$ in the expansions of $\sin(Y)$ and $\cos(Y)$ in \eq{newmark1}, the non-dimensional meridional velocity component equation takes the form \eq{Bdiffeqanew1a1}. To obtain bounded non-trivial solutions of problem \eq{Bdiffeqanew1a1}, we take $\Lambda>0,$ and look for solutions in the form 
\begin{equation}
\hat{B}(Y) = b^{(1)} \cos(\sqrt{\Lambda} Y) + b^{(2)} \sin(\sqrt{\Lambda} Y),
\end{equation}
where $b^{(1)}$ and $b^{(2)}$ are arbitrary constants. Applying the conditions at $Y=\pm \epsilon$, results in the following exact form of the meridional velocity component
\begin{equation}
\hat{B}_{j}(Y) = \begin{cases}
b^{(1)}_{j} \cos(\sqrt{\Lambda_{j}} Y), ~~ \text{$j=1,3,5,\ldots$}, \\
b^{(2)}_{j} \sin(\sqrt{\Lambda_{j}} Y), ~~ \text{$j=2,4,6,\ldots$},
\end{cases} \label{B(y)exp1}
\end{equation} 
where the solvability condition yields 
\begin{equation}
 \sqrt{ \frac{\omega^2}{H} -   k^2  + \frac{ k \Omega }{\omega }} = \frac{j \pi}{2 \epsilon}, ~~~ j = 1,2,3,\ldots. \label{dispsim1}
\end{equation}
Equation  \eq{dispsim1} is the non-dimensional dispersion relation, describing the connection between the non-dimensional frequency and non-dimensional zonal wavenumber for a chosen meridional mode $j$. For each $j,$ there are generally three roots of the dispersion equation when $H, k, \Omega$ and $\epsilon$ are specified. In particular, two of the three roots of \eq{dispsim1} correspond to inertia-gravity waves, while the third root to a Rossby wave as shown in Fig. \ref{new1a}. The region corresponding to $\Lambda > 0$ is shaded in Fig. \ref{new1a}. The representation of the meridional eigenfunction components \eq{B(y)exp1} shows that the even modes ($j=2,4,6,\ldots$) are skew-symmetric about the narrow equatorial region and the odd modes ($j=1,3,5,\ldots$) are symmetric.  The group velocity of the equatorial waves is calculated from the dispersion relation \eq{dispsim1}, and can be approximated by 
 \begin{equation}
  \frac{d \omega}{d k} = \frac{ H \omega (-2 k \omega + \Omega)}{H \Omega k - 2 \omega^3}. \label{normalisedips1a}
 \end{equation}
 The above equation characterises the direction of propagation of the waveforms in a narrow band. Another useful property in the description of equatorial waves is the phase velocity, which is defined by $\omega/k,$ and indicates the speed at which the phase of the individual wave components propagate in the horizontal direction.

As noted previously, there are three main classes of equatorial waves in the narrow strip, which are inertia-gravity, Rossby and Kelvin waves. The meridionally bounded solutions in the south and north of the equatorial band introduce an additional mode: a non-dispersive, westward-propagating anti-Kelvin wave \cite{cane1979forced, philander1989nino, philander1977effects}. It is noted that the values of the non-dimensional parameters $\Omega,H$ and $\epsilon,$ results in only the eastward-propagating Kelvin mode to be situated in the shaded region, defined by $\Lambda>0$ (where $\Lambda$ is given in \eq{newdefin1}), while the anti-Kelvin mode lies outside the shaded region (see Fig. \ref{new1a}). The inertia-gravity waves and Rossby waves are within the shaded area as shown in Fig. \ref{new1a}.

Taking into account \eq{normalos1a} and substituting \eq{B(y)exp1} into the non-dimensional forms of \eq{CAexpre}, yields the non-dimensional zonal velocity and height field components, respectively, as
\begin{equation}
\hat{A}_{j}(Y) = \begin{cases}
- i b^{(1)}_{j}\frac{\Omega\omega \sin(Y)\cos(\sqrt{\Lambda_{j}} Y) -  k H \sqrt{\Lambda_{j}} \sin(\sqrt{\Lambda_{j}} Y )}{\omega^2 -  H k^2}, ~~ \text{$j=1,3,5,\ldots$}, \\
- i b^{(2)}_{j}\frac{\Omega\omega \sin(Y)\sin(\sqrt{\Lambda_{j}} Y) +  k H \sqrt{\Lambda_{j}} \cos(\sqrt{\Lambda_{j}} Y )}{\omega^2 -  H k^2}, ~~  \text{$j=2,4,6,\ldots$}, \label{zonvel1}
\end{cases}
\end{equation} 
and 
\begin{equation}
\hat{C}_{j}(Y) = \begin{cases}
 i b^{(1)}_{j} H  \frac{\Omega k \sin(Y)\cos(\sqrt{\Lambda_{j}} Y) - \omega \sqrt{\Lambda_{j}}\sin(\sqrt{\Lambda_{j}} Y)}{  \omega^2 - H k^2  }, ~~ \text{$j=1,3,5,\ldots$}, \\
 i b^{(2)}_{j} H  \frac{\Omega k \sin(Y) \sin(\sqrt{\Lambda_{j}} Y) + \omega \sqrt{\Lambda_{j}} \cos(\sqrt{\Lambda_{j}} Y) }{ \omega^2 -  H k^2  }, ~~  \text{$j=2,4,6,\ldots$}, \label{zonvel2a}
\end{cases}
\end{equation} 
where $\hat{A}_{j}(Y) = A_{j}(RY)$ and $\hat{C}_{j}(Y) = \sqrt{\frac{g}{R}}C_{j}(RY).$ By introducing the additional variables $x=R \tilde{X}$ and $t=\sqrt{R/g} \tilde{T}$, where the quantities with the symbol ``$\sim$'' are dimensionless, and substituting the above expressions for $\hat{A}_{j}(Y), \hat{B}_{j}(Y)$ and $\hat{C}_{j}(Y)$  into \eq{harmonc1}, results in the dimensionless zonal and meridional velocities and height field representations as follows (dropping the ``$\sim$'' for convenience) 
\begin{eqnarray}
\begin{gathered}
\hat{u}_{j} = -\text{Im}\{ \hat{A}_{j}(Y) \}\sin( k X + \omega T), \label{danoc1} \\
\hat{v}_{j} = \hat{B}_{j}(Y) \cos(k X + \omega T),  \label{danoc3} \\
\hat{h}_{j} = -\text{Im}\{ \hat{C}_{j}(Y) \} \sin(k X + \omega T). \label{danoc3}
\end{gathered}
\end{eqnarray}
Equations \eq{danoc1} describe the zonally propagating Rossby and inertia-gravity waves in a narrow equatorial band. We also note that the above non-dimensional eigenfunctions differ from the Kelvin wave solutions presented in Section \ref{nonrotatus}.

 In the subsequent section, we present examples of the eigenfunctions linked to the propagating inertia-gravity and Rossby waves, highlighting the zonal and meridional components of the zonal wave solutions.

 \subsection{Comparison of the dispersion properties between the harmonic wave model and the Matsuno model}\label{Matsuno1ac}
  
    In this section, we present a quantitative comparison between the dispersion properties of the trapped equatorial waves found in Matsuno \cite{matsuno1966quasi} and the dispersion relation of the harmonic waves derived in Section \ref{harmosh1a}. The equatorial channel solutions derived by Matsuno considered wave motions in an unbounded plane which is inconsistent with the position of the poles and the neglect of high-order terms in the expansion of $\sin(y/R)$ (see Section \ref{rotatvec1}). The analysis presented in this section addresses the equatorial channel problem formulation with Dirichlet boundary conditions. 
  
  We consider the meridional velocity mode in a narrow equatorial band, reduced to the dimensional form of the harmonic oscillator equation (see Section \ref{rotatingshall1}): 
\begin{equation}
 \frac{d^2 B}{d y^2} + \Big[\frac{\omega^2}{g H} -  k^2  + \frac{ k  \beta}{\omega} \Big]B(y) = 0, ~~~~ B(\pm a) = 0, \label{harmoprob1}
\end{equation} 
  where $\beta=2\Omega/R$ is the Rossby parameter \cite{rossby1939relation}, and the remaining variables and parameters are defined in Sections \ref{intro1a} and \ref{rotatvec1}. Introducing the dimensionless variables
  \begin{equation}
  \tilde{Y} = \frac{y}{R},~~~~ \tilde{\omega} = \frac{\omega}{\sqrt{\beta\sqrt{g H}}}, ~~~~ \tilde{k} = k \sqrt{\frac{\sqrt{g H}}{\beta}},
  \end{equation}
  where the quantities with the symbol ``$\sim$'' are dimensionless, the eigenvalue problem \eq{harmoprob1} can be re-written as follows (where the ``$\sim$'' has been dropped for convenience): 
  \begin{equation}
 \frac{d^2 \hat{B}}{d Y^2} + \frac{R^2 \beta}{\sqrt{g H}}\Big[\omega^2 -  k^2  + \frac{ k}{\omega}  \Big]\hat{B}(Y) = 0, ~~~~ \hat{B}(\pm a/R) = 0, \label{asdiohad02}
\end{equation} 
 where $\hat{B}(Y) = B(R Y).$ Following a similar analysis to that in Section \ref{harmosh1a}, and assuming the condition 
 \begin{equation}
 \omega^2 -  k^2  + \frac{ k}{\omega} >0, \label{as89da1}
 \end{equation}
 which applies for non-trivial and bounded solutions of the eigenvalue problem \eq{asdiohad02}, the following non-dimensional dispersion equation can be deduced from the above problem: 
 \begin{equation}
 \omega^2 -  k^2  + \frac{ k}{\omega} = \Upsilon j^2 , ~~~~ j=1,2,3,\ldots, \label{disp1asb}
 \end{equation}
 where $\Upsilon = \frac{\pi^2}{4 a^2} \frac{\sqrt{g H}}{\beta},$ which is dependent on the depth $H$ and strip width $2a.$ Accounting for the RHS of the above equation and the $j$-th equatorial mode, equation \eq{disp1asb} is similar to Matsuno's equation ($8$) (see \cite{matsuno1966quasi}), which is given by 
  \begin{equation}
  \omega^2 - k^2 + \frac{k}{\omega} = 2 n +1, ~~~ n=0,1,2,\ldots, \label{matto1as}
  \end{equation}
  where $n$ is the meridional mode number  \cite{wheeler2000large}. We note that the special mode corresponding to $n=0$ is not captured by the dispersion relation \eq{disp1asb}. This mode corresponds to the mixed-Rossby gravity wave, and it approximates an inertia-gravity wave at high frequencies and a Rossby wave at low frequencies  \cite{matsuno1966quasi, wheeler2000large, philander1989nino, leblond1981waves}. The $n=0$ mode is related to the infinite plane approximation used by Matsuno, and its accuracy compared to the narrow channel solution is discussed in \cite{paldor2015shallow}.
  
Taking into account the approximation of the shape of the Earth, we set the Earth's equatorial radius of $R = 6378 \times 10^{3} $ m  \cite{snyder1987map, rapp1967equatorial}. The angular speed of the Earth in inertial space can be approximated by $\Omega = 7.2921159 \times 10^{-5}$ rad/s. Then, noting that the Hadley cells extend from the equator to about $30^{\circ}$ S in the Southern Hemisphere and $30^{\circ}$ N in the Northern Hemisphere \cite{frierson2007width}, and that they feature air rising near the equator, typical within $5^{\circ}$ S to $5^{\circ}$ N, where the Intertropical Convergence Zone is located, we can approximate the latitude of the narrow channel boundaries between $5^{\circ}$ N  to $30^{\circ}$ N in the Northern Hemisphere and $30^{\circ}$ S to $5^{\circ}$ S in the Southern Hemisphere. Thus, we approximate the width parameter $a$ [$m$] as follows 
\begin{equation}
\frac{531500 \pi}{3} ~ (\approx 556585) < a < 1063 \pi \times 10^{3} ~ (\approx 3339513),  
\end{equation}
  where the approximate sphericity of the Earth is assumed.

     \begin{figure}[H]
  \centering
  \begin{minipage}[b]{0.95\textwidth}
    \includegraphics[width=1\linewidth]{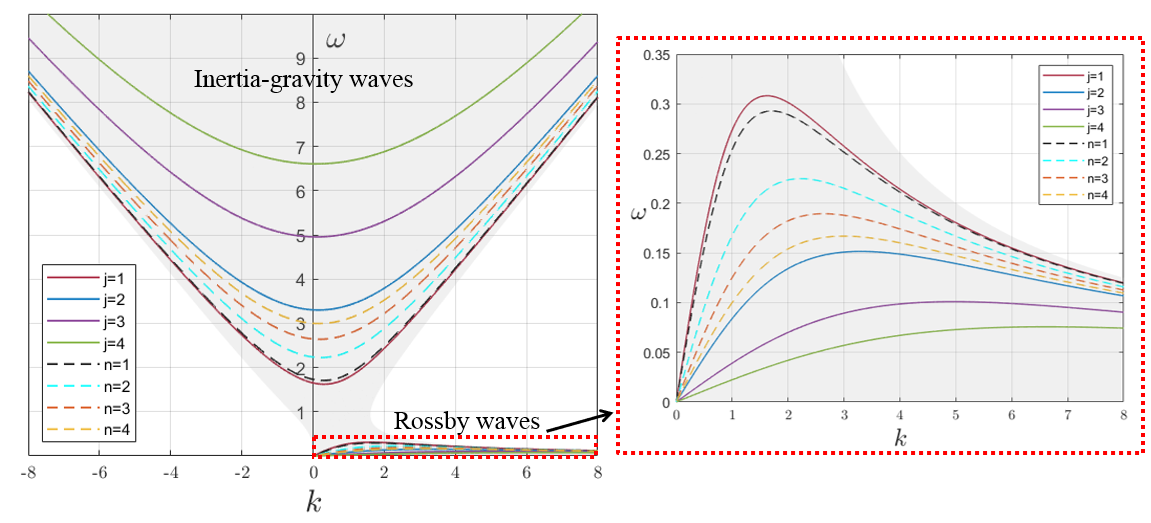}
  \end{minipage}
  \caption{ \footnotesize Dispersion curves of the harmonic (solid) and Matsuno (dashed) models for a range of values of the meridional mode numbers. The Matsuno modes are defined by \eq{matto1as}, while the harmonic modes correspond to \eq{disp1asb} with $\Upsilon = 2.7274.$ The shaded region corresponds to \eq{as89da1}.}
  \label{updsilon1a}
\end{figure}

In Fig. \ref{updsilon1a} we provide the diagram of the dispersion curves obtained by the model of Section \ref{harmosh1a} and the Matsuno model for positive integer values of the meridional mode numbers, which are defined by \eq{disp1asb} and \eq{matto1as}, respectively. The shaded region corresponding to \eq{as89da1} is shown in Fig. \ref{updsilon1a}, which contains the inertia-gravity modes and the Rossby modes.
Experimental analysis, involving measurements of frequencies and wavenumbers, is extremely challenging in the context of acquisition of data, but nevertheless it was possible, and the paper \cite{wheeler1999convectively11} provides the spectral analysis based on the satellite database accumulated over the duration of 18 years.  
It is not surprising that there was no uniform match with the Matsuno model, but certain dispersion curves could be mapped on the spectral peaks for a specific range of the depth values. In \cite{wheeler1999convectively11}, the values of depth were chosen between $12$ m and $200$ m. In our illustrative simulation, presented in Fig. \ref{updsilon1a}, we use the value $H=100$ m approximately in the middle of that range, which is also consistent with the constraints of    
a barotropic model and the linearised shallow water equations. 
Moreover, we assume that the 
channel boundaries are at $10^{\circ}$ S and $10^{\circ}$ N, which gives the approximate value $a = 1113171$ [m],
and thus $\Upsilon = 2.7274.$  This results in the solid dispersion curves of the harmonic model shown in Fig. \ref{updsilon1a}, which are defined by \eq{disp1asb}. The dispersion curves of the inertia-gravity wave and the Rossby wave of the harmonic model corresponding to the first mode ($j=1$) show a very good agreement with the Matsuno dispersion curves for the $n=1$ mode as illustrated in Fig. \ref{updsilon1a}. 
 It is also noted that the lower bound for the ratio of the minimum frequency of the inertia-gravity mode to the maximum frequency of the Rossby mode for the harmonic and Matsuno models are $5.265$ and $5.828,$ respectively, which occur at the first meridional mode.
 Although, this ratio increases for higher modes in both models, it will always be larger in the harmonic model than the Matsuno model for sufficiently large meridional mode numbers, which represents a significant difference between the two models.  
We also emphasise on the difference in approach of Section \ref{harmosh1a}, dealing with the spectral problem in a strip, and the Matsuno model, addressing equations with unbounded coefficients in a plane.

\subsection{Eigenmodes of Rossby and inertia-gravity waves in the narrow equatorial channel}\label{dispequatorial}

In Section \ref{harmosh1a}, we have shown that the non-dimensional dispersion equation, given by \eq{dispsim1}, can be used to determine the frequencies and wavenumbers of the waveforms in a narrow equatorial band. Some examples of the eigenfunctions obtained in Section \ref{harmosh1a} are presented in this section, in connection with various dispersion characteristics of the different types of equatorial waves.

\begin{figure}[ht]
  \centering
  \begin{minipage}[b]{0.95\textwidth}
    \includegraphics[width=1\linewidth]{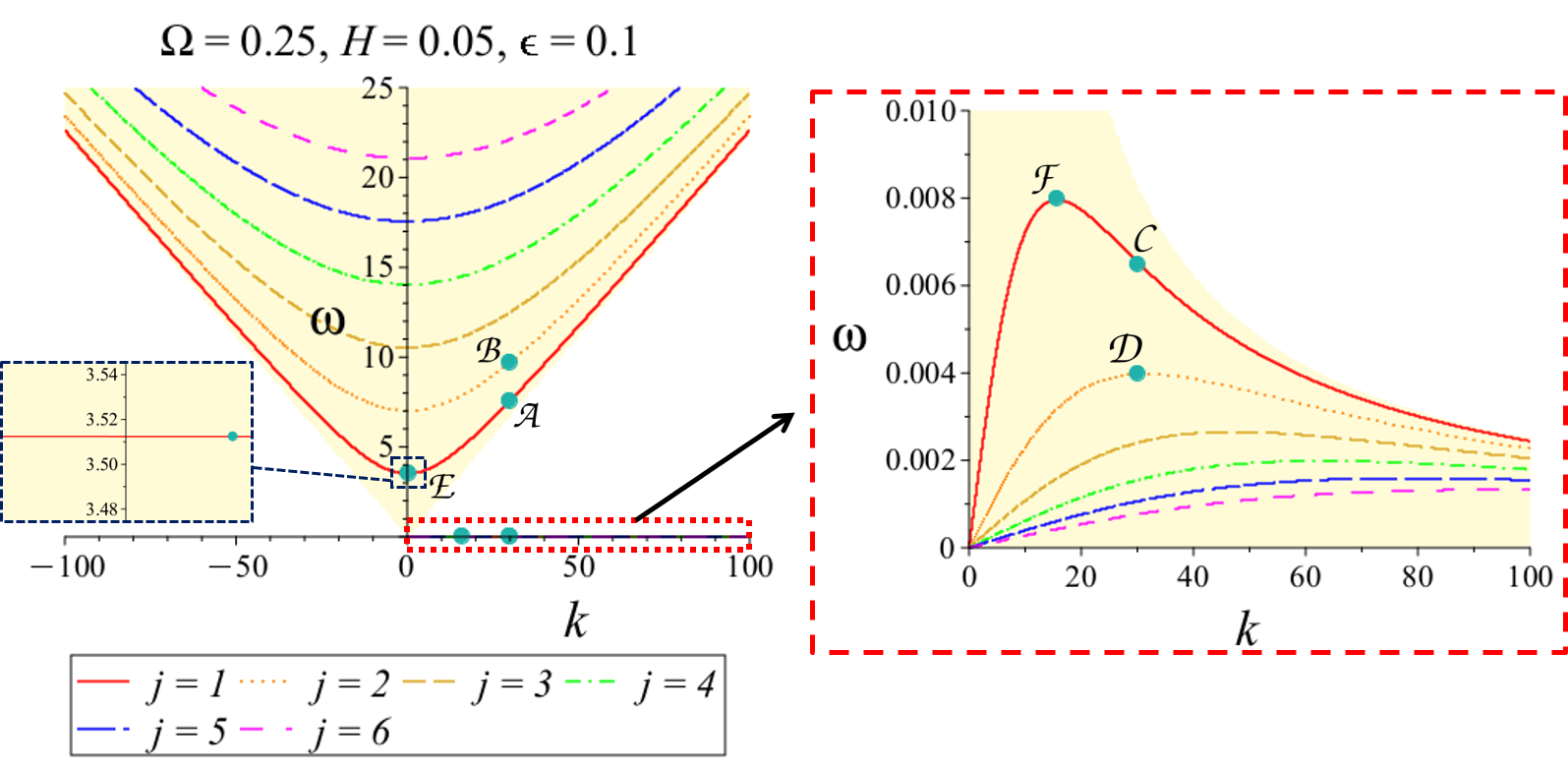}
  \end{minipage}
  \caption{\footnotesize Non-dimensional frequencies as functions of the non-dimensional wavenumbers defined by the dispersion relation \eq{dispsim1} for a range of modes and the non-dimensional parameter values $\Omega=0.25, H=0.05$ and $\epsilon=0.1.$ The points labelled by $\mathpzc{A},\mathpzc{B},\mathpzc{C},\mathpzc{D},\mathpzc{E}$ and $\mathpzc{F},$ shown on the dispersion curves, correspond to the values of $\omega$ and $k$ of the eigenfunctions for the equatorial modes described in the text.}
  \label{neasod}
\end{figure}

 The non-dimensional eigenfunctions of the Kelvin waves, characterised by the zero meridional velocity component, are addressed in Section \ref{nonrotatus}. These waves are non-dispersive and are associated with wind stress anomalies, the Coriolis effect, coastal boundaries and variations in sea surface height and temperature \cite{wang2002kelvin, eriksen1983wind}. Kelvin waves are integral to oceanic-atmospheric interactions and climate phenomena \cite{saha2008earth}. The low frequency modes shown in Fig. \ref{neasod} correspond to the equatorial Rossby waves (also known as planetary waves), which propagate in the atmosphere and oceans. Such waves are characterised by their lower frequencies and longer wavelengths compared to Kelvin waves. At higher frequencies, the equatorial inertia-gravity waves are present, linked to the high-frequency curves displayed in Fig. \ref{neasod}; these waveforms can propagate in various directions. Analysing inertia-gravity waves and Rossby waves is crucial for understanding and predicting atmospheric and oceanic circulation patterns on regional and global scales.

\subsection{Eigenfunctions of the zonally propagating equatorial wave solutions}\label{rotatingshallow2a}
The sinusoidal solutions of the shallow water equations describing zonally propagating waves in a narrow channel have been derived in Section \ref{harmosh1a}.  The set of admissible non-dimensional frequencies $\omega$, and non-dimensional zonal wavenumbers $k$, are defined according to the relation \eq{dispsim1} for a positive integer value of $j.$ In addition, for each $j$, the non-dimensional eigenfunctions of the inertia-gravity and Rossby modes are given by \eq{danoc3}. From the analysis presented in Section \ref{harmosh1a} for inertia-gravity and Rossby waves, if follows that if $j$ is an odd positive integer then $\hat{v}$ is an even function, and $\hat{u}$ and $\hat{h}$ are odd functions with respect to the non-dimensional variable $Y.$ Conversely, if $j$ is an even positive integer, the parities of the velocity and height fields are reversed.

In this section, we illustrate the non-dimensional eigenmodes of the equatorial waveforms in a narrow band, with the focus on the zonal and meridional velocity components. In each figure, one wavelength in the $X$-direction is shown. The group velocity, defined by \eq{normalisedips1a}, is used to determine the speed and direction of motion of the equatorial waves.  In the illustrative examples, the wavenumbers are chosen as either $k=30$ or $k=-30$ to observe one wavelength in the $X$-direction with a comparable order of magnitude to the meridional domain. Additionally, the equator runs through the centre of each diagram, with the horizontal boundaries at $Y=\pm \epsilon.$ The change in frequency values of the waveforms correspond to different points on the dispersion curves. For each illustrative example presented in this section, the parameter values are $H=0.05, \Omega=0.25$ and $\epsilon=0.1.$ 

\begin{figure}[H]
  \centering
  \begin{minipage}[b]{0.95\textwidth}
    \hspace{-0.3cm}\includegraphics[width=1\linewidth]{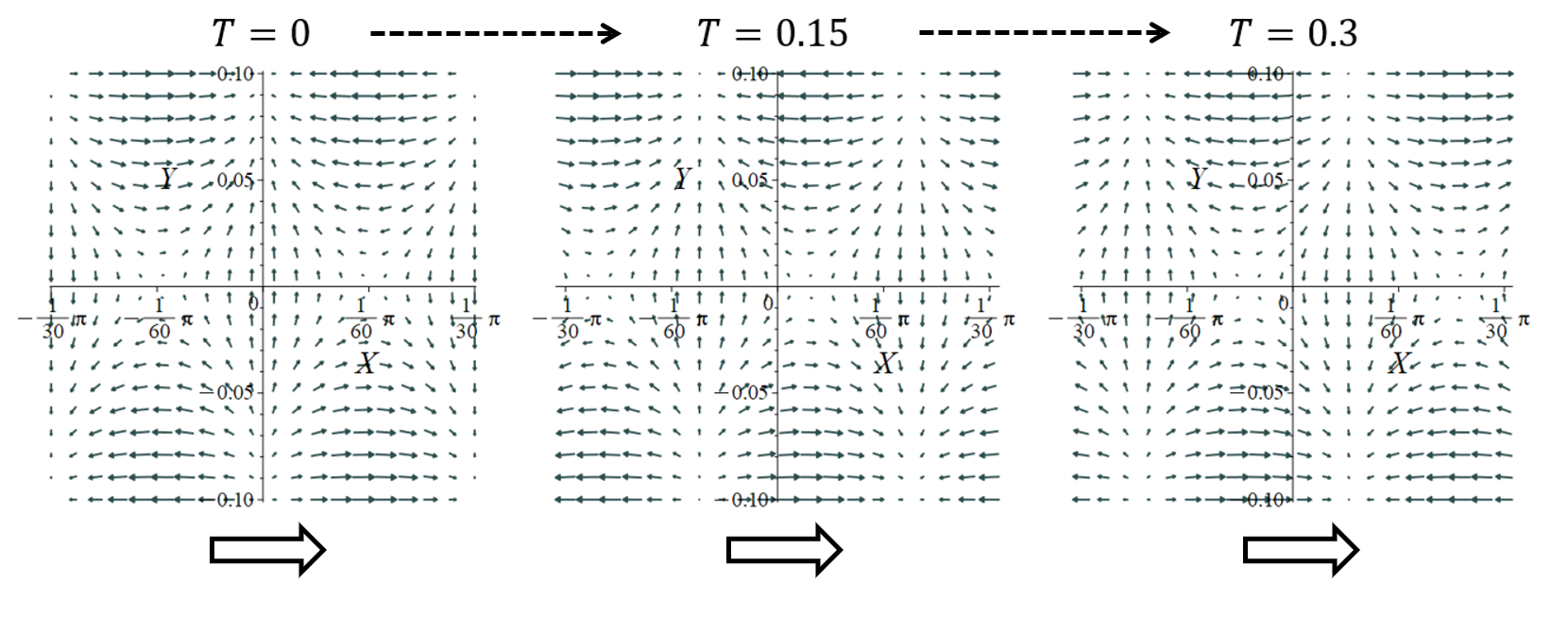}
  \end{minipage}
  \caption{\footnotesize Evolution of the eastward-propagating inertia-gravity eigenmode for the zonal and meridional velocity components at different normalised time intervals corresponding to $j=1$ (see Section \ref{harmosh1a}). The eigenfunctions for the inertia-gravity wave are defined according to \eq{danoc3}. The non-dimensional zonal wavenumber and non-dimensional frequency of the oscillations are respectively $30$ and $7.5688.$ The arrows show the direction of propagation of the waveforms.}
  \label{eigenmodes1b12}
\end{figure}

The eigenfunctions for the zonal and meridional velocities of the eastward-moving inertia-gravity mode for $j=1$ at various normalised time steps are shown in Fig. \ref{eigenmodes1b12}. The direction of the propagating waves is determined by the positive group velocity (see \eq{normalisedips1a}), and is represented by the arrows in Fig. \ref{eigenmodes1b12}. Additionally, the non-dimensional wavenumber and non-dimensional frequency values are $k=30$ and $\omega=7.5688,$ respectively, and correspond to the point $\mathpzc{A}$ in Fig. \ref{neasod}. The eigenfunctions have the sinusoidal form derived in Section \ref{harmosh1a}, reflecting the oscillatory behaviour of these waves in both the zonal and meridional directions.  It is noted that westward-propagating inertia-gravity waves are obtained when $k=-30$ and $\omega=7.5754,$ where the waveforms travel in the negative $X$-direction. A positive group velocity corresponds to an eastward wave propagation, while a negative group velocity indicates a westward-propagating wave. The frequencies of the eastward- and westward-propagating waves differ due to the asymmetry of the inertia-gravity dispersion curves (see Fig. \ref{neasod}), as a result of the Coriolis force. 

\begin{figure}[H]
  \centering
  \begin{minipage}[b]{0.95\textwidth}
    \hspace{-0.3cm}\includegraphics[width=1\linewidth]{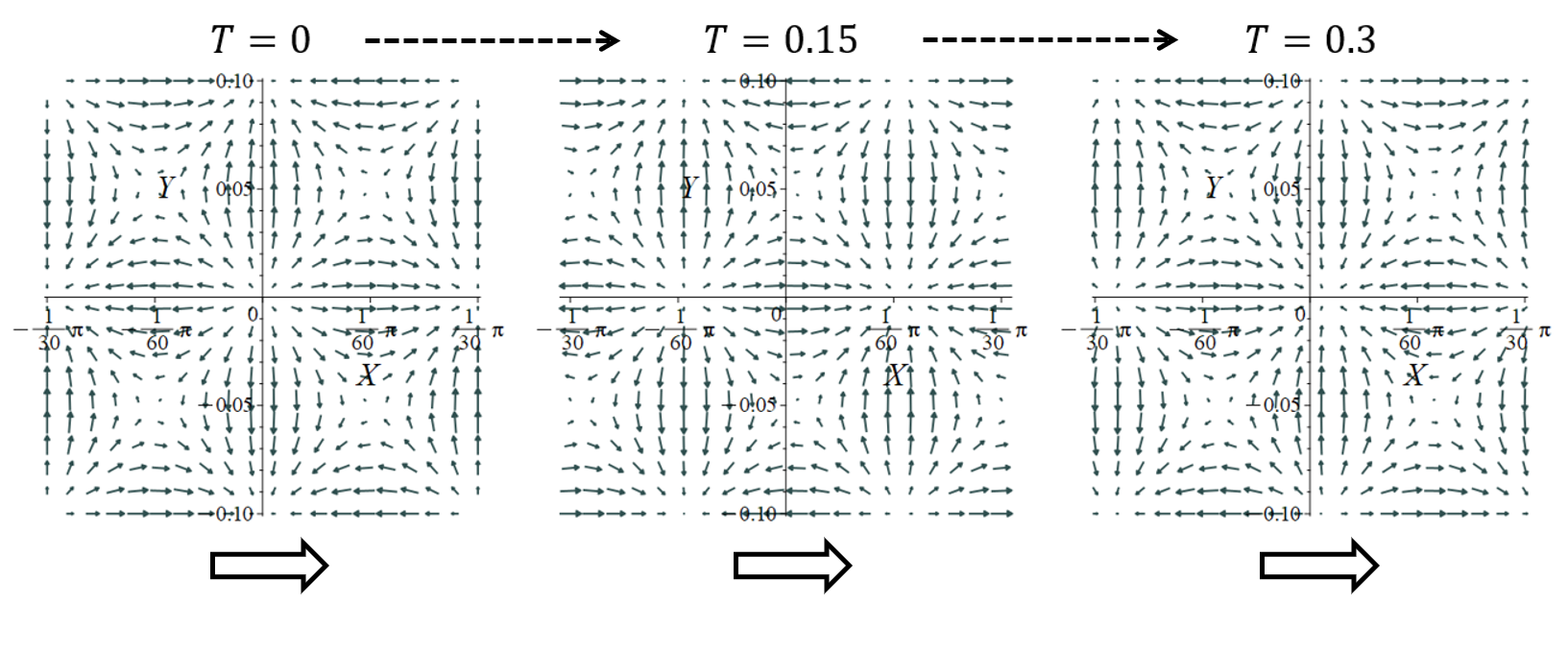}
  \end{minipage}
  \caption{\footnotesize Non-dimensional eigenmode for the eastward-propagating inertia-gravity waveform corresponding to $j=2$, with the non-dimensional parameter values $\Omega=0.25,$ $H=0.05$ and $\epsilon=0.1,$ and a selection of values for the non-dimensional time variable $T$. The non-dimensional wavenumber and non-dimensional frequency values are $30$ and $9.7113,$ respectively. }
  \label{eigenmodes1aasdasd}
\end{figure}

In Fig. \ref{eigenmodes1aasdasd}, we present an example demonstrating the evolution of the eastward-propagating inertia-gravity wave for $j=2$, with the zonal wavenumber $k=30$ and frequency $\omega=9.7113.$ The corresponding dispersion diagram is shown in Fig. \ref{neasod}, with the associated point of the equatorial inertia-gravity mode denoted by $\mathpzc{B}$. Compared to the eigenmodes illustrated in Fig. \ref{eigenmodes1b12}, the parities of the non-dimensional velocity components presented in Fig. \ref{eigenmodes1aasdasd} change: for $j=1,$ $\hat{v}$ and $\hat{u}$ are even and odd in $Y,$ respectively, whereas for $j=2$,  $\hat{v}$ and $\hat{u}$ become odd and even in $Y,$ respectively. In addition, the frequency of the inertia-gravity waves is higher in the example shown in Fig. \ref{eigenmodes1aasdasd} compared to Fig. \ref{eigenmodes1b12}. Although divergent behaviours and meridional velocity variations are present for both modes, the eigenmodes shown in Fig. \ref{eigenmodes1aasdasd} also display a significant zonal velocity component along the centre of the equatorial band. Fundamental differences between the general structures of the inertia-gravity modes can also be observed.  Furthermore, it is noted that eigenfunctions linked to the westward-propagating inertia-gravity waveforms also exist when $k=-30$ and $\omega=9.7153$.

\begin{figure}[H]
  \centering
  \begin{minipage}[b]{0.45\textwidth}
    \hspace{-0.5cm}\includegraphics[width=1\linewidth]{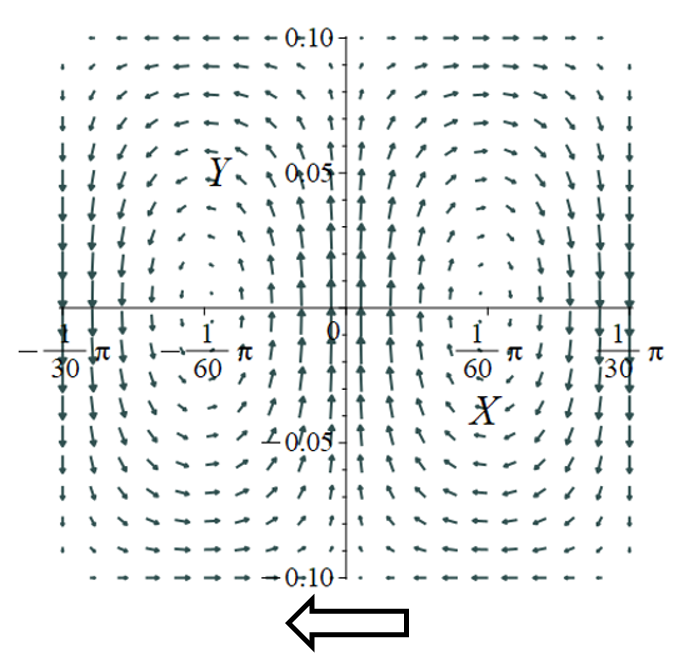}
    \centering\caption*{\footnotesize (a) Westward Rossby wave ($j=1$)}
  \end{minipage}
  \quad
  \begin{minipage}[b]{0.45\textwidth}
    \hspace{-0.5cm}\includegraphics[width=1\linewidth]{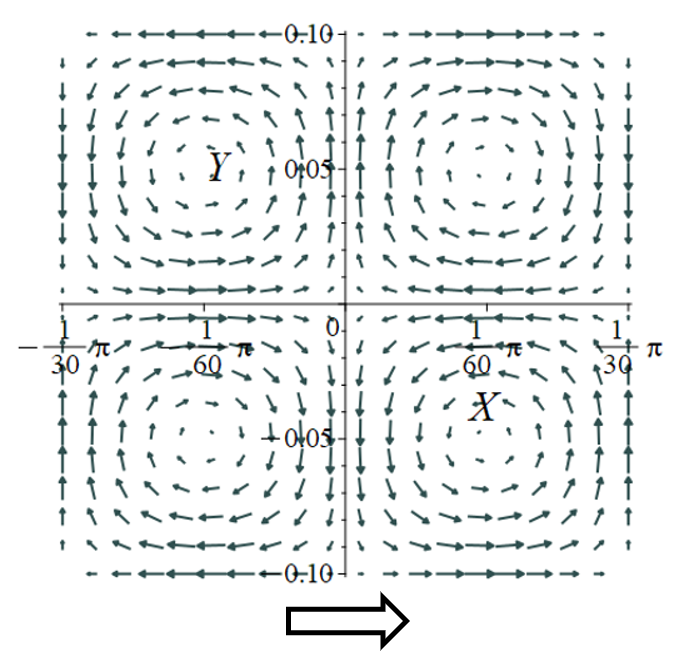}
    \centering\caption*{\footnotesize (b) Eastward Rossby wave ($j=2$)}
  \end{minipage}
  \caption{\footnotesize Eigenfunctions of the non-dimensional velocity components, given by \eq{danoc3}, of the Rossby modes for $j=1$ and $j=2$ in parts (a) and (b), respectively; (a) westward-propagating Rossby wave for $(k,\omega)=(30, 0.0065)$ and (b) eastward-propagating Rossby wave for $(k,\omega)=(30, 0.0040).$ The arrows indicate the direction of the wave propagation.}
  \label{salvol1}
\end{figure}

In Fig. \ref{salvol1}, we present examples of the non-dimensional eigenfunctions of Rossby waves propagating in the narrow equatorial band for two different modes. Fig. \ref{salvol1}(a) displays the westward-propagating Rossby wave for $j=1$ with $(k,\omega)=(30, 0.0065),$ whereas Fig. \ref{salvol1}(b) shows the eastward-propagating Rossby wave for the $j=2$ mode with $(k,\omega)=(30, 0.0040).$ These points are denoted by $\mathpzc{C}$ and $\mathpzc{D}$ in the dispersion diagram in Fig. \ref{neasod}, and correspond to the low-frequency dispersion curves. Vortices are observed in the waveforms for both propagating Rossby modes, which differ from the inertia-gravity modes presented in Fig. \ref{eigenmodes1b12} and Fig. \ref{eigenmodes1aasdasd}. As shown by the examples, higher modes corresponding to higher values of $j$ result in an increase in the number of vortices for the Rossby modes (see Fig. \ref{salvol1}) and greater meridional variations in the flow for the inertia-gravity modes (see Fig. \ref{eigenmodes1b12} and Fig. \ref{eigenmodes1aasdasd}), associated with changes in zonal wavelengths, frequencies and energy distributions. Rossby waves are characterised by rotational flows, while inertia-gravity waves display a more divergent behaviour in connection with their shorter horizontal wavelengths and higher frequencies than Rossby waves. The dominant features of inertia-gravity waveforms in a continuum (e.g. eastward or westward propagation and asymmetric properties of the dispersion curves) are similar to the elastic chiral gravitational waves in a chiral discrete lattice strip presented in Section \ref{discretelatstrip}.

\subsection{Standing mode patterns for Rossby and inertia-gravity waves}
 Examples of standing mode patterns for both Rossby and inertia-gravity waves in a narrow band are presented in this section. Equatorial standing modes represent localised oscillations of currents along the narrow equatorial channel. In such cases, the waveform patterns remain stationary in space, and compared to the energy-transmitting eigenmodes presented in Section \ref{rotatingshallow2a}, the standing waves are associated with zero net energy transfer. This phenomena is also observed in various wave systems, such as in optics and mechanics; in specific structured systems, waves exhibiting such features can lead to wave trapping or stopping effects  \cite{adamou2007trapped, evans1993edge}. The standing modes of Rossby and inertia-gravity waves are characterised by the vanishing group velocity defined by \eq{normalisedips1a}.

 Kelvin waves and standing Rossby waves significantly influence the dynamics of El Ni\~no events, which may provide insights into the decline of ancient civilisations, such as the Moche civilisation in northern Peru in the late sixth century (see \cite{grove2018nino, moseley1987punctuated}). The climate record suggested that the environmental changes likely included a mega El Ni\~no that caused $30$ years of intense rain and flooding on the coast, followed by $30$ years of drought \cite{grove2018nino, BBB2005Lost}. These extreme weather phenomena disrupted the Moche way of life and damaged field and irrigation systems \cite{caramanica2020nino, quinn1987nino, moseley1983principles, moseley1992doomed}. Consequently, the combined effects of environmental changes and weakened political authority may have resulted in the collapse of the Moche civilisation \cite{grove2018nino}.

\begin{figure}[H]
  \centering
  \begin{minipage}[b]{0.45\textwidth}
    \hspace{-0.5cm}\includegraphics[width=1\linewidth]{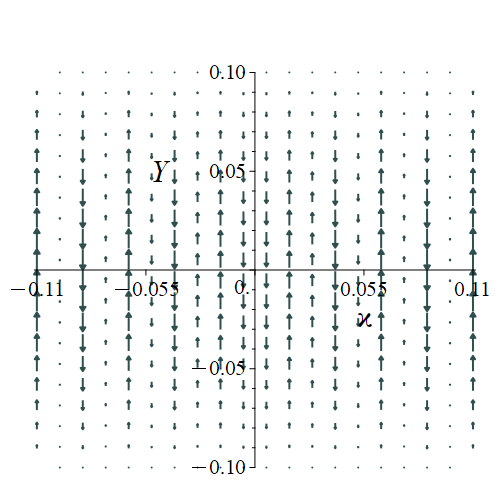}
    \centering\caption*{\footnotesize (a) Standing inertia-gravity mode}
  \end{minipage}
  \begin{minipage}[b]{0.5\textwidth}
    \hspace{-0.5cm}\includegraphics[width=1\linewidth]{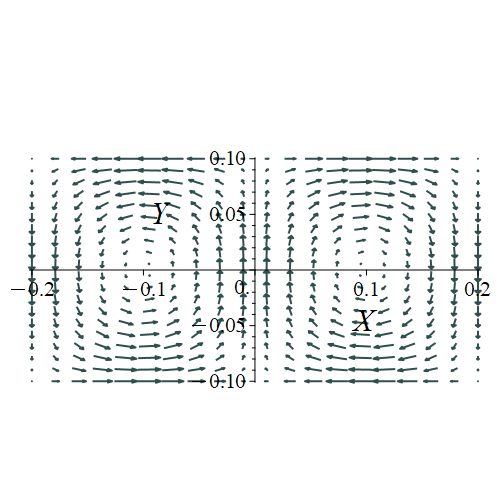}
    \centering\caption*{\footnotesize (b) Standing Rossby mode}
  \end{minipage}
  \caption{\footnotesize Standing inertia-gravity and Rossby modes for $j=1$, with the parameter values $\Omega=0.25,$ $H=0.05$ and $\epsilon=0.1.$ The associated dispersion diagram is shown in Fig. \ref{neasod}, with the points $\mathpzc{E}$ and $\mathpzc{F}$ corresponding to the standing modes in parts (a) and (b), respectively.  (a) Standing inertia-gravity wave for $(k,\omega)=(0.0356, 3.5124)$ with the horizontal scaling $X = 8 \times 10^{2} \chi$, and (b) standing Rossby wave for $(k,\omega)=(15.7079, 0.0080).$  }
  \label{standing_modes}
\end{figure}

Standing wave patterns, described by the non-dimensional eigenfunctions \eq{danoc3} for $j=1,$ corresponding to inertia-gravity and Rossby waves are shown in Fig. \ref{standing_modes}(a) and Fig. \ref{standing_modes}(b), respectively. Similarly to the eigenmode diagrams shown in Section \ref{rotatingshallow2a}, the eigenfunctions in Fig. \ref{standing_modes} are plotted for one wavelength in the $X$-direction. We note that although the group velocities of the waveforms are zero, the phase velocities in both presented examples are positive, indicating that the phase of the wave is moving in the positive $X$-direction. Furthermore, standing wave modes with negative phase velocities and zero group velocities can also be obtained for a suitable value of the non-dimensional parameter $\Omega$. The presence of the Coriolis effect results in asymmetric dispersion curves for the inertia-gravity waves, leading to standing modes with non-zero wavenumbers $k$ (see Fig. \ref{standing_modes}(a)). For higher modes, additional vortices are observed for the standing Rossby modes, and greater meridional variations are noticed in the standing inertia-gravity modes. 
\section{Modes of the meridional velocity equation}\label{rotatusabcd1}
 In this section, we consider the non-dimensional linearised shallow water equations for a horizontal layer of fluid, where the meridional velocity component is described by the Weber-type differential equation \cite{abramowitz1968handbook, ryzhik1963tables, buchholz2013confluent, tricomi1960fonctions, whittaker1920course}.   
When third-order terms in the expansions of $\sin(Y)$ and $\cos(Y)$ are neglected in \eq{newmark1}, we obtain the following meridional velocity equation (see Section \ref{webs1a}):  
 \begin{eqnarray}
 \frac{d^2 \hat{B}}{d Y^2} + \Big[\Lambda - \Phi Y^2 \Big]\hat{B} = 0, ~~~~ \hat{B}(Y=\pm \epsilon) = 0, \label{Bdiffeqanew}
 \end{eqnarray}
  where $\Lambda$ and $\Phi$ are defined in \eq{newdefin1}, and depend on the non-dimensional frequency and non-dimensional zonal wavenumber. The case of $\Phi=0$ was discussed in Section \ref{harmosh1a}. We consider the solutions to the problem \eq{Bdiffeqanew} subject to the conditions
 \begin{equation}
 \Lambda>0, ~~~ \Phi>0. \label{constranto1}
 \end{equation}
Indeed, we assume that the non-dimensional quantities $H$ and $\Omega$ (see Section \ref{rotatvec1}) satisfy the condition $\Omega/\sqrt{H} > 1/2,$ which holds for the corresponding dimensional parameters on the Earth. Based on the analysis of harmonic waves presented in Section \ref{harmosh1a}, we also take $\Lambda>0.$ Then, it follows that 
 \begin{equation}
 \frac{\Omega}{2\omega} - \sqrt{\frac{\omega^2}{H} + \frac{\Omega^2}{4 \omega^2}} < k  < \frac{\Omega}{2\omega} + \sqrt{\frac{\omega^2}{H} + \frac{\Omega^2}{4 \omega^2}},
 \end{equation}
 and taking into account \eq{newdefin1}, it can be deduced that
 \begin{equation}
 \Phi = \frac{\Omega^2}{H} + \frac{k \Omega}{ 2 \omega} > \frac{\Omega^2}{H} + \frac{\Omega^2}{ 4 \omega^2} - \frac{\Omega}{2 \omega}\sqrt{\frac{\omega^2}{H} + \frac{\Omega^2}{4 \omega^2}} = \frac{\Omega^2 ( \omega^2 (4\Omega^2 - H) + 2H\Omega^2 )}{4 H^2 \omega^2 \Big( \frac{\Omega^2}{H} + \frac{\Omega^2}{ 4 \omega^2} + \frac{\Omega}{2 \omega}\sqrt{\frac{\omega^2}{H} + \frac{\Omega^2}{4 \omega^2}} \Big)} > 0.
 \end{equation}

  The system \eq{Bdiffeqanew} represents an equatorial wave problem for the meridional velocity component, and its solutions are given by a linear combination of parabolic cylinder functions \cite{wunsche2003generalized, cane1979forced, ryzhik1963tables, abramowitz1968handbook, whittaker1920course, erdelyi1953higher, oldham2009atlas, miller1955tables} as follows
 \begin{equation}
 \hat{B}(Y) = \mathcal{C}_{1} U\Big(-\frac{\Lambda}{2\sqrt{\Phi}} , \sqrt{2} \Phi^{1/4} Y \Big) + \mathcal{C}_{2} V\Big(-\frac{\Lambda}{2\sqrt{\Phi}}, \sqrt{2} \Phi^{1/4} Y \Big),  \label{okwhita111}
 \end{equation}
 where $\mathcal{C}_{1}$ and $\mathcal{C}_{2}$ are arbitrary constants, while $U$ and $V$ denote the parabolic cylinder functions \cite{abramowitz1968handbook, whittaker1920course}, which form two linearly independent solutions of the differential equation shown in \eq{Bdiffeqanew}. 
  
 Applying the boundary conditions at $Y =\pm \epsilon$ results in the following non-dimensional dispersion relation:
 \begin{equation}
  U\Big(-\frac{\Lambda}{2\sqrt{\Phi}}, - \sqrt{2} \Phi^{1/4} \epsilon \Big) V\Big(-\frac{\Lambda}{2\sqrt{\Phi}}, \sqrt{2} \Phi^{1/4} \epsilon \Big)   - U\Big(-\frac{\Lambda}{2\sqrt{\Phi}}, \sqrt{2} \Phi^{1/4} \epsilon \Big)  V\Big(-\frac{\Lambda}{2\sqrt{\Phi}}, - \sqrt{2} \Phi^{1/4} \epsilon \Big) =0. \label{solvabili1az11}
 \end{equation}
When the condition \eq{solvabili1az11} is satisfied,  there exist non-trivial solutions for the meridional velocity problem \eq{Bdiffeqanew}. Noting the representation \eq{newdefin1},  equation \eq{solvabili1az11} can be solved numerically yielding the approximate non-dimensional eigenvalues associated with the equatorial waveforms.  Moreover, since the boundary value problem \eq{Bdiffeqanew} is of the Sturm-Liouiville form, the eigenvalues form a non-negative increasing sequence in the interval $[-\epsilon, \epsilon].$ The derivation of the asymptotic solutions of the non-dimensional dispersion equation \eq{solvabili1az11} can be found in Section \ref{carabani1a}, where it is shown that the leading order terms of the solutions correspond to the solutions of the dispersion equation \eq{dispsim1} of the harmonic equatorial waves.

  In the papers \cite{matsuno1966quasi, kiladis2009convectively, lighthill1969dynamic}, a dispersion equation was derived for the equatorial waves with the boundary conditions applied at infinity, where the propagating solutions were assumed to decay away from the equatorial region. Such boundary conditions are inconsistent with the neglect of higher-order terms in the expansions of $\sin(y/R)$ and $\cos(y/R)$ (see \eq{Bdiffeqa} in Section \ref{rotatvec1}).
  Solutions presented in \cite{erlick2007linear, paldor2015shallow} use a scaling to derive asymptotic solutions for waves in equatorial channels, with approximations for the waves in narrow and wide channels.

To obtain analytical insights into the solution of the eigenvalue problem \eq{Bdiffeqanew}, we construct an asymptotic approximation of the eigenfunctions and corresponding eigenvalues.

\subsection{Regular perturbation of the Weber-type equation describing equatorial modes}\label{movasymp1}

In this section, we derive the asymptotic approximations of the eigenfunctions and eigenvalues of the meridional velocity mode, corresponding to the Weber-type differential equation described in Section \ref{rotatusabcd1}. 

We examine approximations to the solutions by considering the problem \eq{Bdiffeqanew} as a perturbation eigenvalue problem in the form 
\begin{equation}\label{limaiw1a}
-\hat{B}''_{n}(Y) + \Phi Y^2 \hat{B}_{n}(Y) =  \Lambda_{n}\hat{B}_{n}(Y), ~~~~~ \hat{B}_{n}(\pm\epsilon) =0,
\end{equation} 
where $\Phi>0$ and $\Lambda>0.$
Introducing the variables: 
\begin{equation}\label{vari1abzc}
\xi = \frac{\pi Y}{2 \epsilon}, ~~~ \Psi = \Big(\frac{2 \epsilon}{\pi}\Big)^{4} \Phi, ~~~ \lambda = \Big(\frac{2 \epsilon}{\pi}\Big)^2 \Lambda,
\end{equation}
the eigenvalue problem \eq{limaiw1a} can be re-written as 
\begin{equation}\label{limape1}
-\mathcal{B}''_{n}(\xi) + \Psi \xi^2 \mathcal{B}_{n}(\xi) =  \lambda_{n} \mathcal{B}_{n}(\xi), ~~~~~ \mathcal{B}_{n}\Big(\pm\frac{\pi}{2}\Big) =0,
\end{equation} 
where $\mathcal{B}_{n}(\xi) = \hat{B}_{n}\Big(\frac{2 \epsilon}{\pi} \xi \Big).$ We derive an asymptotic solution for the eigenvalue problem \eq{limape1} for the small perturbation parameter $\Psi.$ Then, the eigenfunctions $\mathcal{B}_{n}(\xi)$ and eigenvalues $\lambda_{n}$ are sought in the asymptotic forms
\begin{eqnarray}
\begin{gathered} \label{asympo12az}
\mathcal{B}_{n}(\xi) = \mathcal{B}_{n}^{(0)}(\xi) + \Psi \mathcal{B}_{n}^{(1)}(\xi) +  O(\Psi^2), \\ 
 \lambda_{n} = \lambda_{n}^{(0)} + \Psi\lambda_{n}^{(1)} + O(\Psi^2),
\end{gathered}
\end{eqnarray}
where $\mathcal{B}_{n}^{(0)}(\xi)$ and $\lambda_{n}^{(0)}$ are the normalised eigenfunction and eigenvalue of the limit problem \eq{limape1} when $\Psi=0,$ respectively, and $\mathcal{B}_{n}^{(1)}(\xi)$ and $\lambda_{n}^{(1)}$ are the correction terms.

\subsection{The limit eigenvalue problem}\label{apprnedix1}
The limit problem of the regular perturbation problem \eq{limape1} takes the form
\begin{equation}
(\mathcal{B}_{n}^{(0)}(\xi))'' + \lambda_{n}^{(0)} \mathcal{B}_{n}^{(0)}(\xi) =  0, ~~~~~ \mathcal{B}_{n}^{(0)}\Big(\pm\frac{\pi}{2}\Big) =0.
\end{equation} 
The eigenvalues and normalised eigenfunctions of the above problem are, respectively, given by 
\begin{equation}
\lambda_{n}^{(0)} = n^2, ~~~~ n = 1, 2, 3, \ldots,
\end{equation}
and 
\begin{equation}
\mathcal{B}_{n}^{(0)}(\xi) =  \begin{cases}
\sqrt{2/\pi} \cos(n \xi), ~~ \text{for $n$ odd}, \\
\sqrt{2/\pi} \sin(n \xi), ~~ \text{for $n$ even}.
\end{cases} \label{limot1}
\end{equation}
The above solution corresponds to the meridional velocity mode component for the harmonic waves in a narrow equatorial band, as discussed in Section \ref{harmosh1a}. It is noted that the solutions of the limit eigenvalue problem also lead to the eigenfunctions in the sinusoidal form.  To simplify our analysis, we re-write the expression \eq{limot1} as follows 
\begin{equation}
\mathcal{B}_{n}^{(0)}(\xi) = \sqrt{\frac{2}{\pi}}\Bigg( \frac{(1 + (-1)^n)}{2} \sin(n \xi) +   \frac{(1 - (-1)^n)}{2} \cos(n \xi) \Bigg), ~~~~ n = 1,2,3,\ldots.
\end{equation}

\subsection{Evaluation of the first-order correction terms for the eigenfunctions and the eigenvalues}

By substituting the asymptotic forms of the eigenfunction and eigenvalue \eq{asympo12az} into \eq{limape1} and using the above representations for the normalised eigenfunction and eigenvalue of the limit eigenvalue problem, it follows that the correction terms $\mathcal{B}_{n}^{(1)}$ and $\lambda_{n}^{(1)}$ satisfy the boundary value problem 
\begin{eqnarray}\label{correct1asbc1}
( \mathcal{B}_{n}^{(1)}(\xi) )'' + n^2 \mathcal{B}_{n}^{(1)}(\xi) = - \lambda_{n}^{(1)} \mathcal{B}_{n}^{(0)}(\xi) + \xi^2 \mathcal{B}_{n}^{(0)}(\xi), ~~~~~ \mathcal{B}_{n}^{(1)}\Big(\pm\frac{\pi}{2}\Big) =0.
\end{eqnarray} 
We look for solutions of the inhomogeneous problem \eq{correct1asbc1} in the form 
\begin{equation}\label{eqna1z}
\mathcal{B}_{n}^{(1)}(\xi) = \sum_{j=1}^{\infty} a_{n j} \mathcal{B}_{j}^{(0)}(\xi) = \sqrt{\frac{2}{\pi}} \sum_{j=1}^{\infty} a_{nj } \Bigg[ \frac{(1 + (-1)^j)}{2} \sin(j \xi) +   \frac{(1 - (-1)^j)}{2} \cos(j \xi) \Bigg].
\end{equation}
By taking into account the forms of  $\mathcal{B}_{n}^{(0)}(\xi)$ and $\lambda_{n}^{(0)}$ in Section \ref{apprnedix1} and the above representation for $\mathcal{B}_{n}^{(1)}(\xi),$ the differential equation in \eq{correct1asbc1} can be written as 
\begin{equation}
\sum_{j=1}^{\infty} (n^2 - j^2)a_{n j} \mathcal{B}_{j}^{(0)}(\xi) = - \lambda_{n}^{(1)} \mathcal{B}_{n}^{(0)}(\xi) + \xi^2 \mathcal{B}_{n}^{(0)}(\xi). \label{anoth1az}
\end{equation}
We note the orthogonality relations of the sinusoidal functions in $\mathcal{B}_{j}^{(0)}(\xi)$:
\begin{equation}\label{eqoaoh1a}
\int_{-\frac{\pi}{2}}^{\frac{\pi}{2}} \mathcal{B}_{j}^{(0)}(\xi) \mathcal{B}_{p}^{(0)}(\xi) d \xi  = \delta_{j  p},
\end{equation}
where $p$ is a positive integer and $\delta_{j p}$ is the Kronecker delta function. As a result, multiplying \eq{anoth1az} by $\mathcal{B}_{ p }^{(0)}(\xi)$ and integrating over the interval $\Big(-\frac{\pi}{2}, \frac{\pi}{2}\Big),$ yields the following
\begin{equation}
a_{n p } (n^2 - p^2) = \int_{-\frac{\pi}{2}}^{\frac{\pi}{2}}(-\lambda_{n}^{(1)} + \xi^2) \mathcal{B}_{n}^{(0)}(\xi)\mathcal{B}_{p}^{(0)}(\xi) d \xi.
\end{equation}
When $p=n,$ we obtain the eigenvalue $\lambda_{n}^{(1)}$:
\begin{equation}
\lambda_{n}^{(1)} = \int_{-\frac{\pi}{2}}^{\frac{\pi}{2}} \xi^2 (\mathcal{B}_{n}^{(0)}(\xi))^2 d \xi = \frac{\pi^2}{12}  - \frac{1}{2 n^2}, ~~~~ n = 1,2,3,\ldots.
\end{equation}
In particular, if $p \neq n$, we derive the coefficients $a_{np}$:
\begin{equation}
a_{n p} =  \frac{1}{n^2 - p^2} \int_{-\frac{\pi}{2}}^{\frac{\pi}{2}}(-\lambda_{n}^{(1)} + \xi^2) \mathcal{B}_{n}^{(0)}(\xi)\mathcal{B}_{p}^{(0)}(\xi) d \xi \\  =  \frac{4  n p (-1)^{\frac{n+p}{2}} }{(n^2 - p^2)^3}( (-1)^{n} +  (-1)^{ p} ).
\end{equation}
To obtain the remaining coefficients $a_{nn},$ we normalise $\mathcal{B}_{n}(\xi)$ so that 
\begin{equation}
\int_{-\frac{\pi}{2}}^{\frac{\pi}{2}} \mathcal{B}_{n}^2(\xi) d \xi =1.
\end{equation}
Then since 
\begin{equation}
\mathcal{B}_{n}^2(\xi) = (\mathcal{B}_{n}^{(0)}(\xi))^2 + 2 \Psi \mathcal{B}_{n}^{(0)}(\xi)\mathcal{B}_{n}^{(1)}(\xi) + O(\Psi^2),
\end{equation}
and 
\begin{equation}
\int_{-\frac{\pi}{2}}^{\frac{\pi}{2}} (\mathcal{B}_{n}^{(0)}(\xi))^2 d \xi = 1,
\end{equation}
it follows that 
\begin{equation}\label{asjo1al}
\int_{-\frac{\pi}{2}}^{\frac{\pi}{2}} \mathcal{B}_{n}^{(0)}(\xi) \mathcal{B}_{n}^{(1)}(\xi) d \xi = 0.
\end{equation}
Finally, equations  \eq{eqna1z}, \eq{eqoaoh1a} and \eq{asjo1al} yield the following 
\begin{equation}
\int_{-\frac{\pi}{2}}^{\frac{\pi}{2}} \mathcal{B}_{n}^{(0)}(\xi)\Bigg(\sum_{j=1}^{\infty} a_{nj} \mathcal{B}_{j}^{(0)}(\xi)\Bigg) d \xi =  0,
\end{equation}
which implies that $a_{nn}=0.$

\subsection{Asymptotic approximation of the dispersion relation}
 In this section, we derive the non-dimensional eigenvalues associated with the frequencies of the equatorial waveforms. 

The asymptotic form of the eigenvalues is given by (see previous sections)
\begin{equation}
\lambda_{n} = n^2 - \frac{1}{2 n^2}\Psi + \frac{\pi^2}{12} \Psi + O(\Psi^2), ~~~~ n =1,2,3,\ldots.
\end{equation}
Noting the variables \eq{vari1abzc}, the above equation can be re-written as 
\begin{equation}\label{eqa1a}
\Lambda = \frac{n^2 \pi^2}{4 \epsilon^2} + \frac{\epsilon^2}{3}\Phi - \frac{2 \epsilon^2}{n^2 \pi^2} \Phi + O\Big( \Big(\frac{2 \epsilon}{\pi}\Big)^6 \Phi^2 \Big).
\end{equation}
Then by substituting the representations of $\Lambda$ and $\Phi$ in terms of the non-dimensional frequency $\omega$ and non-dimensional wavenumber $k,$ defined in \eq{newdefin1}, into \eq{eqa1a}, we obtain the following asymptotic dispersion equation:
\begin{equation}\label{epoisj1a}
\frac{\omega^2}{H} - k^2 + \frac{k \Omega}{\omega} = \frac{n^2 \pi^2}{4 \epsilon^2} + \epsilon^2 \Big(\frac{\Omega^2}{H} + \frac{k \Omega}{2 \omega}\Big)\Big(\frac{1}{3} - \frac{2}{n^2 \pi^2}\Big), ~~~~ n =1,2,3,\ldots,
\end{equation}
where the non-dimensional quantities $\Omega$ and $H$ are defined in Section \ref{rotatvec1}. The dispersion equation \eq{dispsim1} in Section \ref{harmosh1a} is a first-order approximation of the above dispersion relation. Analytical insights of equation \eq{epoisj1a} are provided below. The above dispersion relation can also be derived by considering approximations of the parabolic cylinder functions in \eq{solvabili1az11} as detailed in Section \ref{carabani1a}.

 \subsection{Dispersion properties: waves parallel to a narrow equatorial strip}\label{newdispersprop}
We present the asymptotic solutions of the dispersion equation \eq{epoisj1a}, providing an analytic approximation of the non-dimensional frequencies for the propagating waves in a narrow equatorial band. Approximations of the non-dimensional eigenvalues for equatorial waves in an asymmetric channel were presented in \cite{cane1979forced}.

\begin{figure}[ht]
  \centering
  \begin{minipage}[b]{0.95\textwidth}
    \includegraphics[width=1\linewidth]{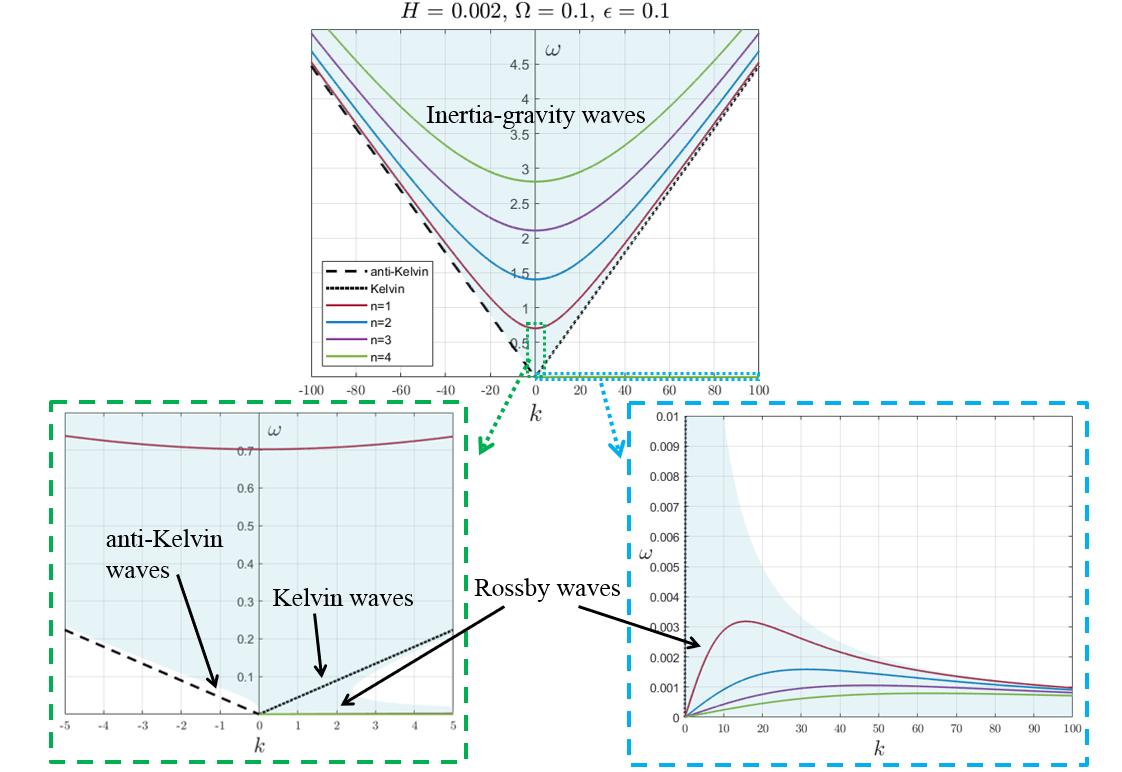}
  \end{minipage}
  \caption{\footnotesize The dispersion diagram corresponding to the non-dimensional dispersion relation \eq{epoisj1a} for $n=1,2, 3$ and $4,$ and the dimensionless parameters $\Omega=0.1, H=0.002$ and $\epsilon=0.1.$ For each value of $n,$ the low-frequency Rossby waves and the high-frequency inertia-gravity are observed, which are situated within the shaded region corresponding to $\Lambda>0$ and $\Phi>0$ (see \eq{constranto1}). The non-dispersive anti-Kelvin and Kelvin waves described in Section \ref{nonrotatus} are also shown; these are defined by the non-dimensional dispersion relation \eq{kelviaw11a}. The curve associated with the anti-Kelvin mode is outside the shaded region defined by \eq{constranto1}.}
  \label{updaznewdisper1}
\end{figure}

Similar to the non-dimensional dispersion relation derived in Section \ref{harmosh1a}, here we also observe that there are three roots of the dispersion equation \eq{epoisj1a} when $H, k, \Omega$ and $\epsilon$ are specified for a given mode $n$: two roots corresponding to inertia-gravity waves and the third to a Rossby wave.  The solutions of \eq{epoisj1a} corresponding to a fixed $n\geq 1$ result in the dispersion curves displayed in Fig. \ref{updaznewdisper1}, where the shaded region is defined by the conditions \eq{constranto1}. In particular, we note that the even modes ($n=2,4,6,\ldots$) of the normalised eigenfunctions derived above are skew-symmetric about the narrow equatorial region, while the odd modes ($n=1,3,5,\ldots$) are symmetric. This behaviour is consistent with the eigenfunctions of the harmonic waves detailed in Section \ref{harmosh1a}. The eigenfunctions of such equatorial modes are also discussed in \cite{cane1979forced, philander1989nino, chart1998study}, in the context of meridionally bounded propagating waves. The set of dispersion curves shown in Fig. \ref{updaznewdisper1} also form a collection of non-intersecting high-frequency inertia-gravity waves and low-frequency Rossby waves that are inside the shaded region, which can be compared to the curves in the dispersion diagram shown in Fig. \ref{new1a}. The main properties of these curves were also captured by the non-dimensional dispersion relation detailed in Section \ref{harmosh1a}.

  Taking into account the above analysis and computations, we identify the equatorial wave classes as follows: 
 \begin{itemize}

 \item  The anti-Kelvin and Kelvin waves that propagate zonally non-dispersively westward and eastward, respectively, in the narrow equatorial band with the normalised phase speed $\sqrt{H}$ (see Section \ref{nonrotatus}).  The corresponding group velocities for the eastward- and westward-propagating waveforms are given by $\pm \sqrt{H},$ respectively. We note that for increasing values of $Y,$ the westward-propagating anti-Kelvin waves have zonally increasing amplitudes, while the eastward-propagating Kelvin waves are trapped in the narrow band with maximum amplitudes at the equator. The values of the non-dimensional parameters $\Omega, H$ and $\epsilon,$ together with the conditions \eq{constranto1}, resulted in the dispersion curve of the anti-Kelvin mode being located outside the shaded region as shown in Fig. \ref{updaznewdisper1}.

   \item For the high-frequency inertia-gravity waves for large $k,$ the dispersion relation \eq{epoisj1a} yields the approximate non-dimensional frequencies, valid to $O(1/k),$
   \begin{eqnarray}
   \begin{gathered}
  \omega_{1,2} = \mp k \sqrt{H} - \frac{\Omega}{2 k} \mp \frac{n^2 \pi^2 \sqrt{H}}{8 k \epsilon^2} + \frac{\Omega (n^2 \pi^2 - 6)(H \mp 2 \Omega\sqrt{H})}{12 H n^2 \pi^2 k}\epsilon^2, ~~~ n=1,2,3\ldots. \label{funcso1avb}
   \end{gathered}
   \end{eqnarray}
The above approximate roots correspond to the equatorial modes associated with the westward- and eastward- propagating inertia-gravity waves, as illustrated in Fig. \ref{updaznewdisper1}. We note that the dispersion curves of the inertia-gravity waves are asymmetric relative to the vertical frequency axis due to the presence of the Coriolis force. The asymmetry becomes more evident for larger values of $|\Omega|,$ and reversing the sign of $\Omega$ results in a change in the directional horizontal shift of the dispersion curves. Additionally, such asymmetry feature allows for short-period inertia-gravity waves with zonal phase velocities in opposite directions compared to the group velocities.  An analogous asymmetry in the dispersion curves was also observed in Section \ref{discretelatstrip} for the infinite discrete chiral lattice strip subjected to gravity, where gyroscopic forces were present instead of the Coriolis force.
   
   \item Low-frequency Rossby waves for large $k$ can be approximated by the following, valid to $O(1/k)$,
   \begin{equation}
   \omega_{3} = \frac{\Omega}{k} - \frac{(n^2 \pi^2 - 6)\Omega }{6 n^2 \pi^2 k}\epsilon^2, ~~~ n=1,2,3\ldots. \label{resso1}
   \end{equation}
By neglecting terms of order $O(\epsilon^2)$, the representation \eq{resso1} shows that the non-dimensional zonal wavenumber $k$ is positive (\emph{resp}. negative) for $\Omega>0$ (\emph{resp}. $\Omega<0$) since the non-dimensional frequency $\omega_3$ is positive. Thus, the phase velocities of the low-frequency Rossby waves in the narrow equatorial strip result in eastward-moving oscillations (\emph{resp}. westward-moving oscillations) for $\Omega>0$ (\emph{resp}. $\Omega<0$).  The Rossby dispersion curves are presented in Fig. \ref{updaznewdisper1}, where it is observed that although the phase velocity of the wave components corresponds to eastward-propagating disturbances for $\Omega>0$ (\emph{resp}. westward-propagating for $\Omega<0$), their zonal group velocity can result in either eastward- or westward-propagating waves. 
 
 \end{itemize}

The above asymptotic analysis shows that the low-frequency Rossby waves can exhibit eastward (\emph{resp}. westward) phase disturbances and westward (\emph{resp}. eastward) wave propagations for $\Omega >0$ (\emph{resp.} $\Omega<0$) and large $k$. Kelvin waves, being non-dispersive, move with phase disturbances at the same speed as their group velocity as detailed in Section \ref{nonrotatus}.  Moreover, inertia-gravity waves can have either eastward or westward phase velocities, as well as either eastward or westward group velocities. For low zonal wavenumbers, the direction of the propagating inertia-gravity waves can vary due to the Coriolis force affecting the phase and group velocities. It is know that inertia-gravity waves propagate much faster than Rossby waves \cite{leblond1981waves, matsuno1966quasi, philander1989nino}.

 \subsection{Dispersion characteristics of the equatorial modes. Asymptotic expansions of Kummer functions}\label{carabani1a}

 The eigenvalue problem \eq{Bdiffeqanew} represents an equatorial wave problem for the meridional velocity component, and its solutions consist of a linear combination of parabolic cylinder functions as detailed in Section \ref{rotatusabcd1}, with the asymptotic approximations provided in Section \ref{movasymp1}. The application of the boundary conditions at $Y=\pm \epsilon$ resulted in the solvability condition \eq{solvabili1az11}, which corresponds to the dispersion relation of the equatorial waveforms. In this section, we present the asymptotic analysis of the non-dimensional dispersion equation \eq{solvabili1az11}. In particular, we show that the solutions of the dispersion equation of the harmonic equatorial waves derived in Section \ref{harmosh1a} provide a good agreement with the asymptotic solutions of \eq{solvabili1az11}.

  To gain some insights into the non-dimensional dispersion relation \eq{solvabili1az11}, we examine approximations of the solutions to derive analytical estimates for the eigenvalues. To advance with the asymptotic analysis, we use the following relations \cite{ abramowitz1968handbook}
  {\footnotesize
 \begin{gather}
 U(\frak{b}, z) = \frac{e^{-\frac{z^2}{4}}\Gamma\Big(\frac{1}{4}-\frac{\frak{b}}{2}\Big)\cos\Big(\pi \Big(\frac{1}{4}+\frac{\frak{b}}{2}\Big)\Big)}{ 2^{\frac{\frak{b}}{2}+\frac{1}{4}} \sqrt{\pi}} M\Big(\frac{\frak{b}}{2}+\frac{1}{4},\frac{1}{2},\frac{z^2}{2}\Big)    - \frac{e^{-\frac{z^2}{4}}\Gamma\Big(\frac{3}{4}-\frac{\frak{b}}{2}\Big)\sin\Big(\pi \Big(\frac{1}{4}+\frac{\frak{b}}{2}\Big)\Big)}{ 2^{\frac{\frak{b}}{2}-\frac{1}{4}} \sqrt{\pi}} z M\Big(\frac{\frak{b}}{2}+\frac{3}{4},\frac{3}{2},\frac{z^2}{2}\Big), \label{relatioab1} \\
 V(\frak{b},z) = \frac{1}{\Gamma\Big(\frac{1}{2}-\frak{b}\Big)}\Bigg\{\frac{e^{-\frac{z^2}{4}}\Gamma\Big(\frac{1}{4}-\frac{\frak{b}}{2}\Big)\sin\Big(\pi \Big(\frac{1}{4}+\frac{\frak{b}}{2}\Big)\Big)}{ 2^{\frac{\frak{b}}{2}+\frac{1}{4}} \sqrt{\pi}} M\Big(\frac{\frak{b}}{2}+\frac{1}{4},\frac{1}{2},\frac{z^2}{2}\Big)  
\nonumber \\ 
+ \frac{e^{-\frac{z^2}{4}}\Gamma\Big(\frac{3}{4}-\frac{\frak{b}}{2}\Big)\cos\Big(\pi \Big(\frac{1}{4}+\frac{\frak{b}}{2}\Big)\Big)}{ 2^{\frac{\frak{b}}{2}-\frac{1}{4}} \sqrt{\pi}} z M\Big(\frac{\frak{b}}{2}+\frac{3}{4},\frac{3}{2},\frac{z^2}{2}\Big)\Bigg\}, \label{relatioab2}
 \end{gather}
 }
  where the variable $z$ and parameter $\frak{b}$ can take complex values. In the above representations, $U(\frak{b}, z)$ and $V(\frak{b}, z)$ are the parabolic cylinder functions, $\Gamma(\mathpzc{z})$ denotes the gamma function and $M(\mathpzc{x},\mathpzc{y},\mathpzc{z})$ is Kummer's function of the first kind  \cite{abramowitz1968handbook, buchholz2013confluent, tricomi1960fonctions, magnus1966formulas, slater1960confluent}. Applying the above relations to the equation \eq{solvabili1az11}, results in the following simplified dispersion equation 
 \begin{eqnarray}
 \begin{gathered}
 \frac{4 \epsilon \Phi^{1/4} e^{- \epsilon^2 \sqrt{\Phi} } M\Big(-\frac{\Lambda}{4 \sqrt{\Phi}}+\frac{1}{4}, \frac{1}{2}, \epsilon^2 \sqrt{\Phi}\Big) M\Big(-\frac{\Lambda}{4 \sqrt{\Phi}}+\frac{3}{4}, \frac{3}{2}, \epsilon^2 \sqrt{\Phi}\Big)}{ \sqrt{\pi}} =0. \label{condi1aa}
 \end{gathered}
 \end{eqnarray} 
By taking into account \eq{newdefin1}, equation \eq{condi1aa} can be solved numerically to obtain the non-dimensional eigenvalues for the meridional modes. To derive the eigenvalues, we consider the asymptotic approximations to the zeros of the Kummer functions in \eq{condi1aa} for large $\Lambda/\sqrt{\Phi}.$

\subsection{Approximations to the zeros of $M\Big(-\frac{\Lambda}{4 \sqrt{\Phi}}+\frac{1}{4}, \frac{1}{2}, \epsilon^2 \sqrt{\Phi}\Big)$ for large $\frac{\Lambda}{\sqrt{\Phi}}$}\label{aisodhasd800a}

In the following analysis, we consider the asymptotic properties of  Kummer's function valid for large $\Lambda/\sqrt{\Phi},$ where the non-dimensional quantities $\Lambda$ and $\Phi$ are defined in \eq{newdefin1}. It is also noted that $\Lambda$ and $\Phi$ are chosen such that the conditions in \eq{constranto1} are satisfied. An asymptotic result for the zeros of $M\Big(-\frac{\Lambda}{4 \sqrt{\Phi}}+\frac{1}{4}, \frac{1}{2}, \epsilon^2 \sqrt{\Phi}\Big)$ with large $\Lambda/\sqrt{\Phi}$ is given by the formula \cite{slater1960confluent, tricomi1947sugli}
 \begin{equation}
\sqrt{\Phi}{\epsilon^2} = \frac{j_{-\frac{1}{2},r}^2 \sqrt{\Phi}}{\Lambda}\Bigg(1 + \frac{(2 j_{-\frac{1}{2}, r}^2 - 3) \Phi}{6 \Lambda^2}\Bigg) +  O\Bigg(\Bigg(\frac{\sqrt{\Phi}}{\Lambda}\Bigg)^{4}\Bigg), ~~~~ r=1,2,3,\ldots, \label{relation1abd}
 \end{equation}
 where $j_{-\frac{1}{2}, r}$ is the $r$-th positive zero of the Bessel function $J_{-\frac{1}{2}}(z)$ of the first kind \cite{abramowitz1968handbook}. Noting that \cite{watson1922treatise, elbert2001some}
 \begin{equation}
 J_{-\frac{1}{2}}(z) = \sqrt{\frac{2}{\pi z}} \cos(z),
 \end{equation}
 we obtain the following representation for $j_{-\frac{1}{2}, r}$:
 \begin{equation}
 j_{-\frac{1}{2}, r} = \frac{(2 r - 1) \pi}{2}, ~~~ r=1,2,3,\ldots.
\end{equation}
   Applying the above asymptotic relation for the Kummer function, we derive the following approximation for the solutions of equation  \eq{condi1aa}: 
 \begin{equation}
 \Lambda = \frac{(2 r - 1)^2 \pi^2}{4 \epsilon^2} + \frac{\epsilon^2}{3}\Phi - \frac{2 \epsilon^2 }{(2r -1)^2 \pi^2}\Phi, ~~~ r=1,2,3,\ldots. \label{dispersion1abc}
 \end{equation}
The dispersion relation \eq{dispersion1abc} corresponds to the approximate odd modes of the dispersion equation \eq{eqa1a} for the equatorial waves in a narrow band.  Taking into account \eq{newdefin1}, the dispersion curves described by the approximation \eq{dispersion1abc} correspond to the odd modes shown in Fig. \ref{updaznewdisper1} for $n=1$ and $n=3$ (i.e. for $r=1$ and $r=2,$ respectively). In particular, for each value of $r,$ there are three types of curves associated with the equatorial waveforms: westward-propagating inertia-gravity waves, eastward-propagating inertia-gravity waves and Rossby waves.

 \subsection{Approximations to the zeros of $M\Big(-\frac{\Lambda}{4 \sqrt{\Phi}}+\frac{3}{4}, \frac{3}{2}, \epsilon^2 \sqrt{\Phi}\Big)$ for large $\frac{\Lambda}{\sqrt{\Phi}}$}
In this section we consider the following asymptotic approximation to the zeros of $M\Big(-\frac{\Lambda}{4 \sqrt{\Phi}}+\frac{3}{4}, \frac{3}{2}, \epsilon^2 \sqrt{\Phi}\Big)$ with large $\Lambda/\sqrt{\Phi}$:
 \begin{equation*}
 \sqrt{\Phi}{\epsilon^2} = \frac{j_{\frac{1}{2},p}^2 \sqrt{\Phi}}{\Lambda}\Bigg(1 + \frac{(2 j_{\frac{1}{2}, p}^2 - 3) \Phi}{6 \Lambda^2}\Bigg) +  O\Bigg(\Bigg(\frac{\sqrt{\Phi}}{\Lambda}\Bigg)^{4}\Bigg), ~~~ p=1,2,3,\ldots,
 \end{equation*}
 where $j_{\frac{1}{2},p}$ is the $p$-th positive zero of the Bessel function $J_{\frac{1}{2}}(z),$ which is given by 
 \cite{elbert2001some}
 \begin{equation*}
 J_{\frac{1}{2}}(z) = \sqrt{\frac{2}{\pi z}} \sin(z).
 \end{equation*}
 By following a similar approach to Section \ref{aisodhasd800a}, we obtain the following asymptotic result of $M\Big(-\frac{\Lambda}{4 \sqrt{\Phi}}+\frac{3}{4}, \frac{3}{2}, \epsilon^2 \sqrt{\Phi}\Big)$ for large $\Lambda/\sqrt{\Phi}$:
 \begin{equation}
 \Lambda = \frac{p^2 \pi^2}{\epsilon^2} + \frac{\epsilon^2}{3}\Phi - \frac{\epsilon^2}{2 p^2 \pi^2}\Phi, ~~~ p=1,2,3,\ldots. \label{dispersion1abc1}
 \end{equation}
Using \eq{newdefin1}, it follows that equation \eq{dispersion1abc1} yields an approximate dispersion relation for the waveforms in a narrow band. Equation \eq{dispersion1abc1} also provides an approximation to the solutions of \eq{condi1aa}, and corresponds to the even modes of the dispersion relation \eq{eqa1a}. The associated dispersion curves are shown in Fig. \ref{updaznewdisper1} for $n=2$ and $n=4$ (i.e. for $p=1$ and $p=2,$ respectively). These curves are related to the inertia-gravity waves and the Rossby waves, and are located inside the shaded region determined by \eq{constranto1}.

By considering \eq{newdefin1} together with  \eq{dispersion1abc} and \eq{dispersion1abc1}, we obtain an approximation for the frequencies and wavenumbers of the equatorial waveforms. From this approximation, the eigenfunctions of the meridional velocity mode can be obtained by using the equation \eq{okwhita111}. The above asymptotic analysis further validates the approximation of the eigenvalues of the harmonic waves detailed in Section \ref{harmosh1a}.

\section{Gyropendulum approximation of the shallow water equations}\label{temporinterfacegyro} 
In this section, we show that a gyropendulum can be used to approximately describe the dynamics of a shallow water ridge in the polar regions of rotating planets. In this approximation, the shallow water equations are considered at high latitudes, where the large magnitude of the Coriolis force contributes to the formation and motion of polar vortices. We demonstrate that starting with a linearised form of the shallow water equations and assuming a purely horizontal flow, one obtains a system of equations resembling the governing equations of an elementary gyropendulum \cite{kandiah2023effect, kandiah2024controlling}. Gyropendulums were also a pivotal structure in the elementary cells of the infinite discrete chiral lattice strips studied in Section \ref{asdjpasdj08212}.

The analysis presented in this section provides a new framework for modelling the fundamental features of the rotating fluid ridge through a discrete gyropendulum structure. Namely, for a fluid particle, the restoring force, primarily due to gravity and buoyancy, returns the particle to its equilibrium position, while the Coriolis force, resulting from the planet's rotation, acts perpendicular to the particle's motion in the rotating frame of reference. Conversely, for a gyropendulum, the restoring force arises due to gravity, whereas rotational effects are introduced through a gyroscopic spinner. Although Coriolis and gyroscopic forces emerge from distinct physical phenomena, they both induce analogous effects that influence the motions of structures in rotating systems. This characteristic forms a fundamental basis for our analysis. 

We will also present typical examples of the gyropendulum trajectories approximating polygonal shapes by following the method presented in  \cite{kandiah2023effect}. The approximate polygonal trajectories, described by the combined action of gyroscopic forces and gravity, are linked to a range of natural phenomena at different scales, such as polygonal shapes in a partially filled cylindrical container subjected to a rotating bottom plate  \cite{vatistas1990note, jansson2006polygons, kadlecsik2023simple} or polygonal patterns in polar observations of rotating planets \cite{ingersoll2020cassini, tabataba2020long, allison1990wave, seviour2017stability}. The illustrative examples of the gyropendulum motions with polygonal approximations are presented in connection with polar jet streams on rotating celestial bodies.

\subsection{Transient motion of a gyropendulum and shallow water ridge}\label{rippler1}

The equations of motion for a linearised shallow water system in the vicinity of the South Pole of the Earth are written as (see Section \ref{rotatingshall1}) 
\begin{equation}
\frac{\partial u}{\partial t} + 2\Omega v = - g\frac{\partial h}{\partial x}, ~~~
\frac{\partial v}{\partial t} - 2\Omega u = -g\frac{\partial h}{\partial y}. \label{whe2aa} 
\end{equation}
It is noted that changing the sign of $\Omega$ results in the corresponding shallow water model at the North Pole and thus, without loss of generality we only consider the system at the South Pole. 

We investigate the instantaneous motion of a layer of incompressible fluid of homogeneous density subject to a clockwise orientation, which is described by the linearised shallow water equations. In our model, we assume that the height deviation $h,$ from the undisturbed fluid surface is negligibly small compared to the fluid depth $H$ (see Section \ref{intro1a}). Then, the fluid element corresponding to the maximum elevation of the perturbed surface can be represented by a ridge. In the following, we show that the motion of the fluid ridge can be approximated by the linearised motion of an elementary gyropendulum. Accordingly, at the fluid height $h$, the flow is approximated as planar. With reference to the above assumptions, we note the following:
\begin{equation}
\frac{\partial h}{\partial x} = \frac{\mathcal{U}}{L}, ~~~ \frac{\partial h}{\partial y} = \frac{\mathcal{V}}{L}, \label{velocias2}
\end{equation}
where $\mathcal{U}$ and $\mathcal{V}$ are the planar displacement components of the fluid particle at the tip of the ridge in the $x$- and $y$-directions, respectively. Assuming that the velocity vectors are functions of time only, with no spatial variations along the $x$- and $y$-axes, linked to a spatially uniform flow, the velocities $u=u(t)$ and $v=v(t)$ in the $x$- and $y$-directions, respectively, are given by  
\begin{equation}
u = \frac{d \mathcal{U}}{d t}, ~~~ v = \frac{d \mathcal{V}}{ d t}. \label{velocias1}
\end{equation}
Substituting the representations \eq{velocias2} and \eq{velocias1} into the system \eq{whe2aa}, and re-arranging, yields the set of coupled differential equations
\begin{equation}
\frac{d^2 \mathcal{U}}{d t^2} + 2 \Omega \frac{d \mathcal{V}}{d t} + g \frac{\mathcal{U}}{L} = 0, ~~~ \frac{d^2 \mathcal{V}}{d t^2} - 2 \Omega \frac{d \mathcal{U}}{d t} + g \frac{\mathcal{V}}{L} = 0. \label{gyroscopicpendo1}
\end{equation}
The system \eq{gyroscopicpendo1} is analogous to the governing equations of a gyropendulum \cite{kandiah2023effect}, describing the linearised motion of a pendulum, attached to a rotating spinner, that swings under gravity. Hence, the shallow water equations subject to the above assumptions provide a connection with the motion of a gyropendulum. We note that the gyropendulum approximation captures some aspects of the shallow water model, but it does not fully account for the complexity of the fluid dynamics. A detailed model of such phenomena would require a comprehensive CFD analysis. The system \eq{gyroscopicpendo1} also describes the trajectory of a Foucault pendulum and the motion of a particle moving on a rotating surface  \cite{kirillov2016rotating, kirillov2021nonconservative}.

\begin{figure}[H]
  \centering
    \begin{minipage}[b]{0.8\textwidth}
    \hspace{-0.6cm}\includegraphics[width=1\linewidth]{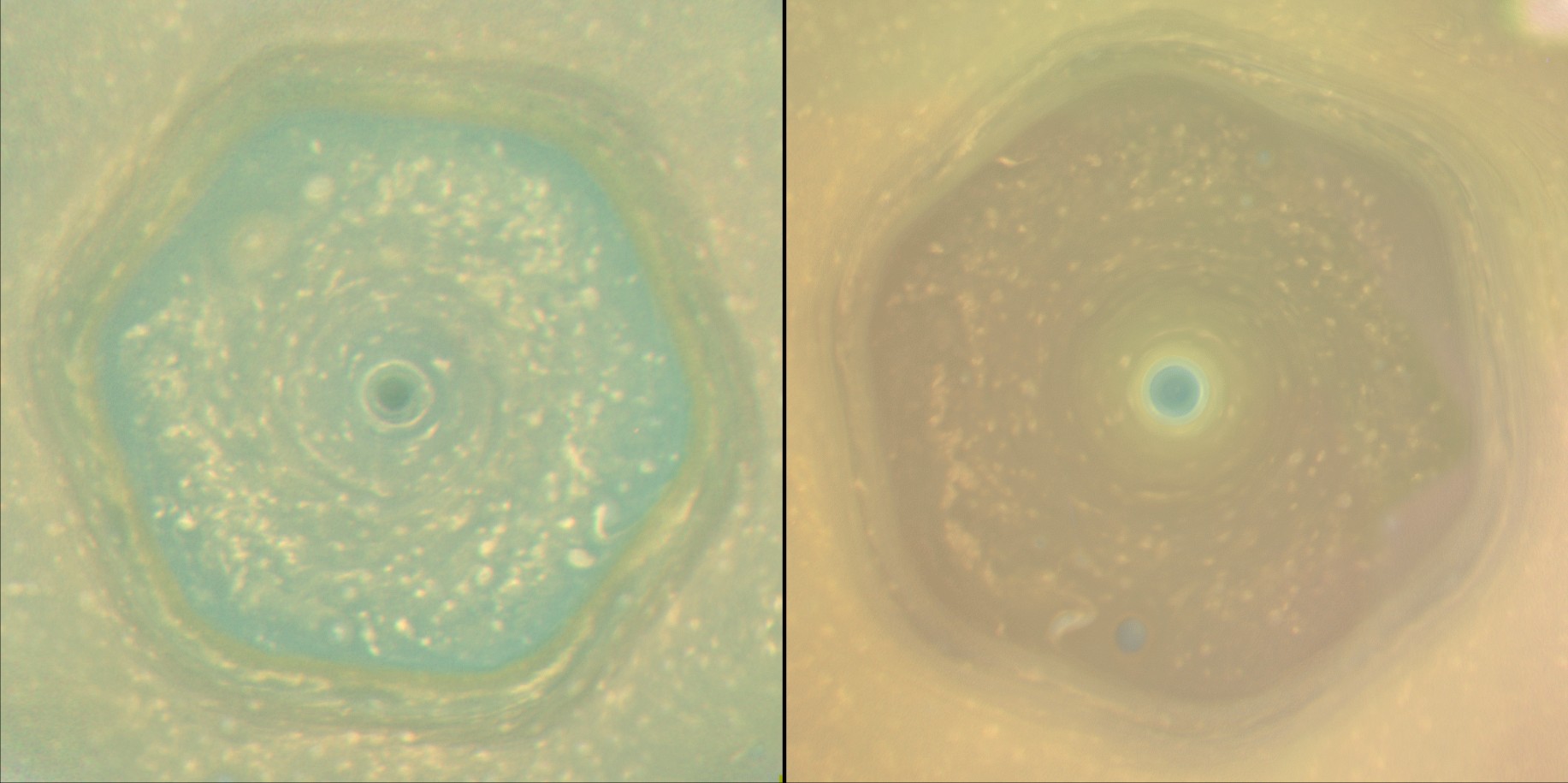}
  \end{minipage}
  \caption{\footnotesize Images from NASA's Cassini mission showing the appearance of Saturn's North Pole in June 2013 (left image) and April 2017 (right image), revealing the distinctive six-sided jet stream known as the ``hexagon."
Image courtesy of NASA, retrieved from \url{https://www.jpl.nasa.gov/images/pia21611-saturns-hexagon-as-summer-solstice-approaches} as of February 2025.}
  \label{planetpoles}
\end{figure}

\begin{figure}[H]
  \centering
    \begin{minipage}[b]{0.45\textwidth}
    \hspace{-0.2cm}\includegraphics[width=1\linewidth]{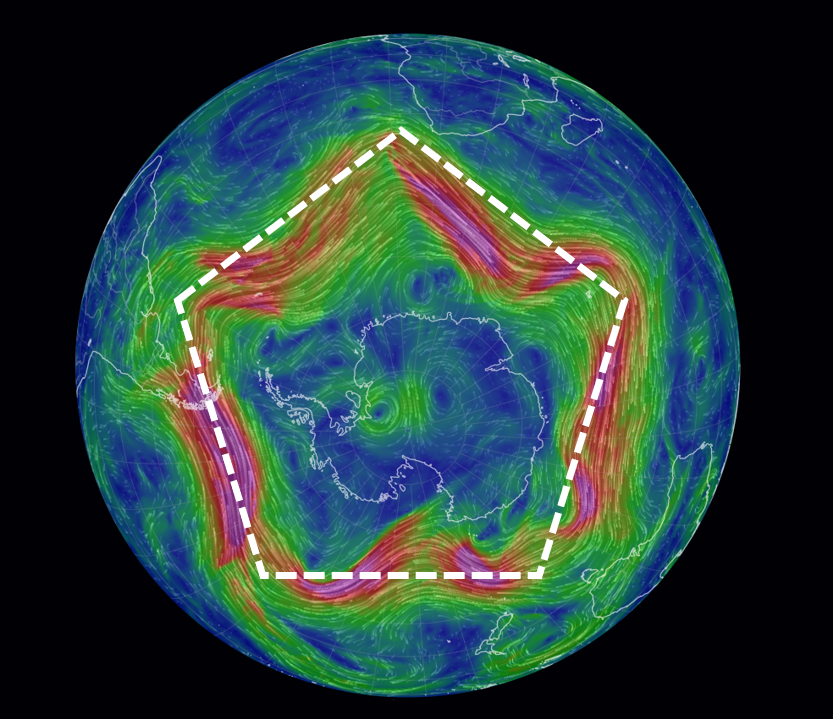}
  \end{minipage}
    \caption{\footnotesize Approximate pentagonal jet stream at the South Pole of Earth accessed from \url{https://earth.nullschool.net/\#2023/01/03/1300Z/wind/isobaric/250hPa/orthographic} taken on 3$^{\text{rd}}$ January, 2023. }
  \label{southearth}
\end{figure}

In the paper \cite{kandiah2023effect}, it was demonstrated that by selecting appropriate initial conditions and physical parameters, the moving gyropendulum can trace a predetermined polygonal trajectory.  The full classification of the gyropendulum trajectories is also provided in \cite{kandiah2023effect}, which includes the discussion of various combinations of initial conditions and parameter values on the motion of the structure. 
Polygonal trajectories, linked to polar jet streams and vortex flows, also exist at the poles of planets, influenced by the combined effects of the planet's rotation and gravity. For example, Saturn's North Pole is surrounded by a six-sided jet stream as illustrated in Fig. \ref{planetpoles}, which was initially discovered by the Voyager spacecraft in the early $1980$s and later observed by Cassini  \cite{ingersoll2020cassini}. An analysis of these observations is detailed in \cite{allison1990wave}. Eight cyclonic vortices appear around a central vortex at the North Pole of Jupiter, while its South Pole consists of five vortices surrounding a central vortex \cite{tabataba2020long}. An approximate pentagonal jet stream is present at the South Pole of Earth as shown in Fig. \ref{southearth}. The numerical analysis of the shallow water equations to investigate the stability of Mars-like annular vortices was discussed in \cite{seviour2017stability}. 
 Laboratory models demonstrating stable polygonal patterns of vortex liquid sloshing in cylindrical containers were presented in \cite{vatistas1990note, aguiar2010laboratory, jansson2006polygons, kadlecsik2023simple}.

\subsection{Dimensionless system of equations}
The dimensionless system of equations of a gyropendulum are expressed as follows\footnote{The dimensionless equations for the transverse displacement components of the gyropendulum can also be obtained by introducing the dimensionless time $\tilde{t}=t\sqrt{g/L}$ into the system \eq{gyroscopicpendo1}, where $\Gamma=2\Omega\sqrt{L/g},$ $\tilde{\mathcal{U}}(\tilde{t}) = \mathcal{U}(\tilde{t}\sqrt{L/g})$ and $\tilde{\mathcal{V}}(\tilde{t}) = \mathcal{V}(\tilde{t}\sqrt{L/g}).$} \cite{CMDS14, kandiah2023effect, kandiah2024controlling} (see for example \cite{kandiah2023effect, kandiah2024controlling})
\begin{equation}
\frac{d^2}{d \tilde{t}^2}\begin{pmatrix} \tilde{\mathcal{U}} \\ \tilde{\mathcal{V}} \end{pmatrix} + \Gamma {\bf R} \frac{d}{d \tilde{t}} \begin{pmatrix} \tilde{\mathcal{U}} \\ \tilde{\mathcal{V}} \end{pmatrix} + \begin{pmatrix} \tilde{\mathcal{U}} \\ \tilde{\mathcal{V}} \end{pmatrix} = \begin{pmatrix} 0 \\ 0 \end{pmatrix}, \label{dim1}
\end{equation}
where $\tilde{t}$ is the dimensionless time variable, $\Gamma$ characterises the combined effects of gravity and gyricity as well as the gyropendulum's geometry, while  $\tilde{\mathcal{U}} = \tilde{\mathcal{U}}(\tilde{t})$ and $\tilde{\mathcal{V}} = \tilde{\mathcal{V}}(\tilde{t})$ represent the dimensionless transverse displacement components in the $x$- and $y$-directions, respectively.  The rotation matrix $\bf{R}$ is given in \eq{90rotation}.

The general solution of \eq{dim1} can be written as a linear combination of the normal mode solutions for each case of the parameter $\Gamma$.  We set the initial conditions 
\begin{equation}
\tilde{\mathcal{U}}(0) = \tilde{\mathcal{U}}_{0}, ~~ \tilde{\mathcal{V}}(0) = \tilde{\mathcal{V}}_{0}, ~~ \frac{d\tilde{\mathcal{U}}}{d \tilde{t}}(0) = \dot{\tilde{\mathcal{U}}}_{0}, ~~ \frac{d\tilde{\mathcal{V}}}{d \tilde{t}}(0) = \dot{\tilde{\mathcal{V}}}_{0}, \label{initcon1}
\end{equation}
where $\tilde{\mathcal{U}}_0,\tilde{\mathcal{V}}_0,\dot{\tilde{\mathcal{U}}}_0$ and $\dot{\tilde{\mathcal{V}}}_0$ are given values of the normalised initial displacements and initial velocities. Assuming time-harmonic solutions with the dimensionless radian frequency $\tilde{\omega}$ of the form $\tilde{\mathcal{U}}= \mathpzc{a} \exp(i\tilde{\omega} \tilde{t}),$ $\tilde{\mathcal{V}}= \mathpzc{b} \exp(i\tilde{\omega} \tilde{t}),$ where $\mathpzc{a}$ and $\mathpzc{b}$ are arbitrary constants, and substituting into the system \eq{dim1}, yields the solutions of \eq{dim1}, satisfying the initial conditions \eq{initcon1}, of the form
\begin{gather}
\begin{pmatrix} \tilde{\mathcal{U}} \\ \tilde{\mathcal{V}} \end{pmatrix} = \frac{-\dot{\tilde{\mathcal{V}}}_{0}+\tilde{\omega}_{2} \tilde{\mathcal{U}}_{0}}{\tilde{\omega}_{1}+\tilde{\omega}_{2}} \begin{pmatrix} \cos(\tilde{\omega}_{1}\tilde{t}) \\ -\sin(\tilde{\omega}_{1}\tilde{t}) \end{pmatrix} + \frac{\dot{\tilde{\mathcal{U}}}_{0} + \tilde{\omega}_{2}\tilde{\mathcal{V}}_{0}}{\tilde{\omega}_{1} + \tilde{\omega}_{2}}\begin{pmatrix} \sin(\tilde{\omega}_{1}\tilde{t}) \\ \cos(\tilde{\omega}_{1}\tilde{t})\end{pmatrix}  \nonumber \\ +\frac{\dot{\tilde{\mathcal{V}}}_{0}+\tilde{\omega}_{1}\tilde{\mathcal{U}}_{0}}{\tilde{\omega}_{1} + \tilde{\omega}_{2}} \begin{pmatrix} \cos( \tilde{\omega}_{2}\tilde{t}) \\ \sin(\tilde{\omega}_{2}\tilde{t}) \end{pmatrix} + \frac{-\dot{\tilde{\mathcal{U}}}_{0}+\tilde{\omega}_{1}\tilde{\mathcal{V}}_{0}}{\tilde{\omega}_{1} + \tilde{\omega}_{2}} \begin{pmatrix} -\sin(\tilde{\omega}_{2} \tilde{t}) \\ \cos(\tilde{\omega}_{2}\tilde{t}) \end{pmatrix}, \label{messsi}
\end{gather}
where 
\begin{eqnarray}
\tilde{\omega}_1 = \frac{1}{2}\Big(-\Gamma+\sqrt{\Gamma^2+4}\Big), ~~ \tilde{\omega}_2  = \frac{1}{2}\Big(\Gamma+\sqrt{\Gamma^2+4}\Big). \label{ogs}
\end{eqnarray}

The analysis presented in \cite{CMDS14, mathematicstoday} shows that the transverse displacements of a gyropendulum can be modelled by the trajectory of a fixed point on the circumference of a circle as the circle rolls along another fixed circle without slipping. A discussion into the formation of polygonal shapes in the trajectory of a gyropendulum, resulting from perturbations of the fundamental circular motion, are provided in \cite{kandiah2023effect, kandiah2024controlling, mathematicstoday}.

\subsection{Approximate polygonal trajectories}\label{gyropolygo}

The solutions \eq{messsi} can be written in the complex form \cite{kandiah2023effect} 
\begin{equation}
z(\tilde{t}) = c_{1} e^{-i \tilde{\omega}_1 \tilde{t}} + c_{2} e^{-i \tilde{\omega}_2 \tilde{t}}, 
\end{equation}
where $z(\tilde{t}) = \tilde{\mathcal{U}}(\tilde{t}) + i \tilde{\mathcal{V}}(\tilde{t})$ and $c_{j},$ for $j=1,2,$ are complex constants. As discussed in \cite{kandiah2023effect}, when $\tilde{\omega}_2$ is chosen as an integer multiple of $\tilde{\omega}_1,$ polygonal trajectories of the gyropendulum can be obtained. In particular, requiring that the ratio $\tilde{\omega}_2/\tilde{\omega}_1=n-1,$ where $n\geq 3$ is the number of sides of the polygon, results in a gyropendulum motion approximating a regular $n$-sided polygon for the appropriately chosen initial conditions. This results in the following value of the quantity $\Gamma$ associated with the approximate polygonal trajectory:
\begin{equation}
\Gamma_{n} = \frac{n-2}{\sqrt{n-1}}. \label{gamos}
\end{equation}
We also note that approximate polygonal trajectories of the gyropendulum can be obtained for $\Gamma<0$ with the appropriate initial conditions. However, here we consider the case of $\Gamma>0.$ 
In this case, the initial conditions at the end of the gyropendulum are chosen as \cite{kandiah2023effect} 
\begin{equation}
\tilde{\mathcal{U}}_{0} = \mathcal{R}, ~~ \tilde{\mathcal{V}}_0 = 0, ~~ \dot{\tilde{\mathcal{U}}}_0 = 0, ~~ \dot{\tilde{\mathcal{V}}}_{0} = \frac{\mathcal{R} (n-1) (2-n)}{2+n(n-1)}, \label{initcon1aaa}
\end{equation}
where $\mathcal{R}$ is linked to the size of the approximate polygon. 
\begin{figure}
\centering
\begin{subfigure}{.48\textwidth}
  \centering
  \includegraphics[width=\linewidth]{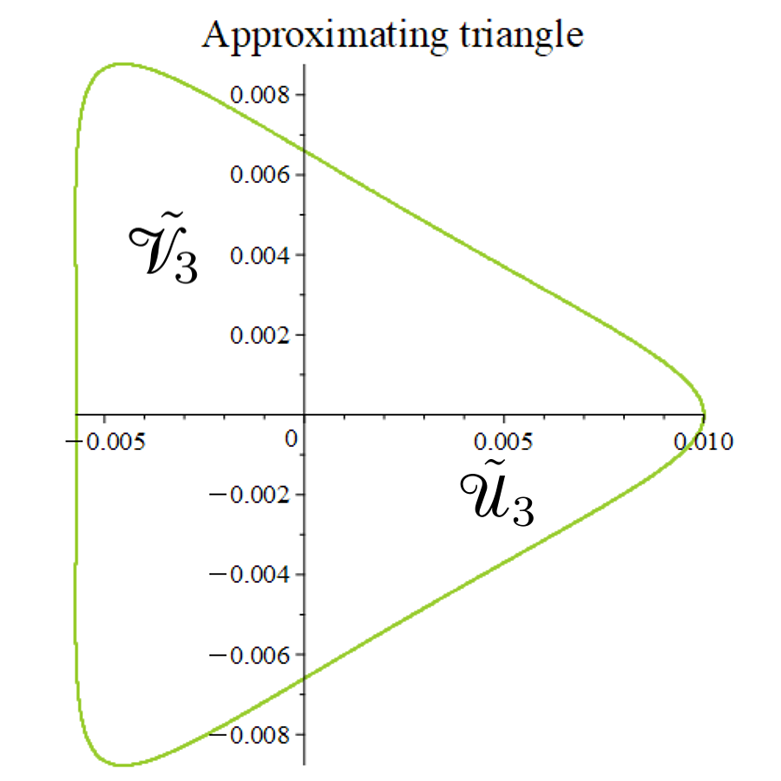}
  \caption{ Approximate triangular trajectory}
\end{subfigure}%
\hspace{0.5cm}
\begin{subfigure}{.48\textwidth}
  \centering
  \includegraphics[width=\linewidth]{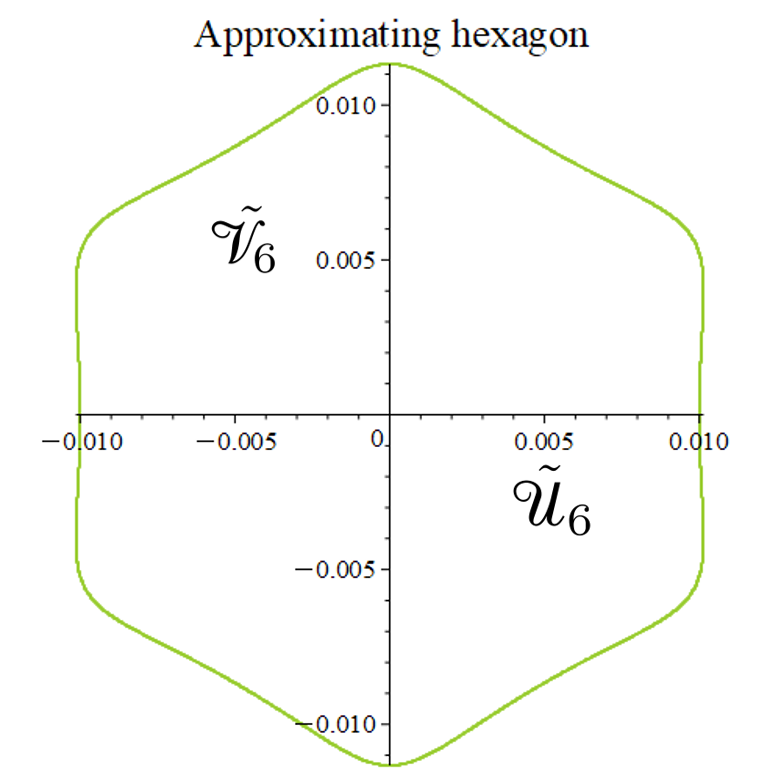}
  \caption{Approximate hexagonal trajectory}
\end{subfigure} 
\caption{\footnotesize The polygonal approximate trajectories of the gyropendulum. (a) The triangle approximation with $n=3,$ $\Gamma_{3}=\sqrt{2}/2,$ $\tilde{\mathcal{U}}_{0}=0.01,$ $\tilde{\mathcal{V}}_0 = 0,$ $\dot{\tilde{\mathcal{U}}}_0 = 0$ and $\dot{\tilde{\mathcal{V}}}_{0}=-0.0025,$ and (b) the hexagon approximation with $n=6$, $\Gamma_{6}=4\sqrt{5}/5,$ $\tilde{\mathcal{U}}_{0}=0.01,$ $\tilde{\mathcal{V}}_0 = 0,$ $\dot{\tilde{\mathcal{U}}}_{0} = 0$ and $\dot{\tilde{\mathcal{V}}}_{0}=-0.00625.$}
\label{changingpogy1a}
\end{figure}

The non-dimensional transverse displacements of the gyropendulum, satisfying the initial conditions \eq{initcon1aaa} and approximating an $n$-sided regular polygon, are then given by 
\begin{eqnarray}
\begin{pmatrix} \tilde{\mathcal{U}}_{n} \\ \tilde{\mathcal{V}}_{n} \end{pmatrix} = \frac{\mathcal{R} ((\tilde{\omega}^{(n)}_1-1)n^2 + (-\tilde{\omega}^{(n)}_1+3)n+2\tilde{\omega}^{(n)}_1-2)}{\tilde{\omega}^{(n)}_1 n (n^2-n+2)} \begin{pmatrix} \cos(\tilde{\omega}^{(n)}_1 (n-1) \tilde{t}) \\ \sin(\tilde{\omega}^{(n)}_1 (n-1) \tilde{t}) \end{pmatrix} \nonumber \\ +\frac{\mathcal{R} (n-1) (n^2 \tilde{\omega}^{(n)}_1 + ( - \tilde{\omega}^{(n)}_1 + 1) n + 2\tilde{\omega}^{(n)}_1 - 2)}{\tilde{\omega}^{(n)}_1 n (n^2-n+2)} \begin{pmatrix} \cos(\tilde{\omega}^{(n)}_1 \tilde{t}) \\ -\sin(\tilde{\omega}^{(n)}_1  \tilde{t}) \end{pmatrix},
\label{dispo12}
\end{eqnarray}
where $\tilde{\omega}^{(n)}_1$ is defined according to the quantity $\Gamma_{n}$ as follows 
\begin{equation}
\tilde{\omega}^{(n)}_1 = \frac{1}{2}\Big(-\Gamma_{n}+\sqrt{\Gamma_{n}^2+4}\Big).
\end{equation}

In Fig. \ref{changingpogy1a}(a) and Fig. \ref{changingpogy1a}(b), we present the approximating polygonal trajectories of the gyropendulum with $\mathcal{R}=0.01$ for a triangular and hexagonal shape, respectively. For the example in Fig. \ref{changingpogy1a}(a), $n=3$ which results in $\tilde{\omega}^{(3)}_1 = \sqrt{2}/2$ and $\Gamma_{3} = \sqrt{2}/2.$ In this case, the transverse displacements are given by 
\begin{equation}
\begin{pmatrix}
\tilde{\mathcal{U}}_{3} \\ \tilde{\mathcal{V}}_{3} \end{pmatrix} = \frac{0.01 (4 - \sqrt{2} )}{12}\begin{pmatrix} \cos(  \tilde{t}\sqrt{2}) \\ \sin(  \tilde{t} \sqrt{2}) \end{pmatrix} + \frac{0.01 (8 + \sqrt{2}) }{12} \begin{pmatrix} \cos( \tilde{t}\sqrt{2}/2) \\ -\sin(\tilde{t}\sqrt{2}/2) \end{pmatrix}, \label{triangulas}
\end{equation}
and the trajectory is periodic with the period $2\sqrt{2}\pi.$ The transverse displacements in Fig. \ref{changingpogy1a}(b) are associated with $n=6,$ resulting in $\tilde{\omega}^{(6)}_1=\sqrt{5}/5$ and $\Gamma_{6}=4\sqrt{5}/5,$ and are given by 
\begin{equation}
\begin{pmatrix}
\tilde{\mathcal{U}}_{6} \\ \tilde{\mathcal{V}}_{6} \end{pmatrix} = \frac{0.01 (8-5\sqrt{5})}{48}\begin{pmatrix} \cos(\tilde{t}\sqrt{5}) \\ \sin(\tilde{t} \sqrt{5}) \end{pmatrix} + \frac{0.01 (40 + 5\sqrt{5})}{48} \begin{pmatrix} \cos(\tilde{t}\sqrt{5}/5) \\ -\sin(\tilde{t}\sqrt{5}/5) \end{pmatrix}. \label{squaras}
\end{equation}
Hence, the period of the approximating hexagonal trajectory is given by $2\pi\sqrt{5}.$ It is noted that the change in the polygonal trace of the gyropendulum from the triangular to the hexagonal trajectory is associated with a change in the parameter $\Gamma$ and a different prescribed initial velocity in the $y$-direction. The same method applies for approximating polygonal shapes of a higher-order symmetry as discussed in \cite{kandiah2023effect}. Moreover, we also note that the polygonal trajectories of the gyropendulum can be obtained for different orientations of the spinner's rotation. 
 The analysis presented in this section demonstrates how the gyropendulum's polygonal motions change with different parameters and initial conditions, providing an approximate comparison with the polygonal formations observed at the poles of rotating planets.

\section{Concluding remarks}\label{concluding6}

 In this paper we have presented an analytical description of a chiral elastic strip under gravity, with either Dirichlet or Neumann boundary conditions, demonstrating both propagating and standing waveforms. We showed that the waveguides in the strip support elastic chiral vortex waves that can be controlled by adjusting the properties of the spinner and the effect of gravity. These characteristics are linked to the dispersion properties of the waveforms. The dynamics of the discrete lattice strip models discussed in the paper can be used as a simple analogy of the high-frequency inertia-gravity guided waves through the equatorial channel. Chiral gravitational waves possess unique directional propagating properties due to their physical chirality, influenced by the presence of spinners and gravity. These properties allowed for the description of eastward- and westward-propagating elastic vortex waves along the discrete lattice strip. In contrast, equatorial atmospheric phenomena occur due to the complex inertial and rotational dynamics near the equator. 
 
This paper also provided an analytical approach to modelling the shallow water equations in narrow bands. The asymptotic method is fully consistent with the boundary conditions of the equatorial band, which allows for a description of the meridionally bounded propagating waves. We established a link between equatorial waves in a continuum and chiral gravitational waves in a discrete structure, with the primary focus on the effects of chirality and gravity in both models. The motion of a single gyropendulum has also been analysed in relation to the approximate planar flow of a fluid ridge. Polygonal pattern formations of the gyropendulum are shown to resemble the flows at the poles of rotating planets, where the gravitational and gyroscopic dynamics influence the fluid motions.

The concepts presented in this paper apply to various physical problems involving discrete and continuous chiral structures subjected to gravity. The mathematical modelling provided here provides key insights into wave dynamics shaped by rotational and gravitational forces. These methods and findings are directly relevant for studying the mechanisms governing atmospheric and oceanic flows.

\vspace{.2in}

{\bf Acknowledgments.} A.K. gratefully acknowledges the financial support of the EPSRC through the Mathematics DTP grant EP/V52007X/1, project reference 2599756.

\end{document}